\documentclass[onecolumn,twoside]{IEEEtran}
\usepackage[T1]{fontenc}
\usepackage[latin9]{inputenc}
\usepackage{amsthm}
\usepackage{amsmath} 
\usepackage{graphicx} 
\usepackage{color} 
\usepackage{cite}
\usepackage{algpseudocode}
\usepackage{algorithm}
\usepackage{amssymb}
\usepackage{subfigure}
\usepackage{esint}
\usepackage{algpseudocode}
\usepackage{epstopdf}
\usepackage{multirow}
\usepackage{makecell}
\usepackage{float}
\usepackage{lipsum}
\usepackage{balance}
\usepackage{bbm}
\usepackage{hyperref}

\newtheorem{definition}{Definition}
\newtheorem{remark}{Remark}

\newtheorem{example}{Example}

\theoremstyle{plain}
\theoremstyle{plain}
\newtheorem{theorem}{Theorem}

\newcommand{\comment}[1]{}

\IEEEoverridecommandlockouts

\begin{document}

\title{The Secret Arithmetic of Patterns: A General Method for Designing Constrained Codes Based on~Lexicographic Indexing}

\author{
   \IEEEauthorblockN{Ahmed Hareedy, \IEEEmembership{Member, IEEE}, Beyza Dabak, and Robert Calderbank, \IEEEmembership{Fellow, IEEE}}\vspace{-1.0em}
   
   \thanks{The authors are with the Department of Electrical and Computer Engineering, Duke University, Durham, NC 27708 USA (e-mail: ahmed.hareedy@duke.edu; beyza.dabak@duke.edu; robert.calderbank@duke.edu). This research was supported in part by NSF under Grant CCF 1717602 and in part by AFOSR under Grant FA 9550-17-1-0291.}
}
\maketitle

\begin{abstract}
Constrained codes are used to prevent errors from occurring in various data storage and data transmission systems. They can help in increasing the storage density of magnetic storage devices, in managing the lifetime of electronic storage devices, and in increasing the reliability of data transmission over wires. Over the years, designing practical (complexity-wise) capacity-achieving constrained codes has been an area of research gaining significant interest. We recently designed various constrained codes based on lexicographic indexing. We introduced binary symmetric lexicographically-ordered constrained (S-LOCO) codes, $q$-ary asymmetric LOCO (QA-LOCO) codes, and a class of two-dimensional LOCO (TD-LOCO) codes. These families of codes achieve capacity with simple encoding and decoding, and they are easy to reconfigure. We demonstrated that these codes can contribute to notable density and lifetime gains in magnetic recording (MR) and Flash systems, and they find application in other systems too. In this paper, we generalize our work on LOCO codes by presenting a systematic method that guides the code designer to build any constrained code based on lexicographic indexing once the finite set of data patterns to forbid is known. In particular, we connect the set of forbidden patterns directly to the cardinality of the LOCO code and most importantly to the rule that uncovers the index associated with a LOCO codeword. By doing that, we reveal the secret arithmetic of patterns, and make the design of such constrained codes significantly easier. We give examples illustrating the method via codes based on lexicographic indexing from the literature. We then design optimal (rate-wise) constrained codes for the new two-dimensional magnetic recording (TDMR) technology. Over a practical TDMR model, we show notable performance gains as a result of solely applying the new codes. Moreover, we show how near-optimal constrained codes for TDMR can be designed and used to further reduce complexity and error propagation. All the newly introduced LOCO codes are designed using the proposed general method, and they inherit all the desirable properties in our previously designed LOCO codes.
\end{abstract}


\section{Introduction}\label{sec_intro}

In 1948, Shannon was the first to represent an infinite sequence in which certain data patterns are not allowed by a finite-state transition diagram (FSTD) \cite{shan_const}. He also used Perron-Frobenius theorem \cite{perron_frobenius} to introduce the notion of capacity, which is the highest achievable rate of a code constrained by forbidding certain patterns, as the graph entropy of the FSTD. As was the case with his result on error-correcting codes, Shannon was so far ahead of his time that his result on constrained codes stayed away from the spotlight until the late 1960s and the early 1970s. By that time, Tang and Bahl \cite{tang_bahl} and Franaszek \cite{franaszek} were among a number of researchers who introduced an important family of constrained codes, named run-length-limited (RLL) codes. Since then, a plethora of research works have investigated constrained codes and their applications.

Mass data storage started with magnetic recording (MR) devices, i.e., hard disk drives (HDDs). Early MR devices adopted peak detection to read the data, where inter-symbol interference (ISI) due to insufficient separation between consecutive transitions is a principal source of error. Binary RLL codes offer control over the minimum and maximum separation between consecutive $1$'s in a stream of bits \cite{tang_bahl}. Associated with transition-based signaling, where a $0$ results in no transition while a $1$ results in a transition ($-$ to $+$ or $+$ to $-$), binary RLL codes can be used to control the separation between consecutive transitions \cite{tang_bahl}. IBM capitalized on this observation, and employed RLL codes in their early HDDs in the 1970s and the 1980s to notably increase the storage density by mitigating ISI and also to maintain self-calibration of the system \cite{siegel_mr, immink_surv}. Modern one-dimensional MR devices adopt sequence detection to read the data, and their underlying channels are modeled as partial-response (PR) channels with certain PR equalization targets \cite{vasic_prc, ahh_prc, ahh_scmr}. Constrained codes are still being employed in these modern MR systems to improve performance and increase density \cite{siegel_const, ahh_loco}.

The introduction of Flash memory by Toshiba in the 1980s as a form of electronic storage eventually changed the landscape of mass data storage since electronic storage is notably faster, albeit more expensive, than magnetic storage. Flash memory devices are currently winning the storage density competition against MR devices. In Flash memory systems, parasitic capacitances within and across floating-gate transistors result in charge propagation during the programming phase \cite{lee_ici}. This charge propagation in turn results in inter-cell interference (ICI), which is a principal source of error in Flash memory systems. Various works introduced constrained codes to forbid data patterns resulting in having an unprogrammed/erased cell surrounded by two adjacent cells programmed to the highest charge level \cite{qin_flash, ravi_const}. More recent research demonstrated that even patterns resulting in the middle cell being programmed to a level less than the highest in the same setup should be forbidden \cite{veeresh_mlc, chee_qlc}. As the Flash device ages, the set of forbidden patterns gets bigger as charges can propagate across non-adjacent cells \cite{ahh_aloco, ahh_qaloco}. Constrained codes for Flash memories are typically associated with level-based signaling, where a codeword symbol is translated to a physical level in the system, e.g., a charge level in Flash.\footnote{The word ``symbols'' subsumes binary ``bits'' when discussing a generic code.} Constrained codes can improve the performance and the lifetime management of Flash devices \cite{veeresh_mlc, ahh_qaloco}.

Two-dimensional magnetic recording (TDMR) \cite{wood_tdmr, chan_tdmr, shayan_tdmr} is a technology that enables magnetic storage to stay competitive with respect to storage density. In TDMR, down (horizontal) tracks are squeezed and are not isoltaed from each other \cite{shayan_tdmr}, which enables a storage density of $10$ terabits per square inch \cite{wood_tdmr, victora_10tb}. This density is about double the maximum achievable density via modern one-dimensional magnetic recording devices \cite{seagate}. Consider a $3 \times 3$ grid in a TDMR system; data patterns resulting in isolating the bit at the center of this grid should be forbidden \cite{mohsen_tdmr}. In particular, patterns having a bit surrounded by $8$ complements--we call them square isolation (SIS) patterns--should be forbidden. Moreover, patterns having a bit surrounded by $4$ complements after ignoring the bits at the $4$ corners--we call them plus isolation (PIS) patterns--should also be forbidden. These patterns significantly exacerbate two-dimensional interference (along down (horizontal) track and cross (vertical) track directions). Preventing the PIS patterns is also called the no-isolated-bit (NIB) constraint \cite{mohsen_tdmr, sharov_TCon}. In the literature, there are works about two-dimensional RLL codes \cite{sharov_TCon, halevy_TD} and other works about two-dimensional (TD) codes preventing isolation patterns \cite{mohsen_tdmr, pituso_tdmr, bd_tdmr}, which offer notably higher code rates for TDMR systems.

There are also other applications for constrained codes in data storage and data transmission. In data storage, constrained codes find application in optical recording systems \cite{immink_opt}. In data transmission, constrained codes are used to mitigate crosstalk between wires or through-silicon vias (TSVs) in integrated circuits \cite{sridhara_ctalk}. Additionally, constrained codes are used in standards such as the universal serial buses (USB) and the peripheral component interconnect express (PCIe) \cite{walker_66, saade_comp}. The primary goals of constrained codes in these standards are to suppress power at frequency zero (at DC), i.e., achieve balancing, and maintain self-calibration of the system, i.e., achieve self-clocking.

Even though designing constrained block codes based on lexicographic indexing started with Tang and Bahl in 1970 \cite{tang_bahl}, the coding theory community deviated from this approach and invested more into designing these codes based on finite-state machines (FSMs). Franaszek introduced the method of state-sequence coding to design FSM-based constrained codes \cite{franaszek}, and many researchers adopted similar ideas in the following years. In 1983, Adler, Coppersmith, and Hassner introduced the method of state splitting and merging, which provided a systematic approach to convert an FSTD into an encoding-decoding FSM of a constrained code \cite{ach_fsm}. Examples of FSM-based constrained codes can be found in \cite{siegel_mr}, \cite{immink_surv}, and \cite{siegel_const}. FSM-based constrained codes typically have a notable gap to capacity, and designing FSM-based codes with rates $\geq 0.9$ gets quite complicated and requires massive storage. Recently, researchers started to look again into lexicographic indexing, especially with the quite high constrained-code rates achievable for modern Flash and TDMR systems \cite{chee_qlc, ahh_qaloco, bd_tdmr}. In 1973, Cover introduced an important result for indexing a sequence within a set of lexicographically-ordered sequences \cite{cover_lex}. Later, this result inspired Immink and others to design enumerative constrained codes \cite{immink_lex, braun_lex}. This result of Cover will play a fundamental role in the general method for designing constrained codes based on lexicographic indexing we present in this paper.

In 2019, we presented binary symmetric lexicographically-ordered constrained (S-LOCO) codes \cite{ahh_loco} to control the separation between consecutive transitions, and thus mitigate ISI and prevent short pulses in MR systems. By protecting only the parity bits of a high performance spatially-coupled (SC) low-density parity-check (LDPC) code designed as in \cite{ahh_scmr}, we showed significant density gains over a practical MR model, with quite limited rate loss \cite{ahh_loco}. Next, we designed binary asymmetric LOCO (A-LOCO) codes \cite{ahh_aloco} and then $q$-ary asymmetric LOCO (QA-LOCO) codes \cite{ahh_qaloco} to minimize charge propagation, and thus mitigate ICI among adjacent and non-adjacent cells, in Flash systems. QA-LOCO codes can contribute to the protection of a Flash device having $q \geq 4$ levels per cell with $< 5\%$ redundancy \cite{ahh_qaloco}. Furthermore, we recently introduced non-binary LOCO codes as TD-LOCO codes for TDMR systems to prevent SIS patterns, and thus enhance the reliability of the device \cite{bd_tdmr}. The idea of all our LOCO codes can be briefly summarized as follows: include all the codewords satisfying a constraint, enumerate them recursively, then find a bijective rule to relate a codeword to its lexicographic index, which we call the encoding-decoding rule.\footnote{For brevity, we call all our constrained codes LOCO codes upon collectively discussing them.} The encoding-decoding rule, which is the core of the code design, allows us to have all the codewords, yet avoid using lookup tables. At most two codewords are then removed from a LOCO code to satisfy self-clocking, which means LOCO codes are capacity-achieving. The encoding-decoding rule is just a summation over cardinalities, which implies simplicity. This property itself also means the codes are reconfigurable; that is, the same hardware can be used to support multiple LOCO codes if the right cardinalities are used as inputs, which is quite helpful to manage the device lifetime. All our LOCO codes are associated with level-based signaling, and more details about them can be found in \cite{ahh_loco}, \cite{ahh_qaloco}, and \cite{bd_tdmr}. The power spectral analysis of S-LOCO and A-LOCO codes can be found in \cite{jes_psd}.

In this paper, we present a general systematic method for designing constrained codes based on lexicographic indexing, i.e., for designing families of LOCO codes. The method works for any one-dimensional finite set of forbidden patterns, and it directly relates this set to the encoding-decoding procedures. In particular, we start with partitioning the codewords of a LOCO code into groups based on the forbidden patterns, and using these groups, we obtain the recursive formula of cardinality (size). Next, and given the forbidden patterns, we determine different cases of existence of a non-zero LOCO codeword symbol according to the adjacent, more significant symbols. For each of these cases, we specify the contribution of a non-zero symbol of a codeword to the index of the codeword through cardinalities via the result of Cover in \cite{cover_lex}, i.e., we derive the encoding-decoding rule of the code. Once the rule is identified, writing the encoding and decoding algorithms becomes a simple task. The details of this direct link between the set of forbidden patterns and the encoding-decoding procedures are~the secret arithmetic of patterns in constrained codes. We illustrate how the method works on two example codes from the literature: binary lexicographically-ordered RLL (LO-RLL) codes and binary symmetric LOCO (S-LOCO) codes.

Moreover, we present new LOCO codes to enhance the performance of TDMR devices. We adopt a TDMR model where the read head is wide, and thus it reads data from three adjacent down tracks simultaneously \cite{chan_tdmr, shayan_tdmr, bd_tdmr}. The codes we present are non-binary LOCO codes, associated with certain mapping-demapping and level-based signaling, designed to prevent the SIS and PIS patterns. While there are efficient TD constrained codes in the literature \cite{mohsen_tdmr, sharov_TCon, halevy_TD, pituso_tdmr}, they either are not customized for TDMR systems, are not systematic, or do not exploit the nature of wide read heads. First, we introduce optimal codes, with respect to the rate, that prevent the SIS and PIS patterns, and we call them optimal square LOCO (OS-LOCO) and optimal plus LOCO (OP-LOCO) codes. We demonstrate notable performance gains by applying OS-LOCO and OP-LOCO codes over a practical TDMR model that is designed based on \cite{mohsen_tdmr}. Next, we show how to further reduce complexity and also error propagation by designing coding schemes that incur a minor capacity loss for the same purpose. We call the codes adopted by these schemes near-optimal square LOCO (NS-LOCO) and near-optimal plus LOCO (NP-LOCO) codes. All the new LOCO codes we propose for TDMR are simple and reconfigurable.\footnote{NS-LOCO codes are the same codes we recently introduced in \cite{bd_tdmr}. The main modification here is that we simplify their encoding-decoding rule via applying the proposed general method.} The new optimal (resp., near-optimal) LOCO codes are capacity-achieving (resp., capacity-approaching) with respect to the system constraint.

The rest of the paper is organized as follows. In Section~\ref{sec_gen}, we describe the new general method in steps. In Section~\ref{sec_exist}, we provide examples from the literature to illustrate the method. In Section~\ref{sec_opt}, we present our optimal constrained codes for TDMR. In Section~\ref{sec_gains}, we show the performance gains in a practical TDMR system. In Section~\ref{sec_compl}, we introduce our near-optimal codes to further reduce complexity. In Section~\ref{sec_conc}, we conclude the paper.

\section{Steps of the General Method}\label{sec_gen}

In this section, we describe the general, systematic method we propose for designing constrained codes based on lexicographic indexing. The method directly links the set of forbidden patterns to the encoding-decoding procedures. We start off with the list of steps, and then we discuss them in detail. \textbf{The steps of our general method are:}
\begin{enumerate}
\item \textit{Use the forbidden patterns to determine a group structure for the code.}
\item \textit{Derive the code cardinality formula using the inherent recursion of the groups and subgroups.}
\item \textit{Specify the codeword patterns that represent special cases given the forbidden patterns.}
\item \textit{Find the contribution of a non-zero codeword symbol to the codeword index in each special/typical case.}
\item \textit{Merge the contributions for all cases in one index equation, which is the encoding-decoding rule.}
\item \textit{Develop the encoding and decoding algorithms of the code based on this rule.}
\end{enumerate}

In order to discuss these six steps in detail, we first introduce some notation. Let $\mathcal{T}$ be a finite set of forbidden patterns. Denote a Galois field of size $q \geq 2$ by GF$(q)$, with $\alpha$ being a primitive element of GF$(q)$. Thus,
\begin{equation}\label{eqn_gfq}
\textup{GF}(q) \triangleq \{0, 1, \alpha, \alpha^2, \dots, \alpha^{q-2}\}.
\end{equation}
In the binary case, $q=2$ and $\textup{GF}(2) = \{0, 1\}$. Let $c$ be a symbol in GF$(q)$. The \textit{integer level-equivalent} of $c$ is $\mathcal{L}(c)$, which is the index of the actual level after signaling is applied, and it is defined as follows: $\mathcal{L}(c) \triangleq \textup{gflog}_\alpha(c) + 1$ if $c \neq 0$, and $\mathcal{L}(c) \triangleq 0$ if $c = 0$. The function $\textup{gflog}_\alpha(\cdot)$ returns the power of the GF element in its argument. The level-equivalent of $\textup{GF}(q)$ is $\{0, 1, 2, 3, \dots, q-1\}$.

A set of sequences is said to be lexicographically ordered if its sequences are ordered ascendingly following the rule $0 < 1 < \alpha < \alpha^2 < \dots < \alpha^{q-2}$ and the symbol significance gets smaller from left to right. In particular, for the two distinct sequences $\bold{c}_u$ and $\bold{c}_v$, we say that $\bold{c}_u < \bold{c}_v$, and thus $\bold{c}_u$ is ordered before $\bold{c}_v$, if at the first position starting from the left where the two sequences differ, the symbol of $\bold{c}_u$ is less than the symbol of $\bold{c}_v$.

We also define a generic LOCO code.

\begin{definition}\label{def_genloco}
A generic lexicographically-ordered $\mathcal{T}$-constrained code, or in short a generic LOCO code, $\mathcal{C}^q_m$ with $q \geq 2$ and $m \geq 1$ is defined by the following properties:
\begin{enumerate}
\item Codewords in $\mathcal{C}^q_m$ are defined over GF$(q)$ and are of length $m$ symbols.
\item Codewords in $\mathcal{C}^q_m$ are ordered lexicographically.
\item A codeword in $\mathcal{C}^q_m$ does not have any pattern in $\mathcal{T}$.
\item All codewords satisfying the above properties are included.
\end{enumerate}
\end{definition}

Different families of LOCO codes can be reached according to the set of forbidden patterns $\mathcal{T}$. Let $N_q(m)$ be the cardinality of the LOCO code $\mathcal{C}^q_m$. Define a codeword $\bold{c}$ in $\mathcal{C}^q_m$ as follows: $\bold{c} \triangleq c_{m-1} c_{m-2} \dots c_0$, with $c_i = z'$ for $i \geq m$, where $z'$ represents out of codeword bounds. The integer level-equivalent of a LOCO codeword symbol $c_i$, $0 \leq i \leq m-1$, is $a_i$, i.e., $a_i \triangleq \mathcal{L}(c_i)$. Denote the lexicographic index of a codeword $\bold{c}$ among all codewords in the LOCO code  $\mathcal{C}^q_m$ by $g_q(m,\bold{c})$, which is typically shorthanded to $g(\bold{c})$ when the context is clear. In general, $g(\bold{c})$ is in $\{0, 1, \dots, N_q(m)-1\}$.

The principal goal of the above six steps is to find a formula for the lexicographic index $g(\bold{c})$ as a function of the codeword symbols and code cardinalities, which is the coding rule. This is what we already did for certain codes recently \cite{ahh_loco, ahh_aloco, ahh_qaloco, bd_tdmr}, but now we aim at doing it in a systematic and general way. Next, we discuss the six steps in detail.\vspace{+0.7em}

\textbf{Step~1) Group Structure:} We first partition the LOCO codewords of $\mathcal{C}^q_m$ into groups based on the set of forbidden patterns $\mathcal{T}$. For ease of analysis, we adopt contiguous partitioning. That is, all codewords belonging to the same group have consecutive lexicographic indices. Let $n_{\textup{f}} \triangleq \vert \mathcal{T} \vert$, and consider the following generic form of $\mathcal{T}$:
\begin{equation}\label{eqn_forbset}
\mathcal{T} \triangleq \{t_{j,p_j-1} t_{j,p_j-2} \dots t_{j,0} \textup{ } | \textup{ } 1 \leq j \leq n_{\textup{f}}\}.
\end{equation}
Each forbidden pattern of index $j$, $1 \leq j \leq n_{\textup{f}}$, is of length $p_j$, $1 \leq p_j \leq m$. For simplicity, we refer to both groups and subgroups as groups in this discussion, and we assume that $m$ is greater than or equal to the maximum length of a pattern in $\mathcal{T}$. We order the forbidden patterns ascendingly according to their lengths, and we access them one by one. For each forbidden pattern $t_{j,p_j-1} t_{j,p_j-2} \dots t_{j,0}$ of index $j$, for all $j$ according to the order, initial groups are specified as follows:
\begin{itemize}
\item There is an initial group having all the codewords starting with $c_{m-1} c_{m-2} \dots c_{m-p_j+1} = t_{j,p_j-1} t_{j,p_j-2} \dots t_{j,1}$ and $c_{m-p_j} \neq t_{j,0}$ from the left, i.e., at their left-most symbol (LMS).
\item There is an initial group or more (only for non-binary) having all the codewords starting with $c_{m-1} c_{m-2} \dots c_{m-p_j+2} \allowbreak = t_{j,p_j-1} t_{j,p_j-2} \dots t_{j,2}$ and $c_{m-p_j+1} \neq t_{j,1}$ from the left, i.e., at their LMSs.
\item There is an initial group or more having all the codewords starting with $c_{m-1} c_{m-2} \dots c_{m-p_j+3} = t_{j,p_j-1} t_{j,p_j-2} \dots t_{j,3}$ and $c_{m-p_j+2} \neq t_{j,2}$ from the left, i.e., at their LMSs.
\item This procedure continues for the rest of symbols in the forbidden pattern until its LMS. For this symbol, there is an initial group or more having all the codewords starting with $c_{m-1} \neq t_{j,p_j-1}$ from the left, i.e., at their LMS.
\end{itemize}

In order to determine the final groups, we need to collectively process these initial groups. If the LMSs of an initial group contain a forbidden pattern in $\mathcal{T}$, this group will be eliminated. If the LMSs defining an initial group appear in one or more other initial groups that have more LMSs defining them, the first group will also be eliminated. In other words, we go with the finer partitioning. If two or more initial groups specified through different patterns end up being identical, only one of them will be left. Handling new forbidden patterns may result in splitting an initial group into multiple final groups. Multiple initial groups forming a contiguous list, with respect to the lexicographic indices of their codewords, can be merged into one final group. Many checks can be performed during the execution of the procedure. The remaining groups at the end are the final ones. We will provide detailed examples in the next section.\vspace{+0.7em}

\textbf{Step~2) Code Cardinality:} Next, we use the (final) group structure determined in Step~1 to derive the cardinality of the code $\mathcal{C}^q_m$, which is $N_q(m)$. Observe that the codewords in different groups have a recursive nature. In other words, if some or all of the LMSs the codewords in a specific group start with are eliminated, the resulting sequence is itself another LOCO codeword in $\mathcal{C}^q_{m'}$, $m' < m$, satisfying the same constraint.

Let $N_{q,i}(m)$ be the cardinality of the group of codewords in $\mathcal{C}^q_m$ that is indexed by $i$. Let also the total number of groups be $n_{\textup{g}}$ and the maximum length of a forbidden pattern be $p_{\textup{max}}$. The typical way of deriving $N_q(m)$ is to express the cardinality of each group as a linear combination with coefficients $\zeta_{\ell,i}$, for all possible $\ell$ and $i$, using recursion as follows:
\begin{equation}\label{eqn_gencard1}
N_{q,i}(m) = \sum_{\ell=1}^{p_{\textup{max}}} \zeta_{\ell,i} N_q(m-\ell).
\end{equation}
Consequently, we get:
\vspace{-0.1em}\begin{equation}\label{eqn_gencard2}
N_q(m) = \sum_{i=1}^{n_{\textup{g}}} N_{q,i}(m) = \sum_{i=1}^{n_{\textup{g}}} \sum_{\ell=1}^{p_{\textup{max}}} \zeta_{\ell,i} N_q(m-\ell) = \sum_{\ell=1}^{p_{\textup{max}}} \zeta'_{\ell} N_q(m-\ell),
\end{equation}
which means $N_q(m)$ is a linear combination of the cardinalities of LOCO codes with lengths smaller than $m$. Observe that the coefficient $\zeta'_{\ell}$ can be zero or negative for certain cardinalities depending on $\mathcal{T}$ \cite{ahh_loco, ahh_qaloco}. Observe also that the maximum value of $\ell$ determines how much we need to go back, i.e., for smaller lengths, to get $N_q(m)$, the cardinality of $\mathcal{C}^q_m$. As expected, this maximum value is the maximum length of a forbidden pattern, which is $p_{\textup{max}}$.

There are also some ideas that can be helpful in deriving $N_q(m)$ in certain cases:
\begin{itemize}
\item In cases where the LOCO code is symmetric, i.e., the number of codewords starting with any symbol in GF$(q)$ from the left is the same for all symbols, it is useful to just derive the cardinality of the group of codewords starting with $0$ (for example) from the left, and multiply by $q$ to find $N_q(m)$.
\item In cases where $q$ is more than $2$, it can be useful to divide groups into subgroups, and find the cardinalities of groups via their subgroups. If a group is symmetric, this can also simplify the calculations.
\item In cases where the normalized capacity approaches $1.00$ (very few or lengthy patterns are forbidden), it might be useful to combine different groups together, and then find $N_q(m)$ by subtracting the number of eliminated sequences from the much bigger cardinality of the less restrictive case.
\end{itemize}

Finally in Step~2, we determine the defined cardinalities, i.e., cardinalities that cannot be derived using the group structure. Some defined cardinalities are directly known or obvious given $\mathcal{T}$. For example, if the shortest forbidden pattern in $\mathcal{T}$ is of length $3$, it is known that $N_q(1) = q$ and also $N_q(2) = q^2$ regardless from the constraint itself. The unknown defined cardinalities, e.g., $N_q(0)$, are obtained from the known cardinalities, the known group cardinalities, and (\ref{eqn_gencard2}).

\begin{remark}
The length $m$ of the LOCO code starting from which the group structure is defined is the smallest length at which different groups can be distinguished. The cardinalities at all lengths smaller than that length are defined cardinalities.
\end{remark}\vspace{+0.2em}

\textbf{Step~3) Special Cases:} The contribution of a LOCO codeword symbol $c_i$ to the overall codeword index $g(\bold{c})$ depends on $c_i$ and the preceding symbols. Here, we specify the different cases of existence of a codeword symbol based on such preceding symbols in order to calculate the symbol contribution in the following step.

Consider the set of forbidden patterns $\mathcal{T}$ in (\ref{eqn_forbset}). We again access the forbidden patterns one by one. For each forbidden pattern $t_{j,p_j-1} t_{j,p_j-2} \dots t_{j,0}$ of index $j$, for all $j$, initial special cases are specified as follows:
\begin{itemize}
\item The case that $c_{i+p_j-1} c_{i+p_j-2} \dots c_{i+1} = t_{j,p_j-1} t_{j,p_j-2} \dots t_{j,1}$ and $c_{i} > t_{j,0}$ according to the lexicographic ordering definition, i.e., $\mathcal{L}(c_{i}) > \mathcal{L}(t_{j,0})$, represents an initial special case unless $t_{j,0} = \alpha^{q-2}$ ($1$ in binary).
\item The case that $c_{i+p_j-2} c_{i+p_j-3} \dots c_{i+1} = t_{j,p_j-1} t_{j,p_j-2} \dots t_{j,2}$ and $c_{i} > t_{j,1}$, i.e., $\mathcal{L}(c_{i}) > \mathcal{L}(t_{j,1})$, represents an initial special case unless $t_{j,1} = \alpha^{q-2}$ ($1$ in binary).
\item The case that $c_{i+p_j-3} c_{i+p_j-4} \dots c_{i+1} = t_{j,p_j-1} t_{j,p_j-2} \dots t_{j,3}$ and $c_{i} > t_{j,2}$, i.e., $\mathcal{L}(c_{i}) > \mathcal{L}(t_{j,2})$, represents an initial special case unless $t_{j,2} = \alpha^{q-2}$ ($1$ in binary).
\item This procedure continues until the symbol to the right of the LMS of the pattern. For this symbol, the case that $c_{i+1} = t_{j,p_j-1}$ and $c_{i} > t_{j,p_j-2}$, i.e., $\mathcal{L}(c_{i}) > \mathcal{L}(t_{j,p_j-2})$, represents an initial special case unless $t_{j,p_j-2} = \alpha^{q-2}$ ($1$ in binary).
\end{itemize}

The case that $c_i$ and the preceding symbols (if any) do not satisfy any of the above conditions in the initial special cases for all patterns in $\mathcal{T}$ is called the \textit{typical case}, which is usually the simplest.

Afterwards, we collectively process these initial special cases to reach the final special cases. Initial special cases implying that a forbidden pattern appears on codeword symbols will be either eliminated or modified such that the sequences where a forbidden pattern appears are eliminated. If there are two special cases characterized by two sequences where one of them is a subsequence of the other starting from the right (from $c_i$), the priority will be given to the special case with the longer sequence. The special case with the shorter sequence will be modified accordingly. Handling new patterns may result in merging multiple initial special cases either partially or totally. Many checks can be performed during the execution of the procedure. The resulting cases at the end are the final special cases.

The goal of specifying these special cases is to appropriately capture the effect of eliminating sequences that violate the constraint by containing forbidden patterns on the contribution of the symbol $c_i$ to the overall index $g(\bold{c})$.\vspace{+0.7em}

\textbf{Step~4) Symbol Contribution:} Next, we find the contribution of each symbol $c_i \neq 0$, $m-1 \geq i \geq 0$ to the index of the LOCO codeword $g(\bold{c})$. Obviously, this contribution is always $0$ for $c_i = 0$. We denote this contribution by $g_i(c_i)$. Recall the codeword $\bold{c} \triangleq c_{m-1} c_{m-2} \dots c_0$ in $\mathcal{C}^q_m$. We define $N_{\textup{symb}}(m, \bold{s})$ as the number of LOCO codewords in $\mathcal{C}^q_m$ that start with the sequence $\bold{s}$ from the left.

According to Cover in \cite{cover_lex}, the contribution of a symbol $c_i \neq 0$ of $\bold{c}$ to the index $g(\bold{c})$ is the number of codewords in $\mathcal{C}^q_m$ starting with the same symbols prior to $c_i$, i.e., $c_{m-1} c_{m-2} \dots c_{i+1}$, from the left and preceding the first codeword starting with $c_{m-1} c_{m-2} \dots c_{i+1} c_{i}$ according to the lexicographic ordering. In other words, this contribution is the number of codewords in $\mathcal{C}^q_m$ starting with $c_{m-1} c_{m-2} \dots c_{i+1} c'_{i}$, for all $c'_{i}$ such that $c'_{i} < c_{i}$ according to the lexicographic ordering definition, i.e., for all $c'_{i}$ such that $\mathcal{L}(c'_{i}) < \mathcal{L}(c_{i})$.

Consequently, and using the aforementioned definition of $N_{\textup{symb}}(m, \bold{s})$, we can mathematically formulate the contribution $g_i(c_i)$ as follows:
\begin{equation}\label{eqn_gengi}
g_i(c_i) = \sum_{c'_{i} < c_{i}} N_{\textup{symb}}(m, c_{m-1} c_{m-2} \dots c_{i+1} c'_{i}).
\end{equation}
Now, we can see that the subscript ``symb'' in $N_{\textup{symb}}(m, \bold{s})$ refers to ``symbol'' contribution. Additionally, in the binary case, i.e., $q=2$, (\ref{eqn_gengi}) reduces to:
\begin{equation}\label{eqn_gengibin}
g_i(c_i) = N_{\textup{symb}}(m, c_{m-1} c_{m-2} \dots c_{i+1} 0), \textup{ } c_i = 1.
\end{equation}
More details can be found in \cite{cover_lex}.

Looking from the right, $N_{\textup{symb}}(m, c_{m-1} c_{m-2} \dots c_{i+1} c'_{i})$, for all $c'_{i}$ such that $c'_{i} < c_{i}$, can be seen as the number of LOCO codewords of length $i+1$ in $\mathcal{C}^q_{i+1}$ that can be concatenated from the right to $c_{m-1} c_{m-2} \dots c_{i+1}$ to generate valid LOCO codewords of length $m$ in $\mathcal{C}^q_m$. Let the total number of (final) special cases from Step~3 be $n_{\textup{c}}-1$, i.e., we have $n_{\textup{c}}$ (final) cases in total after adding the typical case. Denote the contribution of symbol $c_i$ to $g(\bold{c})$ in the case indexed by $i_{\textup{c}}$, $1 \leq i_{\textup{c}} \leq n_{\textup{c}}$, by $g_{i, i_{\textup{c}}}(c_i)$. Based on the observation in this paragraph, $g_{i,i_{\textup{c}}}(c_i)$ can then be written as an arithmetic function, particularly a linear combination, of cardinalities of LOCO codes having lengths at most $i+1$ from Step~2. This means:
\begin{equation}\label{eqn_gengi_2}
g_{i, i_{\textup{c}}}(c_i) = \sum_{c'_{i} < c_{i}} N_{\textup{symb}, i_{\textup{c}}}(m, c_{m-1} c_{m-2} \dots c_{i+1} c'_{i}) = \sum_{\ell=0}^{p_{\textup{max}}} \zeta''_{\ell, i_{\textup{c}}} N_q((i+1)-\ell).
\end{equation}
The reason is that the number of codewords in $\mathcal{C}^q_{i+1}$ we are after can be expressed in a way similar to that of $N_q(m)$ in (\ref{eqn_gencard2}). Observe that the coefficient $\zeta''_{\ell, i_{\textup{c}}}$ can be zero or negative for certain cardinalities in certain cases depending on $\mathcal{T}$. Most important, this coefficient $\zeta''_{\ell, i_{\textup{c}}}$ has to be a function of $\mathcal{L}(c_i) \triangleq a_i$ for one of the cardinalities in all cases.

The end product of this step is $g_{i, i_{\textup{c}}}(c_i)$ expressed as a linear combination of cardinalities, as in (\ref{eqn_gengi_2}), for all the $n_{\textup{c}}$ cases.\vspace{+0.7em}

\textbf{Step~5) Encoding-Decoding Rule:} We are now ready to derive the encoding-decoding rule of the LOCO code $\mathcal{C}^q_m$, which uncovers the index $g(\bold{c})$ associated with a codeword $\bold{c}$. First, we need to merge different expressions of $g_{i, i_{\textup{c}}}(c_i)$, $c_i \neq 0$, for all existence cases into one unified expression representing the contribution $g_i(c_i)$ to the index $g(\bold{c})$.

In order to perform such merging, we introduce some \textit{merging variables}. The aim of these merging variables is to switch on the contribution of a specific case and switch off the contributions of all the other cases in the unified expression of $g_i(c_i)$ given $c_i$ and the preceding symbols (which determine the case to switch on). Observe that the number of (final) merging variables to be used can be notably less than $n_{\textup{c}}$. The reason is that it can happen that the symbol contributions for multiple cases end up being the same, especially if the LOCO code is symmetric, resulting in the same merging variable to switch these contributions on/off in the encoding-decoding rule.

Define $f^{\textup{mer}}_{\ell}(\cdot)$ as the \textit{merging function} for the cardinality $N_q((i+1)-\ell)$ in the unified expression representing the contribution $g_i(c_i)$ to the index $g(\bold{c})$ (see (\ref{eqn_gengi_2})). Let the merging variables of symbol $c_i$ be $y_{i,1}, y_{i,2}, \dots, y_{i,n_{\textup{y}}}$, where $n_{\textup{y}}$ is the number of merging variables with $n_{\textup{y}} < n_{\textup{c}}$. It is clear that the arguments of $f^{\textup{mer}}_{\ell}(\cdot)$ are $y_{i,1}, y_{i,2}, \dots, y_{i,n_{\textup{y}}}$ and $\zeta''_{\ell,1}, \zeta''_{\ell,2}, \dots, \zeta''_{\ell, n_{\textup{c}}}$. Note that the merging variables are determined via $c_i$ and its preceding symbols in $\bold{c}$. The unified expression for $g_i(c_i)$ can then be written as follows:
\begin{equation}\label{eqn_gimerg}
g_i(c_i) = \sum_{\ell=0}^{p_{\textup{max}}} f^{\textup{mer}}_{\ell}(y_{i,1}, y_{i,2}, \dots, y_{i,n_{\textup{y}}}, \zeta''_{\ell,1}, \zeta''_{\ell,2}, \dots, \zeta''_{\ell, n_{\textup{c}}}) N_q((i+1)-\ell).
\end{equation}
Once we find $f^{\textup{mer}}_{\ell}(\cdot)$, for all $\ell$, merging is complete, and the encoding-decoding rule of the LOCO code $\mathcal{C}^q_m$ becomes:
\begin{align}\label{eqn_rulegen}
g(\bold{c}) &= \sum_{i=1}^{m-1} g_i(c_i) = \sum_{i=1}^{m-1} \sum_{c'_{i} < c_{i}} N_{\textup{symb}}(m, c_{m-1} c_{m-2} \dots c_{i+1} c'_{i}) \nonumber \\  &= \sum_{i=1}^{m-1} \sum_{\ell=0}^{p_{\textup{max}}} f^{\textup{mer}}_{\ell}(y_{i,1}, y_{i,2}, \dots, y_{i,n_{\textup{y}}}, \zeta''_{\ell,1}, \zeta''_{\ell,2}, \dots, \zeta''_{\ell, n_{\textup{c}}}) N_q((i+1)-\ell).
\end{align}

Equation (\ref{eqn_rulegen}) is a recap of both Step~4 and Step~5, showing how we employ the result of Cover in \cite{cover_lex} to reach the encoding-decoding rule. Further simplifications to (\ref{eqn_rulegen}) can be performed. The procedure discussed in these two steps serves as a direct proof of the rule. Another way to prove the rule is induction, which is what we followed in \cite{ahh_loco}, \cite{ahh_aloco}, and \cite{ahh_qaloco}.\vspace{+0.7em}

\textbf{Step~6) Code Algorithms:} Before we discuss how the encoding and decoding algorithms are developed, we briefly discuss bridging and self-clocking. Bridging is the process of adding few symbols between each two consecutive LOCO codewords written or transmitted in a stream such that forbidden patterns in $\mathcal{T}$ do not appear at the transition from a codeword to the next one. Bridging solely depends on $\mathcal{T}$. Self-clocking is the process of maintaining self-calibration in the system during reading or receiving the data. For a self-clocked LOCO code, long same-symbol sequences in a stream of LOCO codewords after signaling are not allowed. Self-clocking depends on the GF size $q$ and also on the used bridging. Sometimes one or two codewords need to be removed from the LOCO code to achieve self-clocking. Other times, smart bridging suffices to achieve self-clocking, and no codewords need to be removed. We will discuss both situations in this paper.

For simplicity, assume that no codewords are removed from $\mathcal{C}^q_m$ to achieve self-clocking. Thus, the size of messages $\mathcal{C}^q_m$ encodes is $s = \lfloor \log_2 N_q(m) \rfloor$ in binary bits (see also \cite{ahh_qaloco}), where $N_q(m)$ is derived in Step~2. The idea of the encoding and decoding algorithms was first introduced by Tang and Bahl for RLL codes in \cite{tang_bahl}. A conceptually connected idea is in \cite{laroia_const}.

The encoding algorithm executes the reverse procedure of the encoding-decoding rule in (\ref{eqn_rulegen}). For each incoming binary message of length $s$ bits, the encoder performs binary to decimal conversion to get the index $g(\bold{c})$. This index is the initial value of a variable named residual. At each index $i$, $m-1 \geq i \geq 0$, the merging function $f^{\textup{mer}}(\cdot)$ is set-up using the encoded symbols at positions prior to $i$. For convenience, only here we write $g_i(c_i)$ as $g_i(\mathcal{L}(c_i)) = g_i(a_i)$. Next, the variable residual is compared with different values of $g_i(a_i)$, $a_i \in \{1, 2, 3, \dots, q-1\}$, computed according to (\ref{eqn_gimerg}).
\begin{itemize}
\item If $\textup{residual} < g_i(a_i = 1)$, then $c_i$ is encoded as $0$.
\item If $\textup{residual} \geq g_i(a_i = q-1)$, then $c_i$ is encoded as $\alpha^{q-2}$ and residual is reduced by $g_i(a_i = q-1)$.
\item Otherwise, the comparisons go on until the variable residual satisfies:
\begin{equation}\label{eqn_ineqgen}
g_i(a_i) \leq \textup{residual} < g_i(a_i+1).
\end{equation}
In this case, $c_i$ is encoded as $\mathcal{L}^{-1}(a_i)$ and residual is reduced by $g_i(a_i)$, where $\mathcal{L}^{-1}(\cdot)$ is the inverse function of $\mathcal{L}(\cdot)$.
\end{itemize}
This process continues until all codeword symbols are encoded. Then, bridging is performed, and the process is repeated again for the next binary message.

The decoding algorithm executes the same procedure of the encoding-decoding rule in (\ref{eqn_rulegen}). For each incoming codeword of length $m$, the decoder accesses its symbols one by one. The variable temp\_index is initialized by $0$. For each symbol $c_i$, $m-1 \geq i \geq 0$, the merging function $f^{\textup{mer}}(\cdot)$ is set-up using $c_i$ and the symbols preceding $c_i$. Next, the variable temp\_index is increased by $g_i(c_i)$ (or $g_i(a_i)$) computed according to (\ref{eqn_gimerg}). This process continues until all codeword symbols are decoded, and then  temp\_index becomes $g(\bold{c})$. The decoder performs decimal to binary conversion to get the message from the index $g(\bold{c})$. Then, the decoder skips the few bridging symbols, and the process is repeated again for the next codeword.\vspace{+0.7em}

This general method with its six described steps reveals the secret arithmetic of patterns. In particular, the method reveals how the arithmetic of the allowed patterns in a LOCO codeword, which leads to the index of the codeword, is controlled by the forbidden, or unseen, patterns in $\mathcal{T}$. Observe that with the exception of the codewords removed for self-clocking (if any), all LOCO codewords are included in the code. Moreover, the number of symbols used for bridging does not grow with the length $m$. Thus, our LOCO codes are capacity-achieving. Additionally, having an encoding-decoding rule that is just a summation as shown in (\ref{eqn_rulegen}) guarantees both simplicity and reconfigurability. With the right cardinalities used as inputs to the adder, the same hardware can support multiple LOCO codes \cite{ahh_loco, ahh_qaloco}. Machine learning algorithms can be used to reconfigure the constrained coding hardware by collecting errors and learning the changes on the set of patterns to forbid, which contributes to increasing the lifetime of the storage device.

\section{Examples From Existing Codes}\label{sec_exist}

In this section, we provide examples from constrained codes already existing in the literature to illustrate how the general method described in Section~\ref{sec_gen} works. We will apply the steps of the general method one after another, and demonstrate that the end result is the same as what we know from the literature.

We look at LO-RLL codes and S-LOCO codes as examples. Applying the general method for QA-LOCO codes \cite{ahh_qaloco} is left to the interested reader for brevity, and it can build insights regarding merging groups and special cases during the procedures of their respective steps. These insights will also appear in the next section.

\subsection{Lexicographically-Ordered RLL Codes}

Here, we discuss binary $(d,\infty)$ LO-RLL codes introduced in \cite{tang_bahl}, where the constraint is that two $1$'s must be separated by at least $d$ $0$'s. Denote a $(d,\infty)$ LO-RLL code of length $m$ by $\mathcal{RC}^2_{m,d}$. The definition of the code is exactly the definition of a generic LOCO code, which is Definition~\ref{def_genloco}, with $q=2$, $\mathcal{C}^q_m = \mathcal{RC}^2_{m,d}$, and $\mathcal{T}$ given by:
\begin{equation}\label{eqn_tlorll}
\mathcal{T} = \mathcal{R}^2_d \triangleq \{11, 101, 1001, \dots, 1\bold{0}^{d-1}1\},
\end{equation}
where the notation $\bold{w}^r$ refers to a sequence of $r$ consecutive $w$ symbols. Both $\bold{c}$ in $\mathcal{C}^q_m = \mathcal{RC}^2_{m,d}$ and $g(\bold{c})$ are used as they were in Section~\ref{sec_gen} (actually throughout the paper). The cardinality of $\mathcal{RC}^2_{m,d}$ is $N_q(m) = N_2(m, d)$. Table~\ref{table_1} shows multiple $(d,\infty)$ LO-RLL codes with $d = 1$ and $m$ in $\{1, 2, \dots, 5\}$.

\begin{table*}
\caption{All the Codewords of Five Binary $(d,\infty)$ LO-RLL Codes, $\mathcal{RC}^2_{m,1}$, $m \in \{1, 2, \dots, 5\}$. The Two Different Groups of Codewords Are Explicitly Illustrated for the Code $\mathcal{RC}^2_{5,1}$. The Two Groups Can Be Distinguished Starting From $m=1$.}
\vspace{-0.5em}
\centering
\scalebox{1.00}
{
\begin{tabular}{|c|c|c|c|c|c c|}
\hline
\multirow{2}{*}{Codeword index $g(\bold{c})$} & \multicolumn{6}{|c|}{\makecell{Codewords of the code $\mathcal{RC}^2_{m,1}$}} \\
\cline{2-7}
{} & \makecell{$m=1$} & \makecell{$m=2$} & \makecell{$m=3$} & \makecell{$m=4$} & \multicolumn{2}{|c|}{\makecell{$m=5$}} \\
\hline
$0$ & $0$ & $00$ & $000$ & $0000$ & $00000$ & \multirow{8}{*}{Group~1} \\
\cline{1-1}\cline{2-2}
$1$ & $1$ & $01$ & $001$ & $0001$ & $00001$ & \\
\cline{1-1}\cline{2-2}\cline{3-3}
$2$ &  & $10$ & $010$ & $0010$ & $00010$ & \\
\cline{1-1}\cline{3-3}\cline{4-4}
$3$ &  &  & $100$ & $0100$ & $00100$ & \\
\cline{1-1}
$4$ &  &  & $101$ & $0101$ & $00101$ & \\
\cline{1-1}\cline{4-4}\cline{5-5}
$5$ &  &  &  & $1000$ & $01000$ & \\
\cline{1-1}
$6$ &  &  &  & $1001$ & $01001$ & \\
\cline{1-1}
$7$ &  &  &  & $1010$ & $01010$ & \\
\cline{1-1}\cline{5-5}\cline{6-7}
$8$ &  &  &  &  & $10000$ & \multirow{5}{*}{Group~2} \\
\cline{1-1}
$9$ &  &  &  &  & $10001$ & \\
\cline{1-1}
$10$ &  &  &  &  & $10010$ & \\
\cline{1-1}
$11$ &  &  &  &  & $10100$ & \\
\cline{1-1}
$12$ &  &  &  &  & $10101$ & \\
\hline
Code cardinality & $N_2(1, 1) = 2$  & $N_2(2, 1) = 3$ & $N_2(3, 1) = 5$ & $N_2(4, 1) = 8$ & \multicolumn{2}{|c|}{$N_2(5, 1) = 13$} \\
\hline
\end{tabular}}
\label{table_1}
\end{table*}

Now, we will apply the steps of the general method to find out how to encode and decode $(d,\infty)$ LO-RLL codes using a simple encoding-decoding rule. We will go through the steps of the method in detail for LO-RLL codes.\vspace{+0.7em}

\textbf{Step~1)} Using the patterns in $\mathcal{R}^2_d$, we determine initial groups of $\mathcal{RC}^2_{m,d}$ as follows:
\begin{itemize}
\item For the pattern $11$, there is an initial group having all the codewords starting with $10$ from the left. Then, there is another initial group having all the codewords starting with $0$ from the left.
\item For the pattern $101$, there is an initial group having all the codewords starting with $100$ from the left. There does not exist a group with codewords starting with $11$ from the left since this is a forbidden pattern. Then, there is another initial group having all the codewords starting with $0$ from the left.
\item This procedure continues for the rest of patterns in $\mathcal{R}^2_d$ until the pattern $1\bold{0}^{d-1}1$. For this pattern, there is an initial group having all the codewords starting with $1\bold{0}^{d-1}0 = 1\bold{0}^d$ from the left. There does not exist any group with codewords starting with $1\bold{0}^{d'}1$, for $d-2 \geq d' \geq 0$, from the left since these are all forbidden patterns. Then, there is another initial group having all the codewords starting with $0$ from the left.
\end{itemize}

After eliminating all redundant groups and less restrictive groups, we end up with only two (final) groups covering all the LO-RLL codewords in $\mathcal{RC}^2_{m,d}$: \textit{Group~1}, which contains all the codewords starting with $0$ from the left, and \textit{Group~2}, which contains all the codewords starting with $1\bold{0}^d$ from the left. The groups are defined for $m \geq 1$. The two groups are illustrated for the code $\mathcal{RC}^2_{5,1}$ in Table~\ref{table_1}.\vspace{+0.7em}

\textbf{Step~2)} As for Group~1 of $\mathcal{RC}^2_{m,d}$, each codeword in this group corresponds to a codeword in $\mathcal{RC}^2_{m-1,d}$ (of length $m-1$) such that they share the $m-1$ right-most bits (RMBs). Since this correspondence is bijective, the cardinality of Group~1 is:
\begin{equation}\label{eqn_lorllgr1}
N_{2,1}(m, d) = N_2(m-1,d).
\end{equation}

As for Group~2 of $\mathcal{RC}^2_{m,d}$, each codeword in this group corresponds to a codeword in $\mathcal{RC}^2_{m-d-1,d}$ (of length $m-d-1$) such that they share the $m-d-1$ RMBs. Since this correspondence is also bijective, the cardinality of Group~2 is:
\begin{equation}\label{eqn_lorllgr2}
N_{2,2}(m, d) = N_2(m-d-1,d).
\end{equation}

From (\ref{eqn_lorllgr1}) and (\ref{eqn_lorllgr2}), the cardinality of the code $\mathcal{RC}^2_{m,d}$ is given by:
\begin{equation}\label{eqn_lorllcard}
N_2(m, d) = \sum_{i=1}^2 N_{2,i}(m, d) = N_2(m-1,d) + N_2(m-d-1,d), \textup{ } m \geq 1.
\end{equation}
As for the defined cardinalities, we know that the cardinality of Group~2 for $1 \leq m \leq d+1$ is always $1$. This means using (\ref{eqn_lorllgr2}), $N_2(m-d-1,d) = 1$, for $1 \leq m \leq d+1$. Consequently, the defined cardinalities are:
\begin{equation}\label{eqn_lorllcdef}
N_2(m, d) \triangleq 1, \textup{ } -d \leq m \leq 0.
\end{equation}
With that, we managed to use the inherent recursion of the groups of the $(d,\infty)$ LO-RLL code to compute its cardinality. The result in (\ref{eqn_lorllcard}) and (\ref{eqn_lorllcdef}) is consistent with \cite{tang_bahl} and \cite{ahh_loco}. The cardinalities of $\mathcal{RC}^2_{m,1}$, $m \in \{1, 2, \dots, 5\}$, are given in the last row of Table~\ref{table_1}.\vspace{+0.7em}

\textbf{Step~3)} We now specify the special cases. Using the patterns in $\mathcal{R}^2_d$, we determine initial special cases for $\mathcal{RC}^2_{m,d}$ as follows:
\begin{itemize}
\item For the pattern $11$, there does not exist any special cases since there does not exist a symbol greater than $1$ in GF$(2)$ according to the lexicographic ordering definition.
\item For the pattern $101$, following the same logic in the previous item results in that the only initial case to investigate is for $11$. However, $11$ is a forbidden pattern, which means there does not exist any special cases.
\item This procedure continues for the rest of patterns in $\mathcal{R}^2_d$ until the pattern $1\bold{0}^{d-1}1$. For this pattern, following the same logic in the previous items results in that the initial cases to investigate are for $1\bold{0}^{d'}1$, for $d-2 \geq d' \geq 0$. However, all of these patterns are forbidden patterns, which means there does not exist any special cases.
\end{itemize}

Based on the above discussion, we do not have special cases in $(d,\infty)$ LO-RLL codes. Thus, we only have one (final) case that is the \textit{typical case}. In other words, all $1$'s, i.e., all non-zero symbols here, will have the same contribution to $g(\bold{c})$ regardless from the preceding bits in $\bold{c}$.\vspace{+0.7em}

\textbf{Steps~4 and 5)} Since we do not have any special cases, there will be no need for a merging function as we have a single expression for $g_i(c_i)$. Consequently, Step~5 will be straightforward, and can be combined with Step~4.

Given all the bits prior to $c_i$, i.e., $c_{m-1} c_{m-2} \dots c_{i+1}$, we already know that $g_i(c_i)$, $c_i = 1$, is the number of codewords in $\mathcal{RC}^2_{m,d}$ starting with $c_{m-1} c_{m-2} \dots c_{i+1} 0$ from the left. Looking from the right, this number can be seen as the number of LO-RLL codewords of length $i+1$ in $\mathcal{RC}^2_{i+1,d}$ that can be concatenated from the right to $c_{m-1} c_{m-2} \dots c_{i+1}$ to generate valid LO-RLL codewords of length $m$ in $\mathcal{RC}^2_{m,d}$. These codewords in $\mathcal{RC}^2_{i+1,d}$ are all the codewords starting with $0$ from the left, and thus are all the codewords in Group~1 of $\mathcal{RC}^2_{i+1,d}$. The reason is that prior to $c_i = 1$, there must be a (guaranteed) run of $d$ consecutive $0$'s, i.e., $\bold{0}^d$, in $\bold{c}$ because of the constraint, resulting in no limitations on the codewords starting with $0$ from the left in $\mathcal{RC}^2_{i+1,d}$ for the concatenation. Consequently,
\begin{equation}\label{eqn_lorllrule1}
g_i(c_i) = N_{2,1}(i+1, d) = N_2(i, d), \textup{ } c_i = 1,
\end{equation}
where (\ref{eqn_lorllgr1}) was used to get $N_{2,1}(i+1, d)$.

To account also for the case of $c_i = 0$ (then $g_i(c_i) = 0$), we use $a_i$, which is $\mathcal{L}(c_i)$, as follows:
\begin{equation}\label{eqn_lorllrule2}
g_i(c_i) = a_i N_2(i, d).
\end{equation}
The encoding-decoding rule of a binary $(d,\infty)$ LO-RLL code of length $m$ is then:\begin{equation}\label{eqn_lorllrule3}
g(\bold{c}) = \sum_{i=0}^{m-1} g_i(c_i) = \sum_{i=0}^{m-1} a_i N_2(i, d).
\end{equation}
The result in (\ref{eqn_lorllrule3}) is also consistent with \cite{tang_bahl}.

\begin{example}\label{example_1}
Consider the LO-RLL code $\mathcal{RC}^2_{5,1}$ ($m=5$ and $d=1$) given in Table~\ref{table_1}. Using (\ref{eqn_lorllcard}) and (\ref{eqn_lorllcdef}), we get $N_2(0,1) \triangleq 1$, $N_2(1,1) = 2$, $N_2(2,1) = 3$, $N_2(3,1) = 5$, and $N_2(4,1) = 8$. Consider the codeword $\bold{c} = 10101$ in $\mathcal{RC}^2_{5,1}$. Using (\ref{eqn_lorllrule3}), we get:
\vspace{-0.5em}\begin{align}
g(\bold{c}=10101) &= \sum_{i=0}^4 a_i N_2(i,1) = N_2(4,1) + N_2(2,1) + N_2(0,1) \nonumber \\ & = 8 + 3 + 1 = 12, \nonumber
\end{align}
which is consistent with the index in the table.
\end{example}\vspace{+0.2em}

\textbf{Step~6)} We bridge in $(d,\infty)$ LO-RLL codes as follows. Between each two consecutively written or transmitted codewords in $\mathcal{RC}^2_{m,d}$, we write or transmit $d$ consecutive $0$'s, i.e., the bridging pattern is $\bold{0}^d$. As for self-clocking, recall that $(d,\infty)$ LO-RLL codes are followed by transition-based signaling. Thus, the only codeword that should be removed from $\mathcal{RC}^2_{m,d}$ is $\bold{0}^m$, and by doing so, a transition is guaranteed for each codeword after signaling.

The rate of a self-clocked $(d,\infty)$ LO-RLL code of length $m$ is $\lfloor \log_2 (N_2(m,d)-1) \rfloor/(m+d)$. We follow the procedures described in Step~6 in Section~\ref{sec_gen} to develop the encoding and decoding algorithms based on the rule in (\ref{eqn_lorllrule3}). These algorithms can be found in \cite{tang_bahl}.

\subsection{Binary Symmetric LOCO Codes}

Here, we discuss binary symmetric LOCO (S-LOCO) codes introduced in \cite{ahh_loco}, where the minimum separation between~two consecutive transitions, $0-1$ or $1-0$, is controlled by $x$. Denote an S-LOCO code of length $m$ and having parameter $x$ by $\mathcal{SC}^2_{m,x}$. The definition of the code is exactly the definition of a generic LOCO code, which is Definition~\ref{def_genloco}, with $q=2$, $\mathcal{C}^q_m = \mathcal{SC}^2_{m,x}$, and $\mathcal{T}$ given by:
\begin{equation}\label{eqn_tsloco}
\mathcal{T} = \mathcal{S}^2_x \triangleq \{010, 101, 0110, 1001, \dots, 0\bold{1}^x0, 1\bold{0}^x1\}.
\end{equation}
Both $\bold{c}$ in $\mathcal{C}^q_m = \mathcal{SC}^2_{m,x}$ and $g(\bold{c})$ are used as they were in Section~\ref{sec_gen}. The cardinality of $\mathcal{SC}^2_{m,x}$ is $N_q(m) = N_2(m, x)$. Table~\ref{table_2} shows multiple S-LOCO codes with $x = 2$ and $m$ in $\{1, 2, \dots, 5\}$.

\begin{table*}
\caption{All the Codewords of Five Binary S-LOCO Codes, $\mathcal{SC}^2_{m,2}$, $m \in \{1, 2, \dots, 5\}$. The Four Different Groups of Codewords Are Explicitly Illustrated for the Code $\mathcal{SC}^2_{5,2}$  The Four Groups Can Be Distinguished Starting From $m=2$.}
\vspace{-0.5em}
\centering
\scalebox{1.00}
{
\begin{tabular}{|c|c|c|c|c|c c|}
\hline
\multirow{2}{*}{Codeword index $g(\bold{c})$} & \multicolumn{6}{|c|}{\makecell{Codewords of the code $\mathcal{SC}^2_{m,2}$}} \\
\cline{2-7}
{} & \makecell{$m=1$} & \makecell{$m=2$} & \makecell{$m=3$} & \makecell{$m=4$} & \multicolumn{2}{|c|}{\makecell{$m=5$}} \\
\hline
$0$ & $0$ & $00$ & $000$ & $0000$ & $00000$ & \multirow{4}{*}{Group~1} \\
\cline{1-1}\cline{2-2}\cline{3-3}
$1$ & $1$ & $01$ & $001$ & $0001$ & $00001$ & \\
\cline{1-1}\cline{2-2}\cline{3-3}\cline{4-4}
$2$ &  & $10$ & $011$ & $0011$ & $00011$ & \\
\cline{1-1}\cline{3-3}\cline{4-4}\cline{5-5}
$3$ &  & $11$ & $100$ & $0111$ & $00111$ & \\
\cline{1-1}\cline{3-3}\cline{4-4}\cline{5-5}\cline{6-7}
$4$ &  &  & $110$ & $1000$ & $01110$ & \multirow{2}{*}{Group~4} \\
\cline{1-1}\cline{5-5}
$5$ &  &  & $111$ & $1100$ & $01111$ & \\
\cline{1-1}\cline{6-7}
$6$ &  &  &  & $1110$ & $10000$ & \multirow{2}{*}{Group~3} \\
\cline{1-1}
$7$ &  &  &  & $1111$ & $10001$ & \\
\cline{1-1}\cline{5-5}\cline{6-7}
$8$ &  &  &  &  & $11000$ & \multirow{4}{*}{Group~2} \\
\cline{1-1}
$9$ &  &  &  &  & $11100$ & \\
\cline{1-1}
$10$ &  &  &  &  & $11110$ & \\
\cline{1-1}
$11$ &  &  &  &  & $11111$ & \\
\hline
Code cardinality & $N_2(1, 2) \triangleq 2$  & $N_2(2, 2) = 4$ & $N_2(3, 2) = 6$ & $N_2(4, 2) = 8$ &  \multicolumn{2}{|c|}{$N_2(5, 2) = 12$} \\
\hline
\end{tabular}}
\label{table_2}
\end{table*}

Now, we will apply the steps of the general method to find out how to encode and decode S-LOCO codes using a simple encoding-decoding rule.\vspace{+0.7em}

\textbf{Step~1)} Using the patterns in $\mathcal{S}^2_x$, we determine initial groups of $\mathcal{SC}^2_{m,x}$ as follows:
\begin{itemize}
\item For the pattern $010$ (resp., $101$), there is an initial group having all the codewords starting with $011$ (resp., $100$) from the left. There is another initial group having all the codewords starting with $00$ (resp., $11$) from the left. There is a third initial group having all the codewords starting with $1$ (resp., $0$) from the left.
\item For the pattern $0110$ (resp., $1001$), there is an initial group having all the codewords starting with $0\bold{1}^3$ (resp., $1\bold{0}^3$) from the left. There is another initial group having all the codewords starting with $00$ (resp., $11$) from the left. There is a third initial group having all the codewords starting with $1$ (resp., $0$) from the left.
\item This procedure continues for the rest of patterns in $\mathcal{S}^2_x$ until the patterns $0\bold{1}^x0$ and $1\bold{0}^x1$. For the pattern $0\bold{1}^x0$ (resp., $1\bold{0}^x1$), there is an initial group having all the codewords starting with $0\bold{1}^{x+1}$ (resp., $1\bold{0}^{x+1}$) from the left. There is another initial group having all the codewords starting with $00$ (resp., $11$) from the left. There is a third initial group having all the codewords starting with $1$ (resp., $0$) from the left.
\end{itemize}

After eliminating all redundant groups and less restrictive groups, we end up with four (final) groups covering all the S-LOCO codewords in $\mathcal{SC}^2_{m,x}$: \textit{Group~1}, which contains all the codewords starting with $00$ from the left, \textit{Group~2}, which contains all the codewords starting with $11$ from the left, \textit{Group~3}, which contains all the codewords starting with $1\bold{0}^{x+1}$ from the left, and \textit{Group~4}, which contains all the codewords starting with $0\bold{1}^{x+1}$ from the left. The groups are defined for $m \geq 2$. The four groups are illustrated for the code $\mathcal{SC}^2_{5,2}$ in Table~\ref{table_2}.\vspace{+0.7em}

\textbf{Step~2)} Observe that the symmetry of $\mathcal{S}^2_x$ implies the symmetry of the code $\mathcal{SC}^2_{m,x}$. Thus, the number of codewords starting with $0$ from the left equals the number of codewords starting with $1$ from the left.

We follow the same logic adopted in the previous subsection for LO-RLL codes to derive the cardinalities of groups. The details of this part can be found in \cite{ahh_loco}. The cardinality of Group~1 is:
\begin{equation}\label{eqn_slocogr1}
N_{2,1}(m, x) = \frac{1}{2} N_2(m-1,x).
\end{equation}
The cardinality of Group~4 is:
\begin{equation}\label{eqn_slocogr2}
N_{2,4}(m, x) = \frac{1}{2} N_2(m-x-1,x).
\end{equation}

From (\ref{eqn_slocogr1}), (\ref{eqn_slocogr2}), and symmetry, the cardinality of the code $\mathcal{SC}^2_{m,x}$ is given by the recursive formula:
\begin{equation}\label{eqn_slococard}
N_2(m, x) = \sum_{i=1}^4 N_{2,i}(m, x) = N_2(m-1,x) + N_2(m-x-1,x), \textup{ } m \geq 2.
\end{equation}
As for the defined cardinalities, it is clear that $N_2(1, x) \triangleq 2$. We also know that the cardinality of Group~4 for $2 \leq m \leq x+2$ is always $1$ (the group can be distinguished but not enough bits to have more than $1$ codeword in it). This means using (\ref{eqn_slocogr2}), $N_2(m-x-1,x) = 2$, for $2 \leq m \leq x+2$. Consequently,
\begin{equation}\label{eqn_slococdef}
N_2(m, x) \triangleq 2, \textup{ } 1-x \leq m \leq 1.
\end{equation}
The result in (\ref{eqn_slococard}) and (\ref{eqn_slococdef}) for S-LOCO codes is consistent with \cite{ahh_loco}. The cardinalities of $\mathcal{SC}^2_{m,2}$, $m \in \{1, 2, \dots, 5\}$, are given in the last row of Table~\ref{table_2}.\vspace{+0.7em}

\textbf{Step~3)} We now specify the special cases. Using the patterns in $\mathcal{S}^2_x$, we determine initial special cases for $\mathcal{SC}^2_{m,x}$ as follows:
\begin{itemize}
\item For the pattern $010$, the only initial special case is $c_{i+2} c_{i+1} c_i = 011$. For the pattern $101$, the only initial special case is $c_{i+1} c_i = 11$.
\item For the pattern $0110$, the only initial special case is $c_{i+3} c_{i+2} c_{i+1} c_i = 0\bold{1}^3$. For the pattern $1001$, the only initial special case is $c_{i+1} c_i = 11$.
\item This procedure continues for the rest of patterns in $\mathcal{S}^2_x$ until the patterns  $0\bold{1}^x0$ and $1\bold{0}^x1$. For the pattern $0\bold{1}^x0$, the only initial special case is $c_{i+x+1} c_{i+x} \dots c_i = 0\bold{1}^{x+1}$. For the pattern $1\bold{0}^x1$, the only initial special case is $c_{i+1} c_i = 11$.
\end{itemize}

After removing the redundant initial special cases, we end up with $x+2$ (final) cases for $c_i$ based on $c_i$ and its preceding bits. These cases are: $011$, $0\bold{1}^3$, \dots, $0\bold{1}^{x+1}$, and $11$, which are the \textit{special cases} described above, in addition to the \textit{typical case}. The typical case is the case of $c_i = c_{m-1} = 1$ ($1$ at the LMB) or $c_{i+1} c_i = 01$. The priority increases as the sequence length increases, e.g., the case of $c_{i+1} c_i = 11$ is activated only if there does not exist any $0$ in the $x+1$ positions prior to $c_i$.\vspace{+0.7em}

\textbf{Step~4)} We start off with the typical case, where $c_i = c_{m-1} = 1$ or $c_{i+1} c_i = 01$. We index this case by $i_{\textup{c}} = 1$. The contribution of $c_i$ to $g(\bold{c})$ in this case is either the number of codewords in $\mathcal{SC}^2_{m,x}$ starting with $0$ from the left if $i = m-1$ or the number of codewords in $\mathcal{SC}^2_{m,x}$ starting with $c_{m-1} c_{m-2} \dots c_{i+2} 00$ from the left if $i < m-1$. Thus, in both situations and using symmetry, the contribution is:
\begin{equation}\label{eqn_slocorule1}
g_{i,1}(c_i) = \frac{1}{2} N_2(i+1, x).
\end{equation}

The first special case of existence for $c_i = 1$ is $c_{i+2} c_{i+1} c_i = 011$. The contribution of $c_i$ to $g(\bold{c})$ in this case is the number of codewords in $\mathcal{SC}^2_{m,x}$ starting with $c_{m-1} c_{m-2} \dots c_{i+3} 010$ from the left. This number is $0$ since $010$ is a forbidden pattern in $\mathcal{S}^2_x$. This is also true for the case of $c_{i+3} c_{i+2} c_{i+1} c_i = 0\bold{1}^3$. In fact, this is true for all the cases of $011$, $0\bold{1}^3$, \dots, and $0\bold{1}^{x+1}$, which we index by $i_{\textup{c}} = 2$, $3$, \dots, and $x+1$, respectively. Consequently, we get:
\begin{equation}\label{eqn_slocorule2}
g_{i,i_{\textup{c}}}(c_i) = 0, \textup{ } 2 \leq i_{\textup{c}} \leq x+1.
\end{equation}

The last case of existence for $c_i = 1$ is $c_{i+1} c_i = 11$ such that there does not exist any $0$ in the $x+1$ positions prior to $c_i$. We index this case by $i_{\textup{c}} = x+2$. The contribution of $c_i$ to $g(\bold{c})$ in this case is the number of codewords in $\mathcal{SC}^2_{m,x}$ starting with $c_{m-1} c_{m-2} \dots c_{i+2} 10$ from the left. In order to satisfy the constraint, this $10$ must be followed by $\bold{0}^x$ in $\bold{c}$. Thus, the number we are after is the number of codewords in $\mathcal{SC}^2_{m,x}$ starting with $c_{m-1} c_{m-2} \dots c_{i+2} 1\bold{0}^{x+1}$ from the left. Looking from the right, this number is the number of codewords in $\mathcal{SC}^2_{i+1-x,x}$ starting with $0$ from the left. Thus, and using symmetry, the contribution is:
\begin{equation}\label{eqn_slocorule3}
g_{i,x+2}(c_i) = \frac{1}{2} N_2(i+1-x, x).
\end{equation}

\textbf{Step~5)} Since we have three expressions for $g_{i,i_{\textup{c}}}(c_i)$, we need only two merging variables: $y_{i,1}$, for the cases indexed by $i_{\textup{c}} \in \{2, 3, \dots, x+1\}$, and $y_{i,2}$, for the case indexed by $i_{\textup{c}}=x+2$. If the two variables are zeros, the typical case contribution is switched on. These merging variables are set as follows:
\begin{align}\label{eqn_slocorule4}
y_{i,1} &= 1 \textup{ if } c_{i+x'+1} c_{i+x'} \dots c_i = 0\bold{1}^{x'+1}, \textup{ } 1 \leq x' \leq x, \textup { and } y_{i,1} = 0 \textup{ otherwise}, \nonumber \\
y_{i,2} &= 1 \textup{ if } c_{i+1} c_i = 11 \textup{ s.t. } y_{i,1} = 0, \textup{ and } y_{i,2} = 0 \textup{ otherwise}.
\end{align}

Recall Step~5 in Section~\ref{sec_gen}. Now, we pick the merging function $f^{\textup{mer}}_0 (\cdot) = \frac{1}{2} (a_i - y_{i,1} - y_{i,2})$ for $N_2(i+1, x)$. This function results in $1/2$ only if $a_i = 1$, i.e., $c_i = 1$, and $y_{i,1} = y_{i,2} = 0$ (the case indexed by $1$). Otherwise, the function results in $0$. We also pick the merging function $f^{\textup{mer}}_x (\cdot) = \frac{1}{2} y_{i,2}$ for $N_2(i+1-x, x)$. This function results in $1/2$ only if $y_{i,2} = 1$ (the case indexed by $x+2$). Otherwise, the function results in $0$. If $y_{i,1} = 0$, then $y_{i,2} = 0$ automatically from (\ref{eqn_slocorule4}), resulting in both $f^{\textup{mer}}_0 (\cdot)$ and $f^{\textup{mer}}_x (\cdot)$ being zeros (the rest of cases).

Using these two merging functions, the unified expression representing the contribution of a bit $c_i$ to the codeword index $g(\bold{c})$ can be written as:
\begin{equation}\label{eqn_slocorule5}
g_i(c_i) = \frac{1}{2} (a_i - y_{i,1} - y_{i,2}) N_2(i+1, x) + \frac{1}{2} y_{i,2} N_2(i+1-x, x).
\end{equation}
This can be further simplified if $y_{i,2}$ is used inside the cardinality argument as follows:\begin{equation}\label{eqn_slocorule6}
g_i(c_i) = \frac{1}{2} (a_i - y_{i,1}) N_2(i+1-y_{i,2}\hspace{+0.1em}x, x).
\end{equation}
The formula in (\ref{eqn_slocorule6}) also accounts for the case of $c_i = 0$. The encoding-decoding rule of a binary S-LOCO code is then:
\begin{equation}\label{eqn_slocorule7}
g(\bold{c}) = \sum_{i=0}^{m-1} g_i(c_i) = \frac{1}{2} \sum_{i=0}^{m-1} (a_i - y_{i,1}) N_2(i+1-y_{i,2}\hspace{+0.1em}x, x).
\end{equation}

\begin{example}\label{example_2}
Consider the S-LOCO code $\mathcal{SC}^2_{5,2}$ ($m=5$ and $x=2$) given in Table~\ref{table_2}. Using (\ref{eqn_slococard}) and (\ref{eqn_slococdef}), we get $N_2(-1,2) \triangleq 2$, $N_2(0,2) \triangleq 2$, $N_2(1,2) \triangleq 2$, $N_2(2,2) = 4$, $N_2(3,2) = 6$, $N_2(4,2) = 8$, and $N_2(5,2) = 12$. Consider the codeword $\bold{c} = 01111$ in $\mathcal{SC}^2_{5,2}$. Using (\ref{eqn_slocorule7}), we get:
\begin{align}
g(\bold{c}=01111) &= \frac{1}{2} \sum_{i=0}^{4} (a_i - y_{i,1}) N_2(i+1-2y_{i,2}, 2) \nonumber \\ &= \frac{1}{2} [N_2(4,2) + 0 + 0 + N_2(-1,2)] = \frac{1}{2} [8 + 2]= 5, \nonumber
\end{align}
which is consistent with the index in the table.
\end{example}

The only remaining question will be about why the rule in (\ref{eqn_slocorule7}) looks different from the one in \cite{ahh_loco}, which is:
\begin{equation}\label{eqn_slocoold1}
g(\bold{c}) = \frac{1}{2} \left [ a_{m-1} N_2(m, x) + \sum_{i=0}^{m-1} a_i N_2(i+1-x, x) \right ].
\end{equation}
Clearly, for $c_i = c_{m-1} = 1$, $g_i(c_i)$ derived from both equations is the same, which is $\frac{1}{2} N_2(i+1, x)$. Moreover, for $c_{i+1} c_i = 11$ such that there does not exist any $0$ in the $x+1$ positions prior to $c_i$, $g_i(c_i)$ derived from both equations is also the same, which is $\frac{1}{2} N_2(i+1-x, x)$. The first difference appears for $c_{i+1} c_i = 01$, since $g_i(c_i)$ from (\ref{eqn_slocorule7}) is $\frac{1}{2} N_2(i+1, x)$, while $g_i(c_i)$ from (\ref{eqn_slocoold1}) is $\frac{1}{2} N_2(i+1-x, x)$. The second difference appears for $c_{i+x'+1} c_{i+x'} \dots c_i = 0\bold{1}^{x'+1}$, $1 \leq x' \leq x$, , since $g_i(c_i)$ from (\ref{eqn_slocorule7}) is $0$, while $g_i(c_i)$ from (\ref{eqn_slocoold1}) is still $\frac{1}{2} N_2(i+1-x, x)$. Thus, if we can prove that the contribution of the left-most $1$ in the pattern $0\bold{1}^{x+1}$ to $g(\bold{c})$ from (\ref{eqn_slocorule7}) is the sum of the contributions of all $1$'s in the same pattern from (\ref{eqn_slocoold1}), we will demonstrate that the two rules result in exactly the same $g(\bold{c})$. This proof goes as follows using (\ref{eqn_slococard}):
\begin{align}\label{eqn_slocoold2}
N_2(i+1, x) &= N_2(i, x) + N_2(i-x, x) = N_2(i-1, x) + N_2(i-x, x) + N_2(i-x-1, x) \nonumber \\
&= N_2(i-2, x) + N_2(i-x, x) + N_2(i-x-1, x) + N_2(i-x-2, x) \dots \nonumber \\
&= N_2(i-x+1, x) + N_2(i-x, x) + N_2(i-x-1, x) + \dots + N_2(i-2x+1, x) \nonumber \\
&\implies N_2(i+1, x) = \sum_{i'=i-x}^{i} N_2(i'+1-x, x),
\end{align}
which completes our demonstration.\vspace{+0.7em}

\textbf{Step~6)} We bridge in S-LOCO codes as follows. Between each two consecutively written or transmitted codewords in $\mathcal{SC}^2_{m,x}$, we do bridge with $x$ consecutive no-writing or no-transmission symbols, i.e., the bridging pattern is $\bold{z}^x$, where $z$ denotes the no-writing or no-transmission symbol. Other bridging methods are also possible as shown in \cite{ahh_loco}. As for self-clocking, recall that S-LOCO codes are followed by level-based signaling. Thus, the two codewords that should be removed from $\mathcal{SC}^2_{m,x}$ are $\bold{0}^m$ and $\bold{1}^m$, and by doing so, a transition is guaranteed for each codeword after signaling.

The rate of a self-clocked S-LOCO code of length $m$ and having parameter $x$ is $\lfloor \log_2 (N_2(m,x)-2) \rfloor/(m+x)$. We follow the procedures described in Step~6 in Section~\ref{sec_gen} to develop the encoding and decoding algorithms based on the rule in (\ref{eqn_slocorule7}) or (\ref{eqn_slocoold1}). These algorithms can be found in \cite{ahh_loco}.

\section{Optimal Constrained Codes for TDMR}\label{sec_opt}

In this section, we introduce new optimal (rate-wise) LOCO codes for TDMR systems. The new codes prevent certain error-prone patterns from being written on the TDMR medium, increasing the reliability of the system.

\begin{figure}
\vspace{-1.5em}
\center
\includegraphics[trim={0.0in 1.0in 0.0in 0.3in}, width=3.5in]{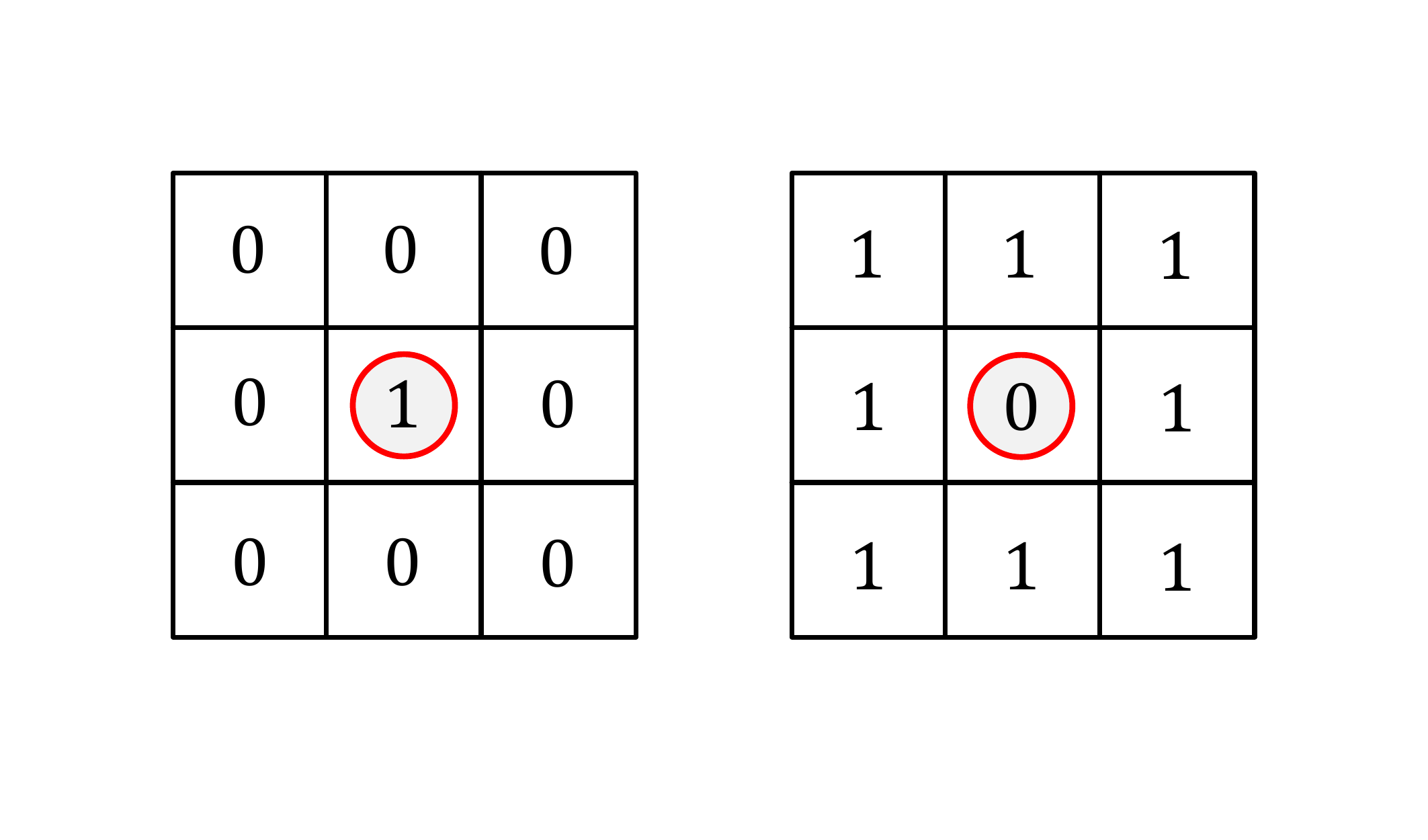}
\vspace{-0.5em}
\caption{The detrimental square isolation patterns (shaped as a square). An error is highly likely to happen on the circled bit at the center.}
\label{fig_1}
\vspace{-0.7em}
\end{figure}

\begin{figure}
\vspace{-0.7em}
\center
\includegraphics[trim={0.0in 1.0in 0.0in 0.3in}, width=3.5in]{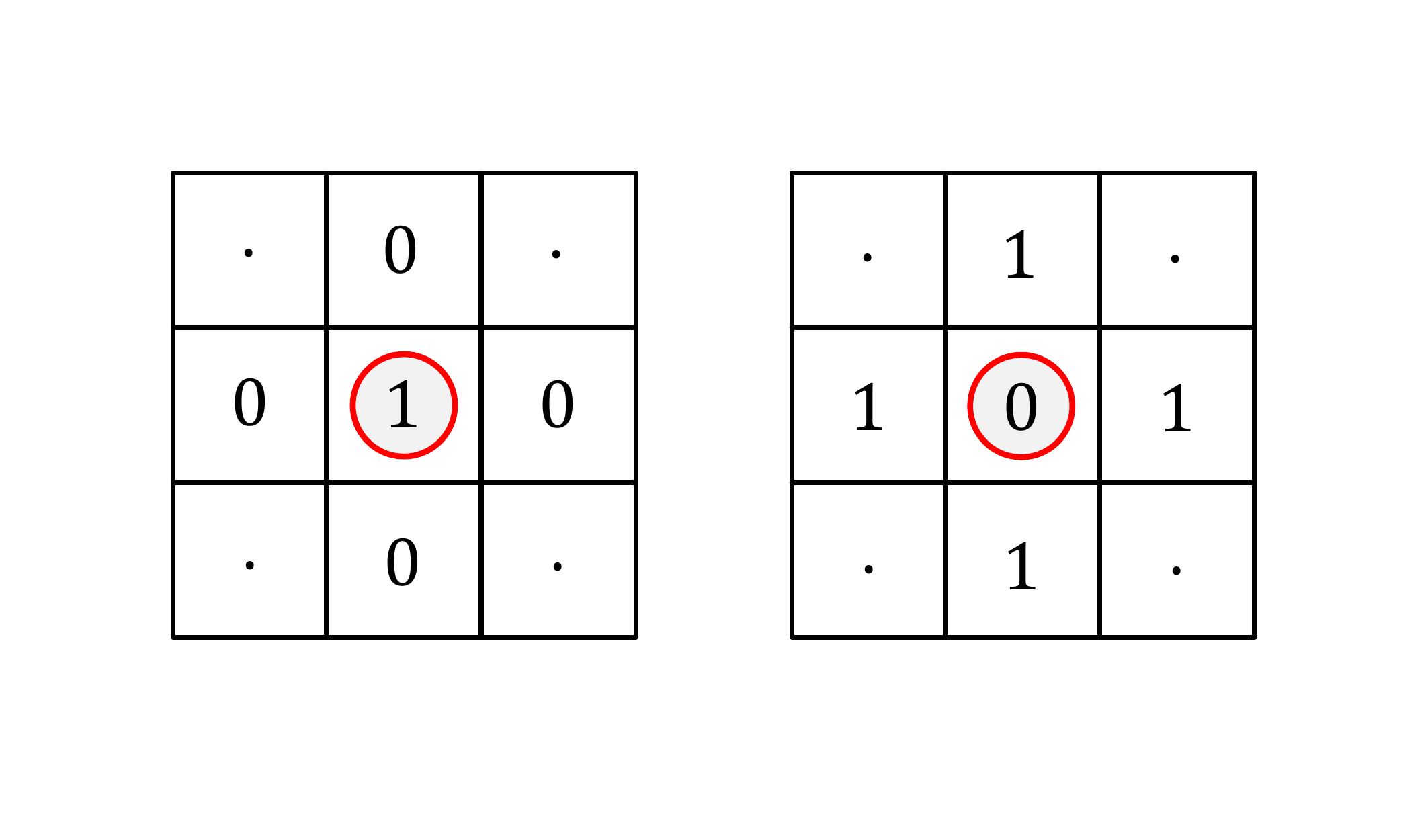}
\vspace{-0.5em}
\caption{The detrimental plus isolation patterns (shaped as a plus sign). An error is highly likely to happen on the circled bit at the center.}
\label{fig_2}
\vspace{-0.5em}
\end{figure}

We discuss two sets of $3 \times 3$ detrimental patterns: the set of square isolation (SIS) patterns and the set of plus isolation (PIS) patterns. SIS patterns are $3 \times 3$ patterns having the (isolated) bit at the center surrounded by $8$ complementary bits. There are only $2$ SIS patterns as shown in Fig.~\ref{fig_1}. PIS patterns are $3 \times 3$ patterns having the (isolated) bit at the center surrounded by $4$ complementary bits at the positions with Manhattan distance $1$ from the center, i.e., non-corner positions. There are $32$ PIS patterns as shown in Fig.~\ref{fig_2} since a ``$\cdot$'' in Fig.~\ref{fig_2} means $0$ or $1$, which means there are $2^4$ patterns with the central bit being $1$ (left panel of Fig.~\ref{fig_2}) and $2^4$ patterns with the central bit being $0$ (right panel of Fig.~\ref{fig_2}).

Since level-based signaling is adopted here, a $0$ is converted into $-A$, indexed by $\mathcal{L}(0) = 0$, and a $1$ is converted into $+A$, indexed by $\mathcal{L}(1) = 1$, upon writing. Consequently, the problem with SIS and PIS patterns is that the level at the center becomes highly likely to change its sign due to two-dimensional interference (along both track directions), resulting in an error during reading. SIS patterns are the most detrimental subclass of PIS patterns. However, they are less likely to occur under unbiased writing since they are $2$ patterns out of $512$ possible ones for the $3 \times 3$ grid. PIS patterns subsume SIS patterns, and they were introduced because of the fact that bits at the corners cause less interference than bits at positions with Manhattan distance $1$ from the center \cite{mohsen_tdmr, sharov_TCon}. PIS patterns are also $16$ times more likely to occur compared with SIS patterns under unbiased writing. However, preventing PIS patterns results in some rate loss compared with preventing SIS patterns. That is why we introduce codes preventing SIS patterns and codes preventing PIS patterns in this section and in Section~\ref{sec_compl}. Error statistics demonstrating the harmfulness of SIS and PIS patterns in a practical TDMR system are presented in Section~\ref{sec_gains}.

As mentioned in the introduction, the practical TDMR model we use adopts a wide read head that reads data from $3$ adjacent down tracks at the same time \cite{chan_tdmr, shayan_tdmr, bd_tdmr}. Suppose that the indices of down tracks in the TD grid are $0$, $1$, $2$, $3$, \dots, $D-1$, where $D$ is the number of down tracks in the TD grid and $3 \mid D$. Then, with that TDMR model, we can partition them into groups, each with $3$ adjacent down tracks to be read together. These groups have the tracks indexed by $(0, 1, 2)$, $(3, 4, 5)$, $(6, 7, 8)$, \dots, $(D-3, D-2, D-1)$. Interference in the cross-track direction from a group into another group is negligible \cite{chan_tdmr, bd_tdmr}. Thus, we can now convert the two-dimensional binary constrained coding problem into a one-dimensional non-binary constrained coding problem. In the new problem, a symbol in GF$(8)$ represents a column with $3$ bits to be written on $3$ adjacent down tracks in the same group. We use the following standard mapping-demapping:
\vspace{-0.1em}\begin{align}\label{eqn_gf8map}
0 &\longleftrightarrow [ 0 \textup{ } 0 \textup{ } 0 ]^{\textup{T}}, \hspace{+3.6em} 1 \longleftrightarrow [ 0 \textup{ } 0 \textup{ } 1 ]^{\textup{T}}, \nonumber \\
\alpha &\longleftrightarrow [ 0 \textup{ } 1 \textup{ } 0 ]^{\textup{T}}, \hspace{+3.0em} \alpha^2 \longleftrightarrow [ 0\textup{ } 1 \textup{ } 1 ]^{\textup{T}}, \nonumber \\
\alpha^3 &\longleftrightarrow [ 1 \textup{ } 0 \textup{ } 0 ]^{\textup{T}}, \hspace{+3.0em} \alpha^4 \longleftrightarrow [ 1 \textup{ } 0 \textup{ } 1 ]^{\textup{T}}, \nonumber \\
\alpha^5 &\longleftrightarrow [ 1 \textup{ } 1 \textup{ } 0 ]^{\textup{T}}, \hspace{+3.0em} \alpha^6 \longleftrightarrow [ 1 \textup{ } 1 \textup{ } 1 ]^{\textup{T}},
\end{align}
We are now ready to build non-binary constrained codes defined over GF$(8)$ for TDMR systems. In this section, we introduce OS-LOCO and OP-LOCO codes.

\begin{remark}
While the NIB constraint forces the elimination of PIS patterns everywhere in the TD grid, our OP-LOCO codes achieve rate gain by focusing only on the PIS patterns within the same group of down tracks as interference in the cross-track direction from a group into another group is of limited significance.
\end{remark}

\subsection{Optimal Square LOCO Codes}

We start with our optimal square LOCO (OS-LOCO) codes, which are codes preventing the SIS patterns shown in Fig.~\ref{fig_1} within each group of three adjacent down tracks. From (\ref{eqn_gf8map}), these two SIS patterns map to the two GF$(8)$ patterns $0\alpha0$ and $\alpha^6\alpha^4\alpha^6$, which have the level-equivalent patterns $020$ and $757$. The FSTD of an infinite $8$-ary constrained sequence in which these two patterns are prevented along with the adjacency matrix are in \cite{bd_tdmr}. The capacity $C$, in input bits per coded symbol, and the normalized capacity $C^{\textup{n}}$, according to the same reference, are:
\begin{equation}\label{eqn_caposloco}
C = 2.9944 \textup{ and } C^{\textup{n}} = \frac{1}{\log_2 8} C = \frac{1}{3} C = 0.9981.
\end{equation}

Denote an OS-LOCO code of length $m$ by $\mathcal{OSC}^8_m$. The definition of the code is exactly the definition of a generic LOCO code, which is Definition~\ref{def_genloco}, with $q=8$, $\mathcal{C}^q_m = \mathcal{OSC}^8_m$, and $\mathcal{T}$ given by:
\begin{equation}\label{eqn_tosloco}
\mathcal{T} = \mathcal{OS}^8 \triangleq \{0\alpha0, \alpha^6\alpha^4\alpha^6\}.
\end{equation}
Both $\bold{c}$ in $\mathcal{C}^q_m = \mathcal{OSC}^8_m$ and $g(\bold{c})$ are used as they were in Section~\ref{sec_gen}. The cardinality of $\mathcal{OSC}^8_m$ is $N_q(m) = N_8(m)$. We could not provide a table as an example listing all the codewords of specific codes because there are way too many codewords for any length $m \geq 4$. However, we will provide an example illustrating the encoding-decoding rule of the code.

Now, we will apply the steps of the general method to find out how to encode and decode OS-LOCO codes using a simple encoding-decoding rule.\vspace{+0.7em}

\textbf{Step~1)} Using the patterns in $\mathcal{OS}^8$, we determine initial groups of $\mathcal{OSC}^8_m$ as follows:
\begin{itemize}
\item For the pattern $0\alpha0$, there is an initial group having all the codewords starting with $0\alpha\beta_1$, $\beta_1 \in \textup{GF}(8) \setminus \{0\}$, from the left. There are seven more initial groups having all the codewords starting with $0\beta_2$, a group for each $\beta_2 \in \textup{GF}(8) \setminus \{\alpha\}$, from the left. There are seven more initial groups having all the codewords starting with $\beta_1$, a group for each $\beta_1 \in \textup{GF}(8) \setminus \{0\}$, from the left. Group merging will be performed.
\item For the pattern $\alpha^6\alpha^4\alpha^6$, there is an initial group having all the codewords starting with $\alpha^6\alpha^4\beta_3$, $\beta_3 \in \textup{GF}(8) \setminus \{\alpha^6\}$, from the left. There are seven more initial groups having all the codewords starting with $\alpha^6\beta_4$, a group for each $\beta_4 \in \textup{GF}(8) \setminus \{\alpha^4\}$, from the left. There are seven more initial groups having all the codewords starting with $\beta_3$, a group for each $\beta_3 \in \textup{GF}(8) \setminus \{\alpha^6\}$, from the left. Group merging will be performed.
\end{itemize}

After operating on these initial groups, we end up with three (final) groups covering all the OS-LOCO codewords in $\mathcal{OSC}^8_m$: \textit{Group~1}, which contains all the codewords starting with $0$ from the left, \textit{Group~2}, which contains all the codewords starting with $\beta_5$, $\beta_5 \in \textup{GF}(8) \setminus \{0,\alpha^6\}$, from the left, and \textit{Group~3}, which contains all the codewords starting with $\alpha^6$ from the left. The groups are defined for $m \geq 2$.

Group~1 is further partitioned into three subgroups: \textit{Subgroup~1.1}, which contains all the codewords starting with $0\beta'_2$, $\beta'_2 \in \{0, 1\}$, from the left, \textit{Subgroup~1.2}, which contains all the codewords starting with $0\alpha\beta_1$ from the left, and \textit{Subgroup~1.3}, which contains all the codewords starting with $0\beta''_2$, $\beta''_2 \in \{\alpha^2, \alpha^3, \dots, \alpha^6\}$, from the left.

Group~2 is further partitioned into six subgroups: \textit{Subgroup~2.$v$}, $v \in \{1, 2, \dots, 6\}$, contains all the codewords starting with $\alpha^{v-1}$ from the left. Observe the symmetry of Group~2; all subgroups within Group~2 have the same size.

Group~3 is further partitioned into three subgroups: \textit{Subgroup~3.1}, which contains all the codewords starting with $\alpha^6\beta'_4$, $\beta'_4 \in \{0, 1, \dots, \alpha^3\}$, from the left, \textit{Subgroup~3.2}, which contains all the codewords starting with $\alpha^6\alpha^4\beta_3$ from the left, and \textit{Subgroup~3.3}, which contains all the codewords starting with $\alpha^6\beta''_4$, $\beta''_4 \in \{\alpha^5, \alpha^6\}$, from the left.\vspace{+0.7em}

\textbf{Step~2)} Theorem~\ref{thm_oslococard} gives the cardinality of an OS-LOCO code.

\begin{theorem}\label{thm_oslococard}
The cardinality of an OS-LOCO code $\mathcal{OSC}^8_m$ is given by:
\begin{equation}\label{eqn_oslococard}
N_8(m) = 8N_8(m-1) - N_8(m-2) + 6N_8(m-3), \text{ } m \geq 2,
\end{equation}
where the defined cardinalities are:
\vspace{-0.1em}\begin{equation}\label{eqn_oslococdef}
N_8(-2) \triangleq \frac{1}{36}, \text{ } N_8(-1) \triangleq \frac{1}{6}, \text{ } N_8(0) \triangleq 1, \text{ and } N_8(1) \triangleq 8.
\end{equation}
\end{theorem}

\begin{IEEEproof}
We first derive recursive cardinality formulae for the three groups of $\mathcal{OSC}^8_m$.

As for Group~1 of $\mathcal{OSC}^8_m$, each codeword in this group is related to a codeword in $\mathcal{OSC}^8_{m-1}$ such that they share the $m-1$ right-most symbols (RMSs). This relation is injective. In order to create a bijective correspondence, all the codewords in $\mathcal{OSC}^8_{m-1}$ that start with $\alpha0$ from the left have to be omitted since $0\alpha0$ is a forbidden pattern in $\mathcal{OS}^8$. Since the number of these codewords to be omitted from $\mathcal{OSC}^8_{m-1}$ for bijective correspondence is the number of codewords starting with $0$ from the left in $\mathcal{OSC}^8_{m-2}$, the cardinality of Group~1 is:
\vspace{-0.1em}\begin{equation}\label{eqn_oslocogr1}
N_{8,1}(m) = \sum_{i=1}^3 N_{8,1.i}(m) = N_8(m-1) - N_{8,1}(m-2),
\end{equation}
where $N_{8,1.i}(m)$ is the cardinality of Subgroup~1.$i$.

As for Group~2 of $\mathcal{OSC}^8_m$, we can just study one subgroup because of symmetry. For Subgroup~2.1 of $\mathcal{OSC}^8_m$, each codeword in this group corresponds to a codeword in $\mathcal{OSC}^8_{m-1}$ such that they share the $m-1$ RMSs. This correspondence is bijective. Consequently, and using symmetry, the cardinality of Group~2 is:
\begin{equation}\label{eqn_oslocogr2}
N_{8,2}(m) = \sum_{i=1}^6 N_{8,2.i}(m) = 6N_{8,2.1}(m) = 6N_8(m-1),
\end{equation}
where $N_{8,2.i}(m)$ is the cardinality of Subgroup~2.$i$.

As for Group~3 of $\mathcal{OSC}^8_m$, it is straightforward to show that its cardinality is exactly the same as the cardinality of Group~1 of $\mathcal{OSC}^8_m$. Thus,
\begin{equation}\label{eqn_oslocogr3}
N_{8,3}(m) = N_{8,1}(m) = N_8(m-1) - N_{8,1}(m-2).
\end{equation}

Next, from (\ref{eqn_oslocogr1}), (\ref{eqn_oslocogr2}), and (\ref{eqn_oslocogr3}), we conclude that the cardinality of $\mathcal{OSC}^8_m$ is:
\begin{equation}\label{eqn_oslocogr4}
N_8(m) = \sum_{i=1}^3 N_{8,i}(m) = 8N_8(m-1) - 2N_{8,1}(m-2).
\end{equation}
The only remaining step is to find $2N_{8,1}(m-2)$. We can write $N_8(m-2)$ as:
\begin{equation}\label{eqn_oslocogr5}
N_8(m-2) = \sum_{i=1}^3 N_{8,i}(m-2) = 2N_{8,1}(m-2) + N_{8,2}(m-2),
\end{equation}
where the second equality is reached using (\ref{eqn_oslocogr3}). Using (\ref{eqn_oslocogr2}), we get:
\begin{equation}\label{eqn_oslocogr6}
2N_{8,1}(m-2) = N_8(m-2) - 6N_8(m-3).
\end{equation}
Substituting (\ref{eqn_oslocogr6}) in (\ref{eqn_oslocogr4}) gives the final recursive formula for cardinality, which is (\ref{eqn_oslococard}):
\begin{equation}\label{eqn_oslocogr7}
N_8(m) = 8N_8(m-1) - N_8(m-2) + 6N_8(m-3), \textup{ } m \geq 2. \nonumber
\end{equation}

As for the defined cardinalities, it is clear that $N_8(1) \triangleq 8$. We also know that $N_8(2) = 8^2 = 64$ since the length of a SIS pattern is $3$, i.e., no sequences to eliminate at that length. Consequently, and using the proved (\ref{eqn_oslococard}),
\begin{equation}\label{eqn_oslocogr8}
8 = 8N_8(0) - N_8(-1) + 6N_8(-2), \textup{ and }
\end{equation}
\vspace{-1.0em}\begin{equation}\label{eqn_oslocogr9}
64 = 8 \times 8 - N_8(0) + 6N_8(-1).
\end{equation}
Furthermore, we know that only two sequences, which are $0\alpha0$ and $\alpha^6\alpha^4\alpha^6$, out of $8^3$ GF$(8)$ sequences are eliminated to arrive at the OS-LOCO code $\mathcal{OSC}^8_3$. Consequently, and again using (\ref{eqn_oslococard}),
\begin{equation}\label{eqn_oslocogr10}
N_8(3) = 8^3-2 = 8 \times 64 - 8 + 6N_8(0),
\end{equation}
resulting in $N_8(0) \triangleq 1$. Substituting this result in (\ref{eqn_oslocogr9}) gives $N_8(-1) \triangleq 1/6$. Substituting that in (\ref{eqn_oslocogr8}) gives $N_8(-2) \triangleq 1/36$, which completes the proof.
\end{IEEEproof}\vspace{+0.7em}

\textbf{Step~3)} We now specify the special cases. Using the patterns in $\mathcal{OS}^8$, we determine initial special cases for the OS-LOCO code $\mathcal{OSC}^8_m$ as follows:
\begin{itemize}
\item For the pattern $0\alpha0$, one initial special case is $c_{i+2} c_{i+1} c_i = 0\alpha\beta_1$, $\beta_1 \in \textup{GF}(8) \setminus \{0\}$. Another initial special case is $c_{i+1} c_i = 0\beta''_2$, $\beta''_2 \in \{\alpha^2, \alpha^3, \dots, \alpha^6\}$. Case merging was performed.
\item For the pattern $\alpha^6 \alpha^4 \alpha^6$, observe that there is no symbol greater than $\alpha^6$ according to the lexicograohic ordering definition. Thus, the only initial special case is $c_{i+1} c_i = \alpha^6\beta''_4$, $\beta''_4 \in \{\alpha^5, \alpha^6\}$. Case merging was performed.
\end{itemize}

Since there are no redundant initial special cases, we end up with four (final) cases for $c_i$ based on $c_i$ and its preceding symbols. These cases are the \textit{three special cases} stated above and the \textit{typical case}. The typical case is simply the case when neither of the three special cases is enabled and $c_i \neq 0$. As usual, the priority of a case increases as its sequence length increases. However, there does not exist a sequence characterizing a special case that is a subsequence (starting from the right) in a longer sequence characterizing another special case here.\vspace{+0.7em}

\textbf{Steps~4 and 5)} Theorem~\ref{thm_oslocorule} gives the encoding-decoding rule of an OS-LOCO code $\mathcal{OSC}^8_m$. Recall that $a_i \triangleq \mathcal{L}(c_i)$.

\begin{theorem}\label{thm_oslocorule}
Let $\bold{c}$ be an OS-LOCO codeword in $\mathcal{OSC}^8_m$. The relation between the lexicographic index $g(\bold{c})$ of this codeword and the codeword itself is given by:
\begin{equation}\label{eqn_oslocorule}
g(\bold{c}) = \sum_{i=0}^{m-1} \left [ \left (a_i - y_{i,1} - \frac{1}{2}y_{i,2} \right ) N_8(i) + \theta_i (1-y_{i,1}) \left ( \left ( 3y_{i,2} - \frac{1}{2} \right ) N_8(i-1) + 3N_8(i-2) \right ) \right ],
\end{equation}
where $y_{i,1}$, $y_{i,2}$, and $\theta_i$ are specified as follows:
\begin{align}\label{eqn_oslocordef}
y_{i,1} &= 1 \text{ if } c_{i+2} c_{i+1} c_i = 0\alpha\beta_1, \text{ } \beta_1 \in \textup{GF}(8) \setminus \{0\}, \text { and } y_{i,1} = 0 \text{ otherwise}, \nonumber \\
y_{i,2} &= 1 \text{ if } c_{i+1} c_i = 0\beta''_2, \text{ } \beta''_2 \in \{\alpha^2, \alpha^3, \dots, \alpha^6\}, \text{ else}, \nonumber \\
y_{i,2} &= 1 \text{ if } c_{i+1} c_i = \alpha^6\beta''_4, \text{ } \beta''_4 \in \{\alpha^5, \alpha^6\}, \text{ and } y_{i,2} = 0 \text{ otherwise}, \nonumber \\
\theta_i &= 1 \text{ if } c_i \neq 0, \text{ and } \theta_i = 0 \text{ otherwise}.
\end{align}
\end{theorem}

\begin{IEEEproof}
First, we perform Step~4 of the method. We aim at computing the contribution of each OS-LOCO codeword symbol $c_i$ to the codeword index $g(\bold{c})$ for the four final cases, i.e., $g_{i,i_{\textup{c}}}(c_i)$ for all $i_{\textup{c}}$.

We start off with the typical case, which we index by $i_{\textup{c}} = 1$. The contribution of $c_i$ to $g(\bold{c})$ in this case is the number of codewords in $\mathcal{OSC}^8_m$ starting with $c_{m-1} c_{m-2} \dots c_{i+1} c'_i$ from the left such that $c'_i < c_i$ according to the lexicographic ordering definition. As usual, the typical case is the unrestricted case. Thus, this number is the number of codewords in $\mathcal{OSC}^8_{i+1}$ starting with $c'_i$, for all $c'_i < c_i$, from the left. Consequently, we can write $g_{i,1}(c_i)$ as:
\begin{equation}\label{eqn_oslocorule1}
g_{i,1}(c_i) = N_{8,1}(i+1) + \sum_{j=1}^{a_i-1} N_{8,2.j}(i+1) = N_{8,1}(i+1) + (a_i-1) N_{8,2.1}(i+1), 
\end{equation}
where the second equality in (\ref{eqn_oslocorule1}) follows from that $N_{8,2.j}(i+1)$ is the same for all $j$. Recall the codeword correspondence in the proof of Theorem~\ref{thm_oslococard}. Substituting from (\ref{eqn_oslocogr2}) and (\ref{eqn_oslocogr6}) in (\ref{eqn_oslocorule1}) gives:
\begin{equation}\label{eqn_oslocorule2}
g_{i,1}(c_i) = \frac{1}{2} N_8(i+1) - 3N_8(i) + (a_i-1)N_8(i) = \frac{1}{2} N_8(i+1) + (a_i-4)N_8(i).
\end{equation}
This can be further expanded using (\ref{eqn_oslococard}) as follows:
\begin{align}\label{eqn_oslocorule2}
g_{i,1}(c_i) &= 4N_8(i) - \frac{1}{2}N_8(i-1) + 3N_8(i-2) + (a_i-4)N_8(i) \nonumber \\ 
&= a_iN_8(i) - \frac{1}{2}N_8(i-1) + 3N_8(i-2).
\end{align}

Next, we study the special case characterized by $c_{i+2} c_{i+1} c_i = 0\alpha\beta_1$, $\beta_1 \in \textup{GF}(8) \setminus \{0\}$, which we index by $i_{\textup{c}} = 2$. The contribution of $c_i$ to $g(\bold{c})$ in this case is the number of codewords in $\mathcal{OSC}^8_m$ starting with $c_{m-1} c_{m-2} \dots c_{i+3} 0\alpha c'_i$ from the left such that $c'_i < c_i = \beta_1$ according to the lexicographic ordering definition. This number is the number of codewords in $\mathcal{OSC}^8_{i+1}$ starting with $c'_i$, for all $c'_i < c_i$ such that $c'_i \neq 0$, from the left. Observe that the codewords starting with $0$ from the left in $\mathcal{OSC}^8_{i+1}$ must be omitted from the count since $c_{i+2} c_{i+1} c_i = 0\alpha0$ is not allowed in $\bold{c}$. Consequently, and using (\ref{eqn_oslocogr2}), we can derive $g_{i,2}(c_i)$ as follows:
\vspace{-0.2em}\begin{equation}\label{eqn_oslocorule3}
g_{i,2}(c_i) = \sum_{j=1}^{a_i-1} N_{8,2.j}(i+1) = (a_i-1) N_8(i).
\end{equation}

Next, we study the special case characterized by $c_{i+1} c_i = 0\beta''_2$, $\beta''_2 \in \{\alpha^2, \alpha^3, \dots, \alpha^6\}$, which we index by $i_{\textup{c}} = 3$. The contribution of $c_i$ to $g(\bold{c})$ in this case is the number of codewords in $\mathcal{OSC}^8_m$ starting with $c_{m-1} c_{m-2} \dots c_{i+2} 0 c'_i$ from the left such that $c'_i < c_i = \beta''_2$ according to the lexicographic ordering definition. This number is the number of codewords in $\mathcal{OSC}^8_{i+1}$ starting with $c'_i$, for all $c'_i < c_i$, from the left except for those starting with $\alpha0$ from the left. Observe that the codewords starting with $\alpha0$ from the left in $\mathcal{OSC}^8_{i+1}$ must be omitted from the count since $c_{i+1} c_i c_{i-1} = 0\alpha0$ is not allowed in $\bold{c}$. The number of these codewords to be omitted is $N_{8,1}(i)$. Consequently, we can write $g_{i,3}(c_i)$ as:
\begin{equation}\label{eqn_oslocorule4}
g_{i,3}(c_i) = N_{8,1}(i+1) + \sum_{j=1}^{a_i-1} N_{8,2.j}(i+1) - N_{8,1}(i).
\end{equation}
Substituting from (\ref{eqn_oslocogr2}) and (\ref{eqn_oslocogr6}) first, and then from (\ref{eqn_oslococard}) in (\ref{eqn_oslocorule4}) gives:
\vspace{-0.1em}\begin{align}\label{eqn_oslocorule5}
g_{i,3}(c_i) &= \frac{1}{2}N_8(i+1) - 3N_8(i) + (a_i-1)N_8(i) - \frac{1}{2}N_8(i) + 3N_8(i-1) \nonumber \\
&= 4N_8(i) - \frac{1}{2}N_8(i-1) + 3N_8(i-2) + \left ( a_i-\frac{9}{2} \right )N_8(i) + 3N_8(i-1) \nonumber \\
&= \left ( a_i-\frac{1}{2} \right )N_8(i) + \frac{5}{2}N_8(i-1) + 3N_8(i-2).
\end{align}

As for the special case characterized by $c_{i+1} c_i = \alpha^6\beta''_4$, $\beta''_4 \in \{\alpha^5, \alpha^6\}$, which we index by $i_{\textup{c}} = 4$, it can be shown that the contribution $g_{i,4}(c_i)$ has the exact same expression as that of $g_{i,3}(c_i)$ in (\ref{eqn_oslocorule5}) because of the symmetry between Group~1 and Group~3 in the OS-LOCO code $\mathcal{OSC}^8_m$.\vspace{+0.7em}

Now, we are ready to perform Step~5 of the method. We want to combine the different contributions for all cases into one expression, which is the OS-LOCO encoding-decoding rule.

Since we have three expressions for $g_{i,i_{\textup{c}}}(c_i)$, we need only two merging variables: $y_{i,1}$, for the case indexed by $i_{\textup{c}} = 2$, and $y_{i,2}$, for the cases indexed by $i_{\textup{c}} \in \{3,4\}$. If the two variables are zeros, the typical case contribution is switched on. These merging variables are set as shown in (\ref{eqn_oslocordef}).

Now, we pick the merging function $f^{\textup{mer}}_1 (\cdot) = a_i - y_{i,1} - \frac{1}{2}y_{i,2}$ for $N_8(i)$. This function results in $a_i$ for the case indexed by $i_{\textup{c}} = 1$, results in $a_i-1$ for the case indexed by $i_{\textup{c}} = 2$, and results in $a_i - \frac{1}{2}$ for the cases indexed by $i_{\textup{c}} \in \{3,4\}$.

We also pick the merging function $f^{\textup{mer}}_2 (\cdot) = \theta_i (1-y_{i,1}) (3y_{i,2} - \frac{1}{2})$ for $N_8(i-1)$, where $\theta$ is specified in (\ref{eqn_oslocordef}) as an indicator function of $c_i$ not being $0$. This function results in $-\frac{1}{2}$ for the case indexed by $i_{\textup{c}} = 1$, results in $0$ for the case indexed by $i_{\textup{c}} = 2$, and results in $\frac{5}{2}$ for the cases indexed by $i_{\textup{c}} \in \{3,4\}$.

We finally pick the merging function $f^{\textup{mer}}_3 (\cdot) = 3\theta_i (1-y_{i,1})$ for $N_8(i-2)$. This function results in $3$ for the case indexed by $i_{\textup{c}} = 1$, results in $0$ for the case indexed by $i_{\textup{c}} = 2$, and results in $3$ for the cases indexed by $i_{\textup{c}} \in \{3,4\}$.

Observe that the values of these merging functions at different cases are quite consistent with (\ref{eqn_oslocorule2}), (\ref{eqn_oslocorule3}), and (\ref{eqn_oslocorule5}). Observe also that if $c_i = 0$, this means $a_i = \theta_i = y_{i,1} = y_{i,2} = 0$, which in turn means $f^{\textup{mer}}_1 (\cdot) = f^{\textup{mer}}_2 (\cdot) = f^{\textup{mer}}_3 (\cdot) = 0$. 

Using these three merging functions, the unified expression representing the contribution of a symbol $c_i$ to the codeword index $g(\bold{c})$ can be written as:
\begin{align}\label{eqn_oslocorule6}
g_i(c_i) &= f^{\textup{mer}}_1 (\cdot) N_8(i) + f^{\textup{mer}}_2 (\cdot) N_8(i-1) + f^{\textup{mer}}_3 (\cdot) N_8(i-1) \nonumber \\
&= \left (a_i - y_{i,1} - \frac{1}{2}y_{i,2} \right ) N_8(i) + \theta_i (1-y_{i,1}) \left ( \left ( 3y_{i,2} - \frac{1}{2} \right ) N_8(i-1) + 3N_8(i-2) \right ).
\end{align}
The encoding-decoding rule of an OS-LOCO code is then:
\begin{equation}\label{eqn_oslocorule7}
g(\bold{c}) = \sum_{i=0}^{m-1} g_i(c_i) = \sum_{i=0}^{m-1} \left [ \left (a_i - y_{i,1} - \frac{1}{2}y_{i,2} \right ) N_8(i) + \theta_i (1-y_{i,1}) \left ( \left ( 3y_{i,2} - \frac{1}{2} \right ) N_8(i-1) + 3N_8(i-2) \right ) \right ], \nonumber
\end{equation}
which completes the proof.
\end{IEEEproof}

\begin{example}\label{example_3}
Consider the OS-LOCO code $\mathcal{OSC}^8_5$ ($m=5$). Using (\ref{eqn_oslococard}) and (\ref{eqn_oslococdef}), we get $N_8(-2) \triangleq 1/36$, $N_8(-1) \triangleq 1/6$, $N_8(0) \triangleq 1$, $N_8(1) \triangleq 8$, $N_8(2) = 64$, $N_8(3) = 510$, and $N_8(4) = 4064$. Consider the codeword $\bold{c} = c_4 c_3 c_2 c_1 c_0 = 0\alpha\alpha^6\alpha^5\alpha^4$ (level-equivalent $02765$) in $\mathcal{OSC}^8_5$. The case indexed by $i_{\textup{c}} = 1$ (the typical case) applies for $c_3$ and $c_0$, which means $y_{3,1} = y_{3,2} = y_{0,1} = y_{0,2} = 0$. The case indexed by $i_{\textup{c}} = 2$ applies for $c_2$, which means $y_{2,1} = 1$ and $y_{2,2} = 0$. The case indexed by $i_{\textup{c}} = 4$ applies for $c_1$, which means $y_{1,2} = 1$ and $y_{1,1} = 0$. Consequently, and using (\ref{eqn_oslocorule}), we get:
\begin{align}
g(\bold{c}=0\alpha\alpha^6\alpha^5\alpha^4) &=  \left [ 2N_8(3) - \frac{1}{2} N_8(2) + 3N_8(1) \right ] + \Big [ 6N_8(2) \Big ] \nonumber \\
&\hspace{+1.0em}+ \left [ \frac{11}{2} N_8(1) + \frac{5}{2} N_8(0) + 3N_8(-1) \right ] + \left [ 5N_8(0) - \frac{1}{2} N_8(-1) + 3N_8(-2) \right ] \nonumber \\
&= 1012 + 384 + 47 + 5 = 1448, \nonumber
\end{align}
which is consistent with the codeword index produced by the program we wrote to exhaustively generate and lexicographically order all OS-LOCO codewords in $\mathcal{OSC}^8_m$, with $m=5$ here. It corresponds to the binary message $00010110101000$ ($s = 14$).
\end{example}\vspace{+0.2em}

\textbf{Step~6)} We bridge in OS-LOCO codes with one GF$(8)$ symbol, i.e., one column of three bits, between each two consecutively written codewords as follows:
\begin{itemize}
\item If the RMS of a codeword and the LMS of the next codeword are both $\alpha^2$'s, bridge with $\alpha^3$.
\item If this is not the case, bridge with $\alpha^2$.
\end{itemize}
The mapping-demapping in (\ref{eqn_gf8map}) illustrates what exactly is written for bridging. This bridging is efficient in terms of low added redundancy, and optimal in terms of maximum protection of edge symbols. Other bridging methods are also possible.

\begin{table}
\caption{Rates, Normalized Rates, and Adder Sizes of OS-LOCO Codes $\mathcal{OSC}_m$ for Different Values of $m$. The Capacity Is $2.9944$, and the Normalized Capacity Is $0.9981$.}
\vspace{-0.5em}
\centering
\scalebox{1.00}
{
\begin{tabular}{|c|c|c|c|}
\hline
\makecell{$m$} & \makecell{$R_{\textup{OS-LOCO}}$} & \makecell{$R_{\textup{OS-LOCO}}^{\textup{n}}$}  & \makecell{Adder size} \\
\hline
$13$ & $2.7143$ & $0.9048$ & $38$ bits \\
\hline
$18$ & $2.7895$ & $0.9298$ & $53$ bits \\
\hline
$23$ & $2.8333$ & $0.9444$ & $68$ bits \\
\hline
$39$ & $2.9000$ & $0.9667$ & $116$ bits \\
\hline
$53$ & $2.9259$ & $0.9753$ & $158$ bits \\
\hline
$89$ & $2.9556$ & $0.9852$ & $266$ bits \\
\hline
\end{tabular}}
\label{table_3}
\vspace{-1.0em}
\end{table}

\begin{algorithm}
\caption{Encoding OS-LOCO Codes}
\begin{algorithmic}[1]
\State \textbf{Input:} Incoming stream of binary messages.
\State Use (\ref{eqn_oslococard}) and (\ref{eqn_oslococdef}) to compute $N_8(i)$, $i \in \{2, 3, 4, \dots\}$.
\State Specify $m$, the smallest $i$ in Step~2 to achieve the desired rate. Then, $s = \lfloor \log_2 N_8(m) \rfloor$.
\State \textbf{for} each incoming message $\bold{b}$ of length $s$ \textbf{do}
\State \hspace{2ex} Compute $g(\bold{c})=\textup{decimal}(\bold{b})$.
\State \hspace{2ex} Initialize $\textup{residual}$ with $g(\bold{c})$ and $c_i$ with $z'$ for $i \geq m$. \textit{($z'$ indicates out of codeword bounds)}
\State \hspace{2ex} \textbf{for} $i \in \{m-1, m-2, \dots, 0\}$ \textbf{do} \textit{(in order)}
\State \hspace{4ex} Initialize $\textup{symbol\_found}$ with $0$.
\State \hspace{4ex} Initialize $y_{1,i}(a_i)$, $y_{2,i}(a_i)$, and $\textup{contrib}(a_i)$ with $0$'s for $a_i \in \{1, 2, \dots, 7\}$.
\State \hspace{4ex} \textbf{if} ($c_{i+2} = 0$) $\land$ ($c_{i+1} = \alpha$) \textbf{then}
\State \hspace{6ex} Set $y_{1,i}(a_i) = 1$ for $a_i \in \{1, 2, \dots, 7\}$.
\State \hspace{4ex} \textbf{end if}
\State \hspace{4ex} \textbf{if} ($c_{i+1} = 0$) \textbf{then}
\State \hspace{6ex} Set $y_{2,i}(a_i) = 1$ for $a_i \in \{3, 4, \dots, 7\}$.
\State \hspace{4ex} \textbf{elseif} ($c_{i+1} = \alpha^6$) \textbf{then}
\State \hspace{6ex} Set $y_{2,i}(a_i) = 1$ for $a_i \in \{6, 7\}$.
\State \hspace{4ex} \textbf{end if}
\State \hspace{4ex} \textbf{for} $a_i \in \{1, 2, \dots 7\}$ \textbf{do}
\State \hspace{6ex} $\textup{contrib}(a_i) = \left ( a_i-y_{1,i}(a_i)-\frac{1}{2}y_{2,i}(a_i) \right )N_8(i) + (1-y_{1,i}(a_i)) \left [\left ( 3y_{2,i}(a_i)-\frac{1}{2} \right )N_8(i-1)+3N_8(i-2) \right ]$.
\State \hspace{4ex} \textbf{end for}
\State \hspace{4ex} \textbf{if} $\textup{residual} \geq \textup{contrib}(7)$ \textbf{then}
\State \hspace{6ex} Encode $c_i = \alpha^6$ and set $\textup{symbol\_found} = 1$. \textit{(level $a_i=7$)}
\State \hspace{6ex} $\textup{residual} \leftarrow \textup{residual} - \textup{contrib}(7)$.
\State \hspace{4ex} \textbf{else}
\State \hspace{6ex} \textbf{for} $a_i \in \{6, 5, \dots, 1\}$ \textbf{do}
\State \hspace{8ex} \textbf{if} $\textup{contrib}(a_i) \leq \textup{residual} < \textup{contrib}(a_i+1)$ \textbf{then}
\State \hspace{10ex} Encode $c_i = \mathcal{L}^{-1}(a_i)$ and set $\textup{symbol\_found} = 1$. \textit{(level $a_i = \mathcal{L}(c_i)$)}
\State \hspace{10ex} $\textup{residual} \leftarrow \textup{residual} - \textup{contrib}(a_i)$.
\State \hspace{10ex} \textbf{break}. \textit{(exit current loop)}
\State \hspace{8ex} \textbf{end if}
\State \hspace{6ex} \textbf{end for}
\State \hspace{4ex} \textbf{end if}
\State \hspace{4ex} \textbf{if} $\textup{symbol\_found} = 0$ \textbf{then}
\State \hspace{6ex} Encode $c_i = 0$. \textit{(level $a_i=0$)}
\State \hspace{4ex} \textbf{end if}
\State \hspace{4ex} \textbf{if} (not first codeword) $\land$ ($i = m-1$) \textbf{then}
\State \hspace{6ex} Bridge with either $\alpha^2$ or $\alpha^3$ before $c_{m-1}$ depending on the RMS of the previous codeword and $c_{m-1}$.
\State \hspace{4ex} \textbf{end if}
\State \hspace{2ex} \textbf{end for}
\State \textbf{end for}
\State \textbf{Output:} Outgoing stream of $8$-ary OS-LOCO codewords. \textit{(to be written on $3$ adjacent down tracks in the TDMR device after binary conversion and signaling)}
\end{algorithmic}
\label{alg_osenc}
\end{algorithm}

\begin{algorithm}
\caption{Decoding OS-LOCO Codes}
\begin{algorithmic}[1]
\State \textbf{Inputs:} Incoming stream of $8$-ary OS-LOCO codewords, in addition to $m$ and $s$. \textit{(stream after reading from $3$ adjacent down tracks in the TDMR device and $8$-ary conversion)}
\State Use (\ref{eqn_oslococard}) and (\ref{eqn_oslococdef}) to compute $N_8(i)$, $i \in \{2, 3, 4, \dots, m-1\}$.
\State \textbf{for} each incoming codeword $\bold{c}$ of length $m$ \textbf{do}
\State \hspace{2ex} Initialize $g(\bold{c})$ with $0$ and $c_i$ with $z'$ for $i \geq m$. \textit{($z'$ indicates out of codeword bounds)}
\State \hspace{2ex} \textbf{for} $i \in \{m-1, m-2, \dots, 0\}$ \textbf{do} \textit{(in order)}
\State \hspace{4ex} Initialize $y_{1,i}$, $y_{2,i}$, and $\theta_i$ with $0$'s.
\State \hspace{4ex} \textbf{if} ($c_{i+2} = 0$) $\land$ ($c_{i+1} = \alpha$) $\land$ ($c_i \in \{1, \alpha, \dots, \alpha^6\}$) \textbf{then}
\State \hspace{6ex} Set $y_{1,i} = 1$.
\State \hspace{4ex} \textbf{end if}
\State \hspace{4ex} \textbf{if} ($c_{i+1} = 0$) $\land$ ($c_i \in \{\alpha^2, \alpha^3, \dots, \alpha^6\}$) \textbf{then}
\State \hspace{6ex} Set $y_{2,i} = 1$.
\State \hspace{4ex} \textbf{elseif} ($c_{i+1} = \alpha^6$) $\land$ ($c_i \in \{\alpha^5, \alpha^6\}$) \textbf{then}
\State \hspace{6ex} Set $y_{2,i} = 1$.
\State \hspace{4ex} \textbf{end if}
\State \hspace{4ex} \textbf{if} $c_i \neq 0$ \textbf{then} \textit{(same as $a_i \neq 0$)}
\State \hspace{6ex} Set $\theta_i = 1$.
\State \hspace{4ex} \textbf{end if}
\State \hspace{4ex} Set $a_i = \mathcal{L}(c_i)$.
\State \hspace{4ex} $g(\bold{c}) \leftarrow g(\bold{c}) + \left ( a_i-y_{1,i}-\frac{1}{2}y_{2,i} \right )N_8(i) + \theta_i (1-y_{1,i}) \left [\left ( 3y_{2,i}-\frac{1}{2} \right )N_8(i-1)+3N_8(i-2) \right ]$.
\State \hspace{2ex} \textbf{end for}
\State \hspace{2ex} Compute $\bold{b}=\textup{binary}(g(\bold{c}))$, which has length $s$.
\State \hspace{2ex} Ignore the next bridging symbol.
\State \textbf{end for}
\State \textbf{Output:} Outgoing stream of binary messages.
\end{algorithmic}
\label{alg_osdec}
\end{algorithm}

In our TDMR system, a transition is counted on the level of the $3 \times 1$ column after signaling. With our bridging, the maximum number of consecutive $3 \times 1$ columns with no transition after writing via an OS-LOCO code $\mathcal{OSC}^8_m$ is $m+1$. This finite maximum is achieved without removing any codewords from the OS-LOCO code for self-clocking. Thus, an OS-LOCO code associated with the aforementioned bridging is inherently self-clocked.

Given our bridging method, the rate, in input bits per coded symbol, and the normalized rate of an OS-LOCO code $\mathcal{OSC}^8_m$ are:
\begin{equation}\label{eqn_oslocorate}
R_{\textup{OS-LOCO}} = \frac{s}{m+1} = \frac{\lfloor N_8(m) \rfloor}{m+1}, \textup{ } R^{\textup{n}}_{\textup{OS-LOCO}} = \frac{1}{3} R_{\textup{OS-LOCO}} = \frac{\lfloor N_8(m) \rfloor}{3(m+1)}.
\end{equation}
It is easy to deduce that OS-LOCO codes are capacity-achieving. To demonstrate that, Table~\ref{table_3} gives the rates and normalized rates of OS-LOCO codes with different lengths. Table~\ref{table_3} shows that the rates of OS-LOCO codes are close to capacity even at moderate lengths. The table also shows that OS-LOCO codes incur very limited redundancy.

Next, we introduce the encoding algorithm of OS-LOCO codes, which is Algorithm~\ref{alg_osenc}, and the decoding algorithm of OS-LOCO codes, which is Algorithm~\ref{alg_osdec}.

To reduce runtime processing, all the terms containing a factor multiplied by a cardinality in Algorithm~\ref{alg_osenc} and Algorithm~\ref{alg_osdec} are computed offline and stored in memory, for all possible factors except factors of the form $2$ raised to some power (negative or positive). The same applies for all same-symbol operations. Thus, the main runtime operations in both algorithms are additions, subtractions, and comparisons, which are all performed by adders. That is the reason why the adder size governs the complexity of the encoding and decoding procedures, demonstrating simplicity (see also Table~\ref{table_3}). Observe that both complexity and storage overhead can be further reduced via arithmetic tricks \cite{ahh_qaloco}. These algorithms complete the process of revealing the secret arithmetic of allowed and forbidden patterns in an OS-LOCO code.

\begin{remark}
While designing OS-LOCO codes, we opted to use simple bridging and not to use the bridging symbol/column to encode information. It is important to note that it is possible to bridge for OS-LOCO codes with one symbol out of the set $\{1, \alpha^2, \alpha^3, \alpha^5\}$ that is picked based on two input bits. Thus, the following notable normalized rate gain can be achieved:
\begin{equation}\label{eqn_osrgain}
\overline{R}^{\textup{n}}_{\textup{OS-LOCO}} - R^{\textup{n}}_{\textup{OS-LOCO}} = \frac{2}{3(m+1)}.
\end{equation}
Some changes for self-clocking and modifications to the encoding-decoding algorithms will be required.
\end{remark}

\subsection{Optimal Plus LOCO Codes}

\begin{figure}
\vspace{-0.5em}
\center
\includegraphics[trim={2.3in 0.7in 2.3in 0.7in}, width=3.5in]{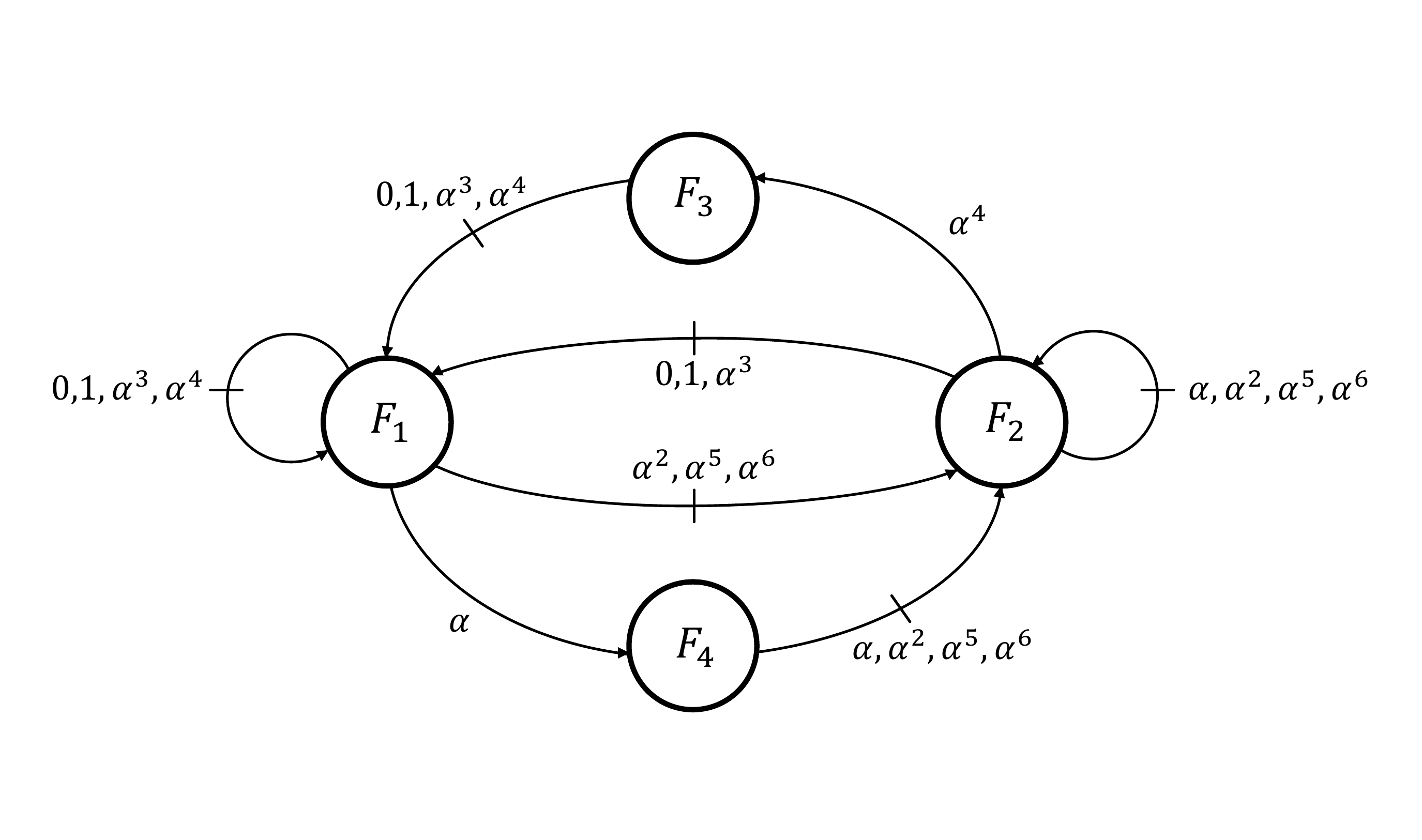}
\vspace{-0.5em}
\caption{An FSTD representing an infinite $\mathcal{OP}^8$-constrained sequence (patterns in $\mathcal{OP}^8$ are prevented).}
\label{fig_3}
\vspace{-0.5em}
\end{figure}

We now present our optimal plus LOCO (OP-LOCO) codes, which are codes preventing the PIS patterns shown in Fig.~\ref{fig_2} within each group of three adjacent down tracks. From (\ref{eqn_gf8map}), these $32$ PIS patterns map to the $32$ GF$(8)$ patterns $\overline{\beta}_1\alpha\beta_1$, for all $\overline{\beta}_1, \beta_1 \in \{0, 1, \alpha^3, \alpha^4\}$, and $\overline{\beta}_2\alpha^4\beta_2$, for all $\overline{\beta}_2, \beta_2 \in \{\alpha, \alpha^2, \alpha^5, \alpha^6\}$, which have the level-equivalent patterns $\mathcal{L}(\overline{\beta}_1)2\mathcal{L}(\beta_1)$, for all $\mathcal{L}(\overline{\beta}_1), \mathcal{L}(\beta_1) \in \{0, 1, 4, 5\}$, and $\mathcal{L}(\overline{\beta}_2)5\mathcal{L}(\beta_2)$, for all $ \mathcal{L}(\overline{\beta}_2), \mathcal{L}(\beta_2) \in \{2, 3, 6, 7\}$. The FSTD of an infinite $8$-ary constrained sequence in which these $32$ patterns are prevented is in Fig.~\ref{fig_3}. The corresponding adjacency matrix is:
\begin{gather*}\label{eqn_oplocoadj}
\bold{F}=
\begin{bmatrix}
4 & 3 & 0 & 1\\
3 & 4 & 1 & 0\\
4 & 0 & 0 & 0\\
0 & 4 & 0 & 0
\end{bmatrix}.
\end{gather*}
The capacity $C$, in input bits per coded symbol, and the normalized capacity $C^{\textup{n}}$ accordingly are:
\begin{equation}\label{eqn_capoploco}
C = \log_2(\lambda_{\textup{max}}(\bold{F})) = \log_2 7.5311 = 2.9129 \textup{ and } C^{\textup{n}} = \frac{1}{3} C = 0.9710,
\end{equation}
where $\lambda_{\textup{max}}(\bold{F})$ is the maximum real positive eigenvalue of $\bold{F}$ \cite{shan_const, perron_frobenius}.

Denote an OP-LOCO code of length $m$ by $\mathcal{OPC}^8_m$. The definition of the code is exactly the definition of a generic LOCO code, which is Definition~\ref{def_genloco}, with $q=8$, $\mathcal{C}^q_m = \mathcal{OPC}^8_m$, and $\mathcal{T}$ given by:
\begin{equation}\label{eqn_toploco}
\mathcal{T} = \mathcal{OP}^8 \triangleq \{\overline{\beta}_1\alpha\beta_1, \overline{\beta}_2\alpha^4\beta_2, \textup{ } \forall \overline{\beta}_1, \beta_1 \in \{0, 1, \alpha^3, \alpha^4\} \textup{ and } \forall \overline{\beta}_2, \beta_2 \in \{\alpha, \alpha^2, \alpha^5, \alpha^6\}\}.
\end{equation}
Both $\bold{c}$ in $\mathcal{C}^q_m = \mathcal{OPC}^8_m$ and $g(\bold{c})$ are used as they were in Section~\ref{sec_gen}. The cardinality of $\mathcal{OPC}^8_m$ is $N_q(m) = N_8(m)$. Again, we could not provide a table as an example listing all the codewords of specific codes because there are way too many codewords for any length $m \geq 4$. However, we will provide an example illustrating the encoding-decoding rule of the code.

Now, we will apply the steps of the general method to find out how to encode and decode OP-LOCO codes using a simple encoding-decoding rule.\vspace{+0.7em}

\textbf{Step~1)} Using the patterns in $\mathcal{OP}^8$, we determine initial groups of $\mathcal{OPC}^8_m$ as shown below. Observe that it is more convenient here to perform some group merging during the procedure.
\begin{itemize}
\item For the patterns $0\alpha\beta_1$, $\beta_1 \in \{0, 1, \alpha^3, \alpha^4\}$, there is an initial group having all the codewords starting with $0\alpha\beta_2$, $\beta_2 \in \{\alpha, \alpha^2, \alpha^5, \alpha^6\}$, from the left. There are seven more initial groups having all the codewords starting with $0\beta_3$, a group for each $\beta_3 \in \textup{GF}(8) \setminus \{\alpha\}$, from the left. There are seven more initial groups having all the codewords starting with non-zero symbols, a group for each element in $\textup{GF}(8) \setminus \{0\}$, from the left. We do the same for the patterns $1\alpha\beta_1$, the patterns $\alpha^3\alpha\beta_1$, and the patterns $\alpha^4\alpha\beta_1$, $\beta_1 \in \{0, 1, \alpha^3, \alpha^4\}$.
\item For the patterns $\alpha^6\alpha^4\beta_2$, $\beta_2 \in \{\alpha, \alpha^2, \alpha^5, \alpha^6\}$, there is an initial group having all the codewords starting with $\alpha^6\alpha^4\beta_1$, $\beta_1 \in \{0, 1, \alpha^3, \alpha^4\}$, from the left. There are seven more initial groups having all the codewords starting with $\alpha^6\beta_4$, a group for each $\beta_4 \in \textup{GF}(8) \setminus \{\alpha^4\}$, from the left. There are seven more initial groups having all the codewords starting with non-$\alpha^6$ symbols, a group for each element in $\textup{GF}(8) \setminus \{\alpha^6\}$, from the left. We do the same for the patterns $\alpha\alpha^4\beta_2$, the patterns $\alpha^2\alpha^4\beta_2$, and the patterns $\alpha^5\alpha^4\beta_2$, $\beta_2 \in \{\alpha, \alpha^2, \alpha^5, \alpha^6\}$.
\end{itemize}

After operating on these initial groups, we end up with eight (final) groups covering all the OP-LOCO codewords in $\mathcal{OPC}^8_m$: \textit{Group~1}, which contains all the codewords starting with $0$ from the left, \textit{Group~2}, which contains all the codewords starting with $1$ from the left, \textit{Group~3}, which contains all the codewords starting with $\alpha$ from the left, \textit{Group~4}, which contains all the codewords starting with $\alpha^2$ from the left, \textit{Group~5}, which contains all the codewords starting with $\alpha^3$ from the left, \textit{Group~6}, which contains all the codewords starting with $\alpha^4$ from the left, \textit{Group~7}, which contains all the codewords starting with $\alpha^5$ from the left, and \textit{Group~8}, which contains all the codewords starting with $\alpha^6$ from the left. The groups are defined for $m \geq 2$.

Group~1 is further partitioned into three subgroups: \textit{Subgroup~1.1}, which contains all the codewords starting with $0\beta'_3$, $\beta'_3 \in \{0, 1\}$, from the left, \textit{Subgroup~1.2}, which contains all the codewords starting with $0\alpha\beta_2$ from the left, and \textit{Subgroup~1.3}, which contains all the codewords starting with $0\beta''_3$, $\beta''_3 \in \{\alpha^2, \alpha^3, \dots, \alpha^6\}$, from the left. The same partitioning to subgroups applies to Group~2, Group~5, and Group~6.

Group~8 is further partitioned into three subgroups: \textit{Subgroup~8.1}, which contains all the codewords starting with $\alpha^6\beta'_4$, $\beta'_4 \in \{0, 1, \dots, \alpha^3\}$, from the left, \textit{Subgroup~8.2}, which contains all the codewords starting with $\alpha^6\alpha^4\beta_1$ from the left, and \textit{Subgroup~8.3}, which contains all the codewords starting with $\alpha^6\beta''_4$, $\beta''_4 \in \{\alpha^5, \alpha^6\}$, from the left. The same partitioning to subgroups applies to Group~3, Group~4, and Group~7.\vspace{+0.7em}

\textbf{Step~2)} Theorem~\ref{thm_oplococard} gives the cardinality of an OP-LOCO code.

\begin{theorem}\label{thm_oplococard}
The cardinality of an OP-LOCO code $\mathcal{OPC}^8_m$ is given by:
\begin{equation}\label{eqn_oplococard}
N_8(m) = 7N_8(m-1) + 4N_8(m-2), \text{ } m \geq 2,
\end{equation}
where the defined cardinalities are:
\begin{equation}\label{eqn_oplococdef}
N_8(0) \triangleq 2 \text{ and } N_8(1) \triangleq 8.
\end{equation}
\end{theorem}

\begin{IEEEproof}
We note that an OP-LOCO code $\mathcal{OPC}^8_m$ is symmetric, i.e., the cardinality of all eight groups in the code is the same, because of the nature of forbidden patterns in $\mathcal{OP}^8$. Thus, we only derive a recursive cardinality formula for Group~1 of $\mathcal{OPC}^8_m$, then multiply by $8$. We work on Group~1 of $\mathcal{OPC}^8_m$.

As for Subgroup~1.1, each codeword starting with $0\beta'_3$ from the left in this subgroup corresponds to a codeword in $\mathcal{OPC}^8_{m-1}$ that starts with the same $\beta'_3$ from the left such that they share the remaining $m-2$ RMSs. This correspondence is bijective. Since $\beta'_3$ is in $\{0, 1\}$ and the code $\mathcal{OPC}^8_{m-1}$ is also symmetric, the cardinality of Subgroup~1.1 is:
\begin{equation}\label{eqn_oplocogr1}
N_{8,1.1}(m) = \frac{2}{8} N_8(m-1) = \frac{1}{4} N_8(m-1).
\end{equation}

As for Subgroup~1.2, each codeword starting with $0\alpha\beta_2$ from the left in this subgroup corresponds to a codeword in $\mathcal{OPC}^8_{m-2}$ that starts with the same $\beta_2$ from the left such that they share the remaining $m-3$ RMSs. This correspondence is bijective. Since $\beta_2$ is in $\{\alpha, \alpha^2, \alpha^5, \alpha^6\}$ and the code $\mathcal{OPC}^8_{m-2}$ is also symmetric, the cardinality of Subgroup~1.2 is:
\begin{equation}\label{eqn_oplocogr2}
N_{8,1.2}(m) = \frac{4}{8} N_8(m-2) = \frac{1}{2} N_8(m-2).
\end{equation}

As for Subgroup~1.3, each codeword starting with $0\beta''_3$ from the left in this subgroup corresponds to a codeword in $\mathcal{OPC}^8_{m-1}$ that starts with the same $\beta''_3$ from the left such that they share the remaining $m-2$ RMSs. This correspondence is bijective. Since $\beta''_3$ is in $\{\alpha^2, \alpha^3, \dots, \alpha^6\}$ and the code $\mathcal{OPC}^8_{m-1}$ is also symmetric, the cardinality of Subgroup~1.3 is:
\begin{equation}\label{eqn_oplocogr3}
N_{8,1.3}(m) = \frac{5}{8} N_8(m-1).
\end{equation}

Using (\ref{eqn_oplocogr1}), (\ref{eqn_oplocogr2}), and (\ref{eqn_oplocogr3}), the cardinality of Group~1 in $\mathcal{OPC}^8_m$ then is:
\begin{equation}\label{eqn_oplocogr4}
N_{8,1}(m) = \sum_{i=1}^3 N_{8,1.i}(m) = \frac{7}{8} N_8(m-1) + \frac{1}{2} N_8(m-2).
\end{equation}
From (\ref{eqn_oplocogr4}) and using the symmetry of the code, the cardinality of $\mathcal{OPC}^8_m$ is:
\begin{equation}\label{eqn_oplocogr5}
N_8(m) = \sum_{i=1}^8 N_{8,i}(m) = 8 N_{8,1}(m) = 7N_8(m-1) + 4N_8(m-2), \textup{ } m \geq 2. \nonumber
\end{equation}

As for the defined cardinalities, it is clear that $N_8(1) \triangleq 8$. We also know that $N_8(2) = 8^2 = 64$ since the length of a PIS pattern is $3$, i.e., no sequences to eliminate at that length. Consequently, and using the proved (\ref{eqn_oplococard}),
\begin{equation}\label{eqn_oplocogr6}
64 = 7 \times 8 + 4N_8(0) \implies N_8(0) \triangleq 2.
\end{equation}
Note that we can also compute $N_8(-1)$. However, it will never be used neither to compute cardinalities nor in the encoding and decoding procedures as we shall see shortly. Computing the defined cardinalities completes the proof.
\end{IEEEproof}\vspace{+0.7em}

\textbf{Step~3)} We now specify the special cases. Using the patterns in $\mathcal{OP}^8$, we determine initial special cases for the OP-LOCO code $\mathcal{OPC}^8_m$ as shown below. Observe that it is more convenient here to also perform some special case processing during the procedure as we did for the groups.
\begin{itemize}
\item For the patterns $0\alpha\beta_1$, $\beta_1 \in \{0, 1, \alpha^3, \alpha^4\}$, one initial special case is $c_{i+2} c_{i+1} c_i = 0\alpha\beta'_2$, $\beta'_2 \in \{\alpha, \alpha^2\}$. Another initial special case is $c_{i+2} c_{i+1} c_i = 0\alpha\beta''_2$, $\beta'_2 \in \{\alpha^5, \alpha^6\}$. A third initial special case is $c_{i+1} c_i = 0\beta''_3$, $\beta''_3 \in \{\alpha^2, \alpha^3, \dots, \alpha^6\}$. We do the same for the patterns $1\alpha\beta_1$, the patterns $\alpha^3\alpha\beta_1$, and the patterns $\alpha^4\alpha\beta_1$, $\beta_1 \in \{0, 1, \alpha^3, \alpha^4\}$.
\item For the patterns $\alpha^6\alpha^4\beta_2$, $\beta_2 \in \{\alpha, \alpha^2, \alpha^5, \alpha^6\}$, one initial special case is $c_{i+2} c_{i+1} c_i = \alpha^6\alpha^4\beta''_1$, $\beta''_1 \in \{\alpha^3, \alpha^4\}$. Another initial special case is $c_{i+1} c_i = \alpha^6\beta''_4$, $\beta''_4 \in \{\alpha^5, \alpha^6\}$. We do the same for the patterns $\alpha\alpha^4\beta_2$, the patterns $\alpha^2\alpha^4\beta_2$, and the patterns $\alpha^5\alpha^4\beta_2$, $\beta_2 \in \{\alpha, \alpha^2, \alpha^5, \alpha^6\}$.
\end{itemize}

After further processing, we end up with six (final) cases for $c_i$ based on $c_i$ and its preceding symbols: a \textit{special case} for $c_{i+2} c_{i+1} c_i = \overline{\beta}_1\alpha\beta'_2$, a \textit{special case} for $c_{i+2} c_{i+1} c_i = \overline{\beta}_1\alpha\beta''_2$, a \textit{special case} for $c_{i+1} c_i = \overline{\beta}_1\beta''_3$, a \textit{special case} for $c_{i+2} c_{i+1} c_i = \overline{\beta}_2\alpha^4\beta''_1$, a \textit{special case} for $c_{i+1} c_i = \overline{\beta}_2\beta''_4$, and the \textit{typical case}. Recall that $\overline{\beta}_1 \in \{0, 1, \alpha^3, \alpha^4\}$ and $\overline{\beta}_2 \in \{\alpha, \alpha^2, \alpha^5, \alpha^6\}$, while the rest of variables are specified above. The typical case is simply the case when neither of the five special cases is enabled and $c_i \neq 0$. Observe that symmetry here enables merging special cases via their LMS, $c_{i+2}$ or $c_{i+1}$. As usual, the priority of a case increases as its sequence length increases.\vspace{+0.7em}

\textbf{Steps~4 and 5)} Theorem~\ref{thm_oplocorule} gives the encoding-decoding rule of an OP-LOCO code $\mathcal{OPC}^8_m$. Recall that $a_i \triangleq \mathcal{L}(c_i)$.

\begin{theorem}\label{thm_oplocorule}
Let $\bold{c}$ be an OP-LOCO codeword in $\mathcal{OPC}^8_m$. The relation between the lexicographic index $g(\bold{c})$ of this codeword and the codeword itself is given by:
\begin{equation}\label{eqn_oplocorule}
g(\bold{c}) = \sum_{i=0}^{m-1} \left [ \frac{1}{8} (a_i -2y_{i,1} - 4y_{i,2} - y_{i,3}) N_8(i+1) + \frac{1}{2}y_{i,3} N_8(i) \right ],
\end{equation}
where $y_{i,1}$, $y_{i,2}$, and $y_{i,3}$ are specified as follows:
\begin{align}\label{eqn_oplocordef}
y_{i,1} &= 1 \text{ if } c_{i+2} c_{i+1} c_i = \overline{\beta}_1\alpha\beta'_2, \text{ } \overline{\beta}_1 \in \{0, 1, \alpha^3, \alpha^4\},\beta'_2 \in \{\alpha, \alpha^2\}, \text{ else}, \nonumber \\
y_{i,1} &= 1 \text{ if } c_{i+2} c_{i+1} c_i = \overline{\beta}_2\alpha^4\beta''_1, \text{ } \overline{\beta}_2 \in \{\alpha, \alpha^2, \alpha^5, \alpha^6\},\beta''_1 \in \{\alpha^3, \alpha^4\}, \text{ and } y_{i,1} = 0 \text{ otherwise}, \nonumber \\
y_{i,2} &= 1 \text{ if } c_{i+2} c_{i+1} c_i = \overline{\beta}_1\alpha\beta''_2, \text{ } \overline{\beta}_1 \in \{0, 1, \alpha^3, \alpha^4\},\beta''_2 \in \{\alpha^5, \alpha^6\}, \text{ and } y_{i,2} = 0 \text{ otherwise}, \nonumber \\
y_{i,3} &= 1 \text{ if } c_{i+1} c_i = \overline{\beta}_1\beta''_3, \text{ } \overline{\beta}_1 \in \{0, 1, \alpha^3, \alpha^4\},\beta''_3 \in \{\alpha^2, \alpha^3, \dots, \alpha^6\} \textup{ s.t. } y_{i,1} = y_{i,2} = 0, \text{ else}, \nonumber \\
y_{i,3} &= 1 \text{ if } c_{i+1} c_i = \overline{\beta}_2\beta''_4, \text{ } \overline{\beta}_2 \in \{\alpha, \alpha^2, \alpha^5, \alpha^6\},\beta''_4 \in \{\alpha^5, \alpha^6\} \textup{ s.t. } y_{i,1} = y_{i,2} = 0, \text{ and } y_{i,3} = 0 \text{ otherwise}.
\end{align}
\end{theorem}

\begin{IEEEproof}
First, we perform Step~4 of the method. We aim at computing the contribution of each OP-LOCO codeword symbol $c_i$ to the codeword index $g(\bold{c})$ for the six final cases, i.e., $g_{i,i_{\textup{c}}}(c_i)$ for all $i_{\textup{c}}$.

We start off with the typical case, which we index by $i_{\textup{c}} = 1$. The contribution of $c_i$ to $g(\bold{c})$ in this case is the number of codewords in $\mathcal{OPC}^8_m$ starting with $c_{m-1} c_{m-2} \dots c_{i+1} c'_i$ from the left such that $c'_i < c_i$ according to the lexicographic ordering definition. As usual, the typical case is the unrestricted case. Thus, this number is the number of codewords in $\mathcal{OPC}^8_{i+1}$ starting with $c'_i$, for all $c'_i < c_i$, from the left. Consequently, and using symmetry, we can write $g_{i,1}(c_i)$ as:
\vspace{+0.1em}\begin{equation}\label{eqn_oplocorule1}
g_{i,1}(c_i) = \sum_{j=1}^{a_i} N_{8,j}(i+1) = a_i N_{8,1}(i+1) = \frac{1}{8} a_i N_8(i+1).
\end{equation}

Next, we study the special case characterized by $c_{i+2} c_{i+1} c_i = \overline{\beta}_1\alpha\beta'_2$, $\overline{\beta}_1 \in \{0, 1, \alpha^3, \alpha^4\}$ and $\beta'_2 \in \{\alpha, \alpha^2\}$, which we index by $i_{\textup{c}} = 2$. The contribution of $c_i$ to $g(\bold{c})$ in this case is the number of codewords in $\mathcal{OPC}^8_m$ starting with $c_{m-1} c_{m-2} \dots c_{i+3} \overline{\beta}_1\alpha c'_i$ from the left such that $c'_i < c_i = \beta'_2$ according to the lexicographic ordering definition. This number is the number of codewords in $\mathcal{OPC}^8_{i+1}$ starting with $c'_i$, for all $c'_i < c_i$ such that $c'_i \notin \{0, 1\}$, from the left. Observe that the codewords starting with $0$ or $1$ from the left in $\mathcal{OPC}^8_{i+1}$ must be omitted from the count since $\overline{\beta}_1\alpha0$ and $\overline{\beta}_1\alpha1$ are forbidden patterns (PIS patterns). Consequently, and using symmetry, we can derive $g_{i,2}(c_i)$ as follows:
\begin{equation}\label{eqn_oplocorule2}
g_{i,2}(c_i) = \sum_{j=1}^{a_i-2} N_{8,j}(i+1) =  \frac{1}{8} (a_i-2) N_8(i+1).
\end{equation}

Next, we study the special case characterized by $c_{i+2} c_{i+1} c_i = \overline{\beta}_1\alpha\beta''_2$, $\overline{\beta}_1 \in \{0, 1, \alpha^3, \alpha^4\}$ and $\beta''_2 \in \{\alpha^5, \alpha^6\}$, which we index by $i_{\textup{c}} = 3$. The contribution of $c_i$ to $g(\bold{c})$ in this case is the number of codewords in $\mathcal{OPC}^8_m$ starting with $c_{m-1} c_{m-2} \dots c_{i+3} \overline{\beta}_1\alpha c'_i$ from the left such that $c'_i < c_i = \beta''_2$ according to the lexicographic ordering definition. This number is the number of codewords in $\mathcal{OPC}^8_{i+1}$ starting with $c'_i$, for all $c'_i < c_i$ such that $c'_i \notin \{0, 1, \alpha^3, \alpha^4\}$, from the left. Observe that the codewords starting with $0$, $1$, $\alpha^3$, or $\alpha^4$ from the left in $\mathcal{OPC}^8_{i+1}$ must be omitted from the count since $\overline{\beta}_1\alpha0$, $\overline{\beta}_1\alpha1$, $\overline{\beta}_1\alpha\alpha^3$, and $\overline{\beta}_1\alpha\alpha^4$ are forbidden patterns (PIS patterns). Consequently, and using symmetry, we can derive $g_{i,3}(c_i)$ as follows:
\begin{equation}\label{eqn_oplocorule3}
g_{i,3}(c_i) = \sum_{j=1}^{a_i-4} N_{8,j}(i+1) =  \frac{1}{8} (a_i-4) N_8(i+1).
\end{equation}

Next, we study the special case characterized by $c_{i+1} c_i = \overline{\beta}_1\beta''_3$, $\overline{\beta}_1 \in \{0, 1, \alpha^3, \alpha^4\}$ and $\beta''_3 \in \{\alpha^2, \alpha^3, \dots, \alpha^6\}$, which we index by $i_{\textup{c}} = 4$. The contribution of $c_i$ to $g(\bold{c})$ in this case is the number of codewords in $\mathcal{OPC}^8_m$ starting with $c_{m-1} c_{m-2} \dots c_{i+2} \overline{\beta}_1 c'_i$ from the left such that $c'_i < c_i = \beta''_3$ according to the lexicographic ordering definition. Looking from the right, these codewords correspond to codewords in $\mathcal{OPC}^8_{i+1}$. We divide such codewords in $\mathcal{OPC}^8_{i+1}$ into two portions. The first portion has the codewords in $\mathcal{OPC}^8_{i+1}$ starting with $c'_i$, for all $c'_i < c_i$ such that $c'_i \neq \alpha$, from the left. Let the number of codewords in this portion be $g'_{i,4}(c_i)$. Consequently, and using symmetry, we can derive $g'_{i,4}(c_i)$ as follows:
\begin{equation}\label{eqn_oplocorule4}
g'_{i,4}(c_i) = \sum_{j=1}^{a_i-1} N_{8,j}(i+1) =  \frac{1}{8} (a_i-1) N_8(i+1).
\end{equation}
The second portion has the codewords in $\mathcal{OPC}^8_{i+1}$ starting with $c'_i = \alpha$ from the left. Let the number of codewords in this portion be $g''_{i,4}(c_i)$. From the set of forbidden patterns $\mathcal{OP}^8$, we know that $\overline{\beta}_1\alpha$ in an OP-LOCO codeword has to be followed by $\beta_2 \in \{\alpha, \alpha^2, \alpha^5, \alpha^6\}$. Recall the codeword correspondence in the proof of Theorem~\ref{thm_oplococard}. Consequently, and aided by (\ref{eqn_oplocogr2}), we can derive $g''_{i,4}(c_i)$ as follows:
\begin{equation}\label{eqn_oplocorule5}
g''_{i,4}(c_i) = \frac{4}{8} N_8(i) = \frac{1}{2} N_8(i).
\end{equation}
Using (\ref{eqn_oplocorule4}) and (\ref{eqn_oplocorule5}), we get:
\begin{equation}\label{eqn_oplocorule6}
g_{i,4}(c_i) = g'_{i,4}(c_i) + g''_{i,4}(c_i) = \frac{1}{8} (a_i-1) N_8(i+1) + \frac{1}{2} N_8(i).
\end{equation}

As for the special case characterized by $c_{i+2} c_{i+1} c_i = \overline{\beta}_2\alpha^4\beta''_1$, $\overline{\beta}_2 \in \{\alpha, \alpha^2, \alpha^5, \alpha^6\}$ and $\beta''_1 \in \{\alpha^3, \alpha^4\}$, which we index by $i_{\textup{c}} = 5$, it can be shown that the contribution $g_{i,5}(c_i)$ has the exact same expression as that of $g_{i,2}(c_i)$ in (\ref{eqn_oplocorule2}) because of the symmetry of the OP-LOCO code $\mathcal{OPC}^8_m$. As for the special case characterized by $c_{i+1} c_i = \overline{\beta}_2\beta''_4$, $\overline{\beta}_2 \in \{\alpha, \alpha^2, \alpha^5, \alpha^6\}$ and $\beta''_4 \in \{\alpha^5, \alpha^6\}$, which we index by $i_{\textup{c}} = 6$, it can be shown that the contribution $g_{i,6}(c_i)$ has the exact same expression as that of $g_{i,4}(c_i)$ in (\ref{eqn_oplocorule6}) because of the symmetry of the OP-LOCO code $\mathcal{OPC}^8_m$.\vspace{+0.7em}

Now, we are ready to perform Step~5 of the method. We want to combine the different contributions for all cases into one expression, which is the OP-LOCO encoding-decoding rule.

Since we have four expressions for $g_{i,i_{\textup{c}}}(c_i)$, we need only three merging variables: $y_{i,1}$, for the cases indexed by $i_{\textup{c}} \in \{2, 5\}$, $y_{i,2}$, for the case indexed by $i_{\textup{c}} = 3$, and $y_{i,3}$, for the cases indexed by $i_{\textup{c}} \in \{4, 6\}$ (lower priority). If the three variables are zeros, the typical case contribution is switched on. These merging variables are set as shown in (\ref{eqn_oplocordef}).

Now, we pick the merging function $f^{\textup{mer}}_0 (\cdot) = \frac{1}{8} (a_i -2y_{i,1} - 4y_{i,2} - y_{i,3})$ for $N_8(i+1)$. This function results in $\frac{1}{8} a_i$ for the case indexed by $i_{\textup{c}} = 1$, results in $\frac{1}{8}(a_i - 2)$ for the cases indexed by $i_{\textup{c}} \in \{2, 5\}$, results in $\frac{1}{8}(a_i - 4)$ for the case indexed by $i_{\textup{c}} = 3$, and results in $\frac{1}{8}(a_i - 1)$ for the cases indexed by $i_{\textup{c}} \in \{4, 6\}$.

We also pick the merging function $f^{\textup{mer}}_1 (\cdot) = \frac{1}{2}y_{i,3}$ for $N_8(i)$. This function results in $0$ for the cases indexed by $i_{\textup{c}} \in \{1, 2, 3, 5\}$, and results in $\frac{1}{2}$ for the cases indexed by $i_{\textup{c}} \in \{4, 6\}$.

Observe that the values of these merging functions at different cases are quite consistent with (\ref{eqn_oplocorule1}), (\ref{eqn_oplocorule2}), (\ref{eqn_oplocorule3}), and (\ref{eqn_oplocorule6}). Observe also that if $c_i = 0$, this means $a_i = y_{i,1} = y_{i,2} = y_{i,3} = 0$, which in turn means $f^{\textup{mer}}_0 (\cdot) = f^{\textup{mer}}_1 (\cdot) = 0$. 

Using these two merging functions, the unified expression representing the contribution of a symbol $c_i$ to the codeword index $g(\bold{c})$ can be written as:
\begin{align}\label{eqn_oplocorule7}
g_i(c_i) &= f^{\textup{mer}}_0 (\cdot) N_8(i+1) + f^{\textup{mer}}_1 (\cdot) N_8(i) \nonumber \\
&= \frac{1}{8} (a_i -2y_{i,1} - 4y_{i,2} - y_{i,3}) N_8(i+1) + \frac{1}{2}y_{i,3} N_8(i).
\end{align}
The encoding-decoding rule of an OP-LOCO code is then:
\begin{equation}\label{eqn_oplocorule8}
g(\bold{c}) = \sum_{i=0}^{m-1} g_i(c_i) = \sum_{i=0}^{m-1} \left [ \frac{1}{8} (a_i -2y_{i,1} - 4y_{i,2} - y_{i,3}) N_8(i+1) + \frac{1}{2}y_{i,3} N_8(i) \right ], \nonumber
\end{equation}
which completes the proof.
\end{IEEEproof}

\begin{table}
\caption{Rates, Normalized Rates, and Adder Sizes of OP-LOCO Codes $\mathcal{OPC}_m$ for Different Values of $m$. The Capacity Is $2.9129$, and the Normalized Capacity Is $0.9710$.}
\vspace{-0.5em}
\centering
\scalebox{1.00}
{
\begin{tabular}{|c|c|c|c|}
\hline
\makecell{$m$} & \makecell{$R_{\textup{OP-LOCO}}$} & \makecell{$R_{\textup{OP-LOCO}}^{\textup{n}}$}  & \makecell{Adder size} \\
\hline
$13$ & $2.7143$ & $0.9048$ & $38$ bits \\
\hline
$18$ & $2.7368$ & $0.9123$ & $52$ bits \\
\hline
$23$ & $2.7917$ & $0.9306$ & $67$ bits \\
\hline
$39$ & $2.8250$ & $0.9417$ & $113$ bits \\
\hline
$53$ & $2.8519$ & $0.9506$ & $154$ bits \\
\hline
$89$ & $2.8778$ & $0.9593$ & $259$ bits \\
\hline
\end{tabular}}
\label{table_4}
\vspace{-0.5em}
\end{table}

\begin{algorithm}
\caption{Encoding OP-LOCO Codes}
\begin{algorithmic}[1]
\State \textbf{Input:} Incoming stream of binary messages.
\State Use (\ref{eqn_oplococard}) and (\ref{eqn_oplococdef}) to compute $N_8(i)$, $i \in \{2, 3, 4, \dots\}$.
\State Specify $m$, the smallest $i$ in Step~2 to achieve the desired rate. Then, $s = \lfloor \log_2 N_8(m) \rfloor$.
\State \textbf{for} each incoming message $\bold{b}$ of length $s$ \textbf{do}
\State \hspace{2ex} Compute $g(\bold{c})=\textup{decimal}(\bold{b})$.
\State \hspace{2ex} Initialize $\textup{residual}$ with $g(\bold{c})$ and $c_i$ with $z'$ for $i \geq m$. \textit{($z'$ indicates out of codeword bounds)}
\State \hspace{2ex} \textbf{for} $i \in \{m-1, m-2, \dots, 0\}$ \textbf{do} \textit{(in order)}
\State \hspace{4ex} Initialize $\textup{symbol\_found}$ with $0$.
\State \hspace{4ex} Initialize $y_{1,i}(a_i)$, $y_{2,i}(a_i)$, $y_{3,i}(a_i)$, and $\textup{contrib}(a_i)$ with $0$'s for $a_i \in \{1, 2, \dots, 7\}$.
\State \hspace{4ex} \textbf{if} ($c_{i+2} \in \{0, 1, \alpha^3, \alpha^4\}$) $\land$ ($c_{i+1} = \alpha$) \textbf{then}
\State \hspace{6ex} Set $y_{1,i}(a_i) = 1$ for $a_i \in \{2, 3\}$, and set $y_{2,i}(a_i) = 1$ for $a_i \in \{6, 7\}$.
\State \hspace{4ex} \textbf{elseif} ($c_{i+2} \in \{\alpha, \alpha^2, \alpha^5, \alpha^6\}$) $\land$ ($c_{i+1} = \alpha^4$) \textbf{then}
\State \hspace{6ex} Set $y_{1,i}(a_i) = 1$ for $a_i \in \{4, 5\}$.
\State \hspace{4ex} \textbf{end if}
\State \hspace{4ex} \textbf{if} ($c_{i+1} \in \{0, 1, \alpha^3, \alpha^4\}$) \textbf{then}
\State \hspace{6ex} Set $y_{3,i}(a_i) = 1 - (y_{1,i}(a_i)+y_{2,i}(a_i))$ for $a_i \in \{3, 4, \dots, 7\}$.
\State \hspace{4ex} \textbf{elseif} ($c_{i+1} \in \{\alpha, \alpha^2, \alpha^5, \alpha^6\}$) \textbf{then}
\State \hspace{6ex} Set $y_{3,i}(a_i) = 1 - (y_{1,i}(a_i)+y_{2,i}(a_i))$ for $a_i \in \{6, 7\}$.
\State \hspace{4ex} \textbf{end if}
\State \hspace{4ex} \textbf{for} $a_i \in \{1, 2, \dots 7\}$ \textbf{do}
\State \hspace{6ex} $\textup{contrib}(a_i) = \frac{1}{8} (a_i -2y_{i,1}(a_i) - 4y_{i,2}(a_i) - y_{i,3}(a_i)) N_8(i+1) + \frac{1}{2}y_{i,3}(a_i) N_8(i)$.
\State \hspace{4ex} \textbf{end for}
\State \hspace{4ex} \textbf{if} $\textup{residual} \geq \textup{contrib}(7)$ \textbf{then}
\State \hspace{6ex} Encode $c_i = \alpha^6$ and set $\textup{symbol\_found} = 1$. \textit{(level $a_i=7$)}
\State \hspace{6ex} $\textup{residual} \leftarrow \textup{residual} - \textup{contrib}(7)$.
\State \hspace{4ex} \textbf{else}
\State \hspace{6ex} \textbf{for} $a_i \in \{6, 5, \dots, 1\}$ \textbf{do}
\State \hspace{8ex} \textbf{if} $\textup{contrib}(a_i) \leq \textup{residual} < \textup{contrib}(a_i+1)$ \textbf{then}
\State \hspace{10ex} Encode $c_i = \mathcal{L}^{-1}(a_i)$ and set $\textup{symbol\_found} = 1$. \textit{(level $a_i = \mathcal{L}(c_i)$)}
\State \hspace{10ex} $\textup{residual} \leftarrow \textup{residual} - \textup{contrib}(a_i)$.
\State \hspace{10ex} \textbf{break}. \textit{(exit current loop)}
\State \hspace{8ex} \textbf{end if}
\State \hspace{6ex} \textbf{end for}
\State \hspace{4ex} \textbf{end if}
\State \hspace{4ex} \textbf{if} $\textup{symbol\_found} = 0$ \textbf{then}
\State \hspace{6ex} Encode $c_i = 0$. \textit{(level $a_i=0$)}
\State \hspace{4ex} \textbf{end if}
\State \hspace{4ex} \textbf{if} (not first codeword) $\land$ ($i = m-2$) \textbf{then}
\State \hspace{6ex} Bridge with $\alpha^2$, $\alpha^3$, or $z$ before $c_{m-1}$ depending on the two RMSs of the previous codeword and $c_{m-1} c_{m-2}$.
\State \hspace{4ex} \textbf{end if}
\State \hspace{2ex} \textbf{end for}
\State \textbf{end for}
\State \textbf{Output:} Outgoing stream of $8$-ary OP-LOCO codewords. \textit{(to be written on $3$ adjacent down tracks in the TDMR device after binary conversion and signaling)}
\end{algorithmic}
\label{alg_openc}
\end{algorithm}

\begin{algorithm}
\caption{Decoding OP-LOCO Codes}
\begin{algorithmic}[1]
\State \textbf{Inputs:} Incoming stream of $8$-ary OP-LOCO codewords, in addition to $m$ and $s$. \textit{(stream after reading from $3$ adjacent down tracks in the TDMR device and $8$-ary conversion)}
\State Use (\ref{eqn_oplococard}) and (\ref{eqn_oplococdef}) to compute $N_8(i)$, $i \in \{2, 3, 4, \dots, m\}$.
\State \textbf{for} each incoming codeword $\bold{c}$ of length $m$ \textbf{do}
\State \hspace{2ex} Initialize $g(\bold{c})$ with $0$ and $c_i$ with $z'$ for $i \geq m$. \textit{($z'$ indicates out of codeword bounds)}
\State \hspace{2ex} \textbf{for} $i \in \{m-1, m-2, \dots, 0\}$ \textbf{do} \textit{(in order)}
\State \hspace{4ex} Initialize $y_{1,i}$, $y_{2,i}$, and $y_{3,i}$ with $0$'s.
\State \hspace{4ex} \textbf{if} ($c_{i+2} \in \{0, 1, \alpha^3, \alpha^4\}$) $\land$ ($c_{i+1} = \alpha$) \textbf{then}
\State \hspace{6ex} \textbf{if} $c_i \in \{\alpha, \alpha^2\}$ \textbf{then}
\State \hspace{8ex} Set $y_{1,i} = 1$
\State \hspace{6ex} \textbf{elseif}  $c_i \in \{\alpha^5, \alpha^6\}$ \textbf{then}
\State \hspace{8ex} Set $y_{2,i} = 1$
\State \hspace{6ex} \textbf{end if}
\State \hspace{4ex} \textbf{elseif} ($c_{i+2} \in \{\alpha, \alpha^2, \alpha^5, \alpha^6\}$) $\land$ ($c_{i+1} = \alpha^4$) $\land$ ($c_i = \{\alpha^3, \alpha^4\}$) \textbf{then}
\State \hspace{6ex} Set $y_{1,i} = 1$.
\State \hspace{4ex} \textbf{end if}
\State \hspace{4ex} \textbf{if} ($c_{i+1} \in \{0, 1, \alpha^3, \alpha^4\}$) $\land$ ($c_i = \{\alpha^2, \alpha^3, \dots, \alpha^6\}$) \textbf{then}
\State \hspace{6ex} Set $y_{3,i} = 1 - (y_{1,i}+y_{2,i})$.
\State \hspace{4ex} \textbf{elseif} ($c_{i+1} \in \{\alpha, \alpha^2, \alpha^5, \alpha^6\}$) $\land$ ($c_i = \{\alpha^5, \alpha^6\}$) \textbf{then}
\State \hspace{6ex} Set $y_{3,i} = 1 - (y_{1,i}+y_{2,i})$.
\State \hspace{4ex} \textbf{end if}
\State \hspace{4ex} Set $a_i = \mathcal{L}(c_i)$.
\State \hspace{4ex} $g(\bold{c}) \leftarrow g(\bold{c}) + \frac{1}{8} (a_i -2y_{i,1} - 4y_{i,2} - y_{i,3}) N_8(i+1) + \frac{1}{2}y_{i,3} N_8(i)$.
\State \hspace{2ex} \textbf{end for}
\State \hspace{2ex} Compute $\bold{b}=\textup{binary}(g(\bold{c}))$, which has length $s$.
\State \hspace{2ex} Ignore the next bridging symbol.
\State \textbf{end for}
\State \textbf{Output:} Outgoing stream of binary messages.
\end{algorithmic}
\label{alg_opdec}
\end{algorithm}

\begin{example}\label{example_4}
Consider the OP-LOCO code $\mathcal{OPC}^8_5$ ($m=5$). Using (\ref{eqn_oplococard}) and (\ref{eqn_oplococdef}), we get $N_8(0) \triangleq 2$, $N_8(1) \triangleq 8$, $N_8(2) = 64$, $N_8(3) = 480$, $N_8(4) = 3616$, and $N_8(5) = 27232$. Consider the codeword $\bold{c} = c_4 c_3 c_2 c_1 c_0 = \alpha^3\alpha^3\alpha\alpha^5\alpha^6$ (level-equivalent $44267$) in $\mathcal{OPC}^8_5$. The case indexed by $i_{\textup{c}} = 1$ (the typical case) applies for $c_4$ and $c_2$, which means $y_{4,1} = y_{4,2} = y_{4,3} = y_{2,1} = y_{2,2} = y_{2,3} = 0$. The case indexed by $i_{\textup{c}} = 4$ applies for $c_3$, which means $y_{3,3} = 1$ and $y_{3,1} = y_{3,2} = 0$. The case indexed by $i_{\textup{c}} = 3$ applies for $c_1$, which means $y_{1,2} = 1$ and $y_{1,1} = y_{1,3} = 0$. The case indexed by $i_{\textup{c}} = 6$ applies for $c_0$, which means $y_{0,3} = 1$ and $y_{0,1} = y_{0,2} = 0$. Consequently, and using (\ref{eqn_oplocorule}), we get:
\vspace{-0.1em}\begin{align}
g(\bold{c}=\alpha^3\alpha^3\alpha\alpha^5\alpha^6) &= \left [ \frac{1}{8} \times 4N_8(5) \right ] + \left [ \frac{1}{8}\times 3N_8(4) + \frac{1}{2} N_8(3) \right ] + \left [ \frac{1}{8} \times 2N_8(3) \right ] \nonumber \\
&\hspace{+1.0em}+ \left [ \frac{1}{8} \times 2N_8(2) \right ] + \left [ \frac{1}{8}\times 6N_8(1) + \frac{1}{2} N_8(0) \right ] \nonumber \\
&= 13616 + 1596 + 120 + 16 + 7 = 15355, \nonumber
\end{align}
which is consistent with the codeword index produced by the program we wrote to exhaustively generate and lexicographically order all OP-LOCO codewords in $\mathcal{OPC}^8_m$, with $m=5$ here. It corresponds to the binary message $11101111111011$ ($s = 14$).
\end{example}\vspace{+0.2em}

\textbf{Step~6)} Recall that $\beta_1$ is in $\{0, 1, \alpha^3, \alpha^4\}$, $\beta_2$ is in $\{\alpha, \alpha^2, \alpha^5, \alpha^6\}$,  $\beta_3$ is in $\textup{GF}(8) \setminus \{\alpha\}$, and $\beta_4$ is in $\textup{GF}(8) \setminus \{\alpha^4\}$. We bridge in OP-LOCO codes with one GF$(8)$ symbol, i.e., one column of three bits, or one no-writing symbol $z$, i.e., one column of three unmagnetized grid entries, between each two consecutively written codewords as shown below. We use the notation ``$\textup{RMS(s)} - \textup{LMS(s)}$'' to represent the RMS(s) of a codeword and the LMS(s) of the next codeword for brevity.
\begin{itemize}
\item If the $\textup{RMSs} - \textup{LMSs}$ are $\beta_1\alpha - \alpha^4\beta_2$, bridge with one no-writing symbol $z$.
\item \textit{Else if the $\textup{RMS(s)} - \textup{LMS(s)}$ are $\beta_1\alpha - \alpha^4\beta_1$, $\alpha - \beta_4$, $\beta_2\alpha - \alpha^4\beta_1$, or $\alpha^3 - \alpha^3$, bridge with $\alpha^2$.}
\item Else if the $\textup{RMS(s)} - \textup{LMS(s)}$ are $\beta_2\alpha - \alpha^4\beta_2$, $\alpha^4 - \beta_3$, or $\alpha^2 - \alpha^2$, bridge with $\alpha^3$.
\item For any other scenario, bridge with $\alpha^2$.
\end{itemize}
Obviously, the second item above in italic could be removed as it is included it in the last item. However, we keep it because of the importance of the stated $\textup{RMS(s)} - \textup{LMS(s)}$. The mapping-demapping in (\ref{eqn_gf8map}) illustrates what exactly is written for bridging. This bridging is optimal in terms of minimum added redundancy, and it offers near-maximum protection of edge symbols. Other bridging methods are also possible.

Recall that a transition is counted on the level of the $3 \times 1$ column after signaling. With our bridging, the maximum number of consecutive $3 \times 1$ columns with no transition after writing via an OP-LOCO code $\mathcal{OPC}^8_m$ is $m+1$. This finite maximum is achieved without removing any codewords from the OP-LOCO code for self-clocking. Thus, an OP-LOCO code associated with the aforementioned bridging is inherently self-clocked.

Given our bridging method, the rate, in input bits per coded symbol, and the normalized rate of an OP-LOCO code $\mathcal{OPC}^8_m$ are:
\begin{equation}\label{eqn_oplocorate}
R_{\textup{OP-LOCO}} = \frac{s}{m+1} = \frac{\lfloor N_8(m) \rfloor}{m+1}, \textup{ } R^{\textup{n}}_{\textup{OP-LOCO}} = \frac{1}{3} R_{\textup{OS-LOCO}} = \frac{\lfloor N_8(m) \rfloor}{3(m+1)}.
\end{equation}
It is easy to deduce that OP-LOCO codes are capacity-achieving. To demonstrate that, Table~\ref{table_4} gives the rates and normalized rates of OP-LOCO codes with different lengths. Table~\ref{table_4} shows that the rates of OP-LOCO codes are close to capacity even at moderate lengths. The table also shows that OP-LOCO codes incur very limited redundancy, albeit a little bit more than the redundancy incurred by OS-LOCO codes.

Next, we introduce the encoding algorithm of OP-LOCO codes, which is Algorithm~\ref{alg_openc}, and the decoding algorithm of OP-LOCO codes, which is Algorithm~\ref{alg_opdec}.

To reduce runtime processing, all the terms containing a factor multiplied by a cardinality in Algorithm~\ref{alg_openc} and Algorithm~\ref{alg_opdec} are computed offline and stored in memory, for all possible factors except factors of the form $2$ raised to some power (negative or positive). The same applies for all same-symbol operations. Observe that a multiplication by $2^{-\eta}$ (resp., $2^{\eta}$), $\eta \in \{1, 2, 3, \dots\}$, is executed as a right-shift (resp., left-shift) by $\eta$. Thus, the main runtime operations in both algorithms are also additions, subtractions, and comparisons, which are all performed by adders. That is the reason why the adder size governs the complexity of the encoding and decoding procedures, demonstrating simplicity (see also Table~\ref{table_4}). Observe that both complexity and storage overhead can be further reduced via arithmetic tricks \cite{ahh_qaloco}. These algorithms complete the process of revealing the secret arithmetic of allowed and forbidden patterns in an OP-LOCO code.

\section{Performance Gains in TDMR}\label{sec_gains}

In this section, we provide experimental results demonstrating the performance gains achieved by our optimal LOCO codes described in Section~\ref{sec_opt} in a practical TDMR system. We start off with defining some TDMR channel parameters.
\begin{itemize}
\item $TW$ or track width is the width of the bit in the cross-track direction, i.e., the width of the down track.
\item $BP$ or bit period is the width of the bit in the down-track direction, i.e., the width of the cross track.
\item $PW_{50,\textup{CT}}$ is the TD read-head impulse response duration at half the amplitude in the cross-track direction.
\item $PW_{50,\textup{DT}}$ is the TD read-head impulse response duration at half the amplitude in the down-track direction.
\item $D_{\textup{TD}}$ is the TD channel density \cite{shayan_tdmr}, which is given by:
\begin{equation}\label{eqn_tddensity}
D_{\textup{TD}} = \frac{PW_{50,\textup{CT}} \times PW_{50,\textup{DT}}}{TW \times BP}.
\end{equation}
Increasing the TD channel density exacerbates TD interference, which is interference in both down and cross track directions, resulting in degraded system performance.
\item The TD channel (read-head) impulse response is a $3 \times 3$ matrix, which represents the intersection of $3$ adjacent down tracks in the same group with $3$ consecutive cross tracks.
\end{itemize}

Next, we discuss our TDMR system setup. We have the writing setup, the channel setup, and the reading setup.

\textbf{Writing setup:} We generate random input messages. Then, we use Algorithm~\ref{alg_osenc} or Algorithm~\ref{alg_openc} to encode each message $\bold{b}$ into the corresponding $8$-ary OS-LOCO or OP-LOCO codeword $\bold{c}$, respectively. Each GF$(8)$ symbol in the LOCO codeword is converted into a $3 \times 1$ column of binary bits according to the mapping-demapping in (\ref{eqn_gf8map}). Consequently, a codeword of length $m$ will be written over a grid of size $3 \times m$ spanning $3$ down tracks in the TDMR system. A bridging column separates each two consecutive LOCO codewords. Before writing, level-based signaling is applied, which converts each $0$ into $-1$, each $1$ into $+1$, and each $z$ into no magnetization or zero (handled in a special way). Upon writing, these $-1$ and $+1$ values will be updated to values depending on $TW$ and $BP$.

We use the following three LOCO codes in the simulations (see also Tables~\ref{table_3} and \ref{table_4}):
\begin{itemize}
\item The OS-LOCO code $\mathcal{OSC}^8_{23}$ with codeword length $m = 23$, message length $s = 68$, and normalized rate $R^{\textup{n}}_{\textup{LOCO}} = 0.9444$.\footnote{We use a unified notation, $R^{\textup{n}}_{\textup{LOCO}}$, to express the normalized rate of an OS-LOCO or an OP-LOCO code for simplicity.}
\item The OP-LOCO code $\mathcal{OPC}^8_{18}$ with codeword length $m = 18$, message length $s = 52$, and normalized rate $R^{\textup{n}}_{\textup{LOCO}} = 0.9123$.
\item The OP-LOCO code $\mathcal{OPC}^8_{23}$ with codeword length $m = 23$, message length $s = 67$, and normalized rate $R^{\textup{n}}_{\textup{LOCO}} = 0.9306$.
\end{itemize}
In order to keep the energy per input message bit in the coded setting the same as it is in the uncoded setting, we obtain $TW$ and $BP$ of the coded setting via scaling both $TW$ and $BP$ of the uncoded setting by $\sqrt{R^{\textup{n}}_{\textup{LOCO}}}$, respectively.

\textbf{Channel setup:} Our baseline channel model is the TDMR model in \cite{mohsen_tdmr}, which is a Voronoi model. Here, we only consider media noise/interference (electronic noise is not included). We modify this model such that it is suitable for a wide read head that reads data from $3$ adjacent down tracks simultaneously. In particular, in each group of $3$ adjacent down tracks, the upper and lower tracks in our model have additional protection from interference in the cross track direction. Thus, only the middle down track in each group suffers from notable interference from the two surrounding tracks \cite{chan_tdmr, bd_tdmr}.

\begin{figure}
\vspace{-0.3em}
\center
\includegraphics[trim={0.0in 0.0in 0.0in 0.0in}, width=3.5in]{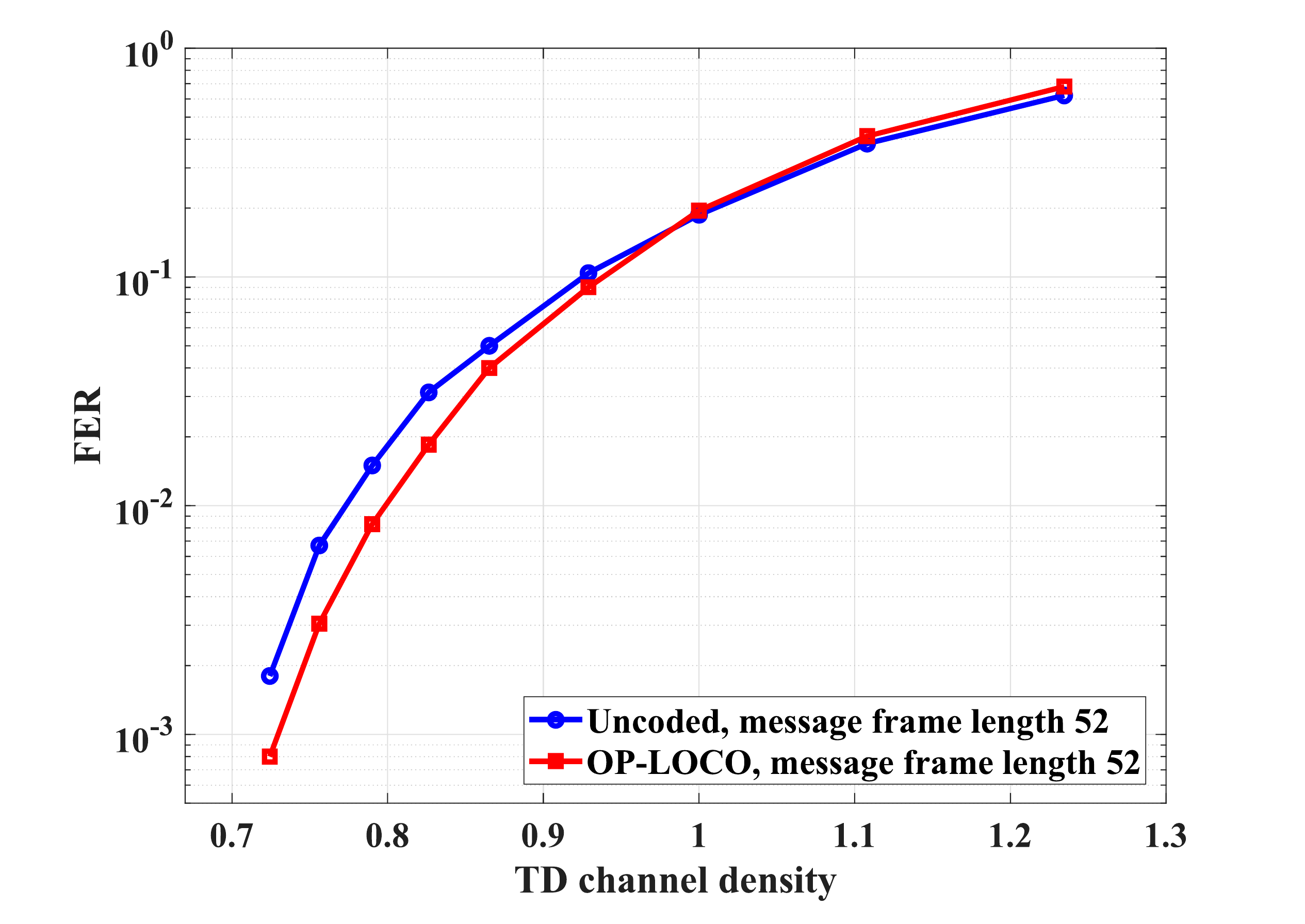}
\includegraphics[trim={0.0in 0.0in 0.0in 0.0in}, width=3.5in]{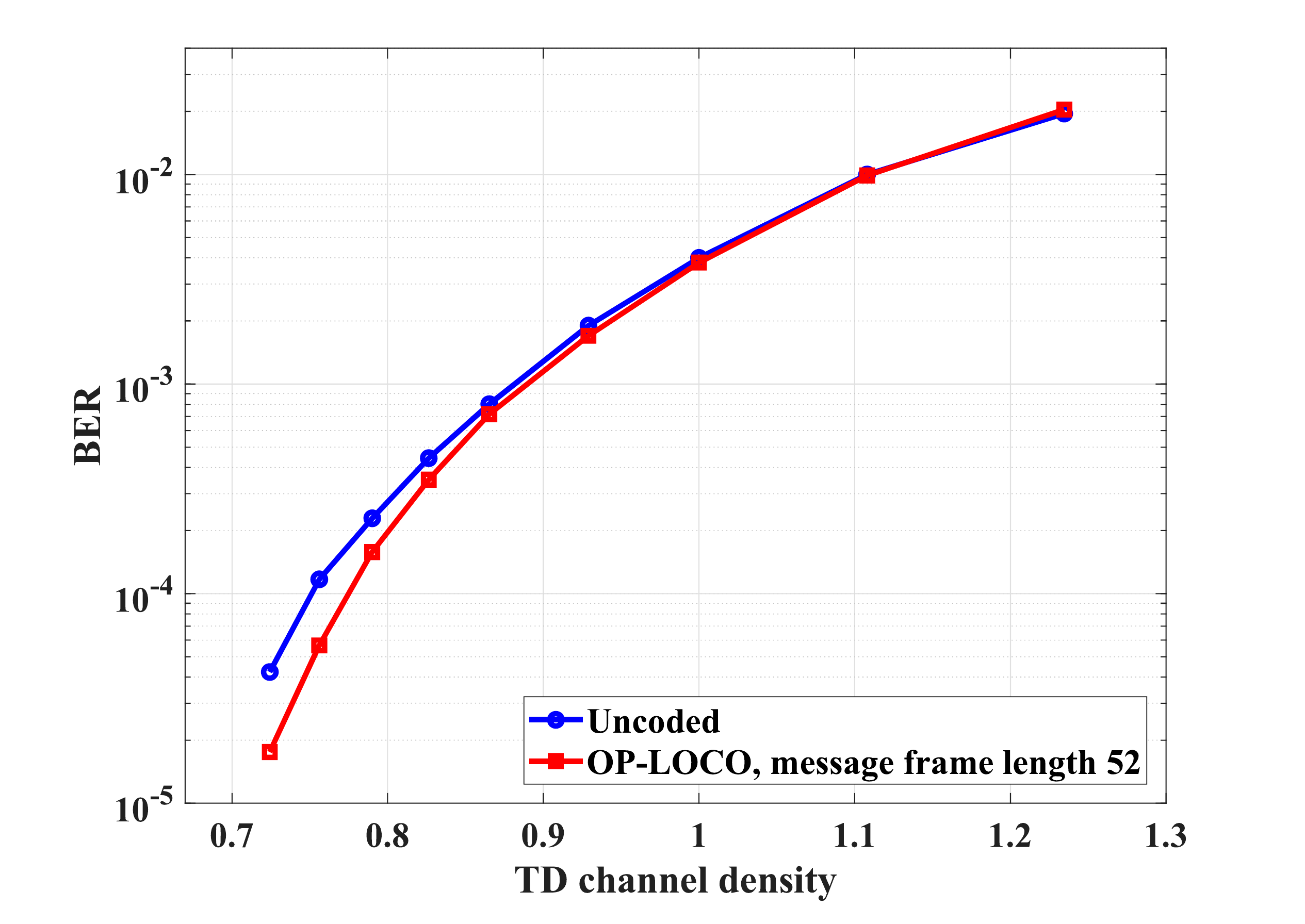}
\vspace{-0.8em}
\caption{FER (left) and BER (right) comparisons between the uncoded setting and the setting adopting the OP-LOCO code $\mathcal{OPC}^8_{18}$ ($m = 18$).}
\label{fig_4}
\vspace{-0.3em}
\end{figure}

\begin{figure}
\vspace{-0.3em}
\center
\includegraphics[trim={0.0in 0.0in 0.0in 0.0in}, width=3.5in]{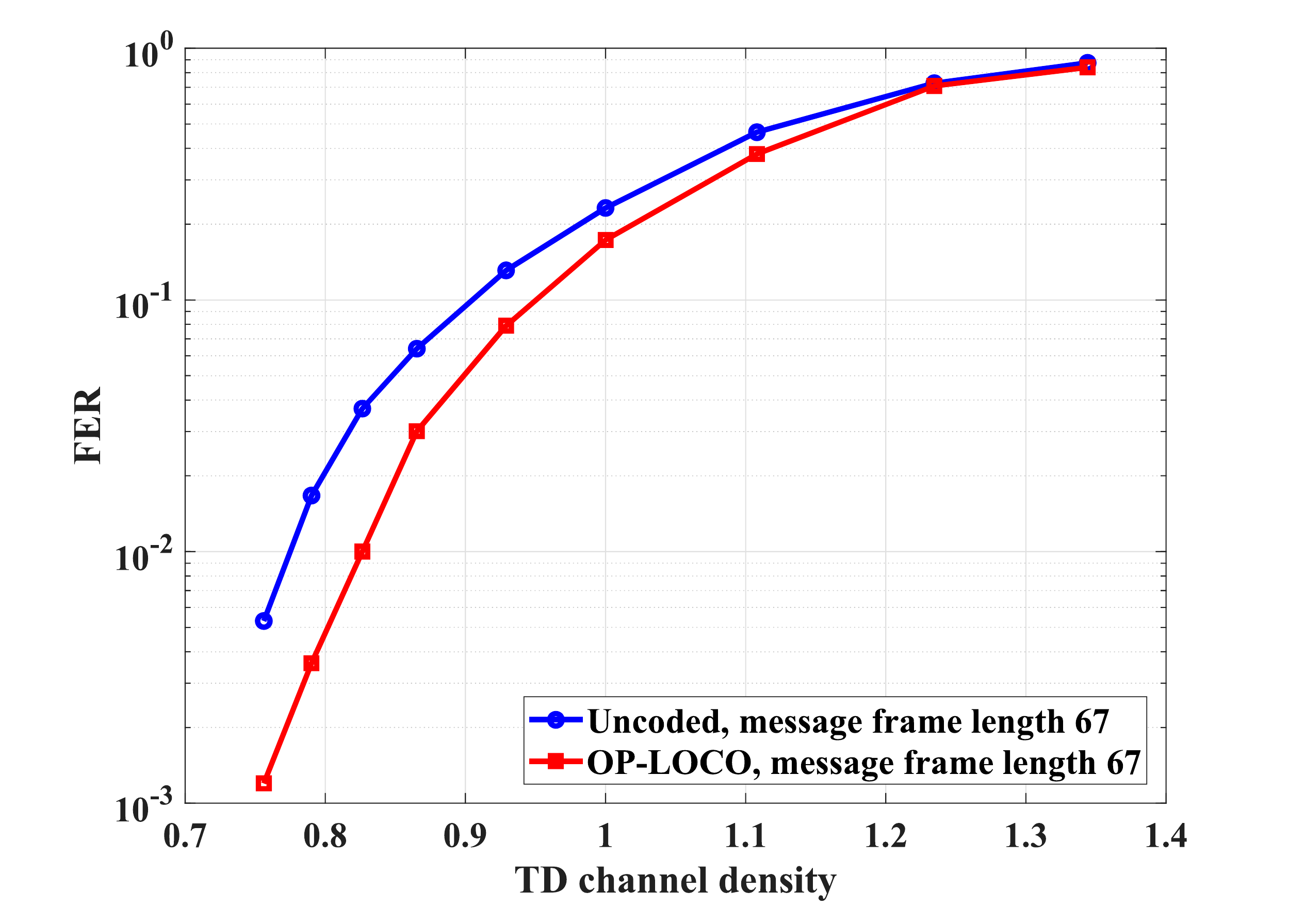}
\includegraphics[trim={0.0in 0.0in 0.0in 0.0in}, width=3.5in]{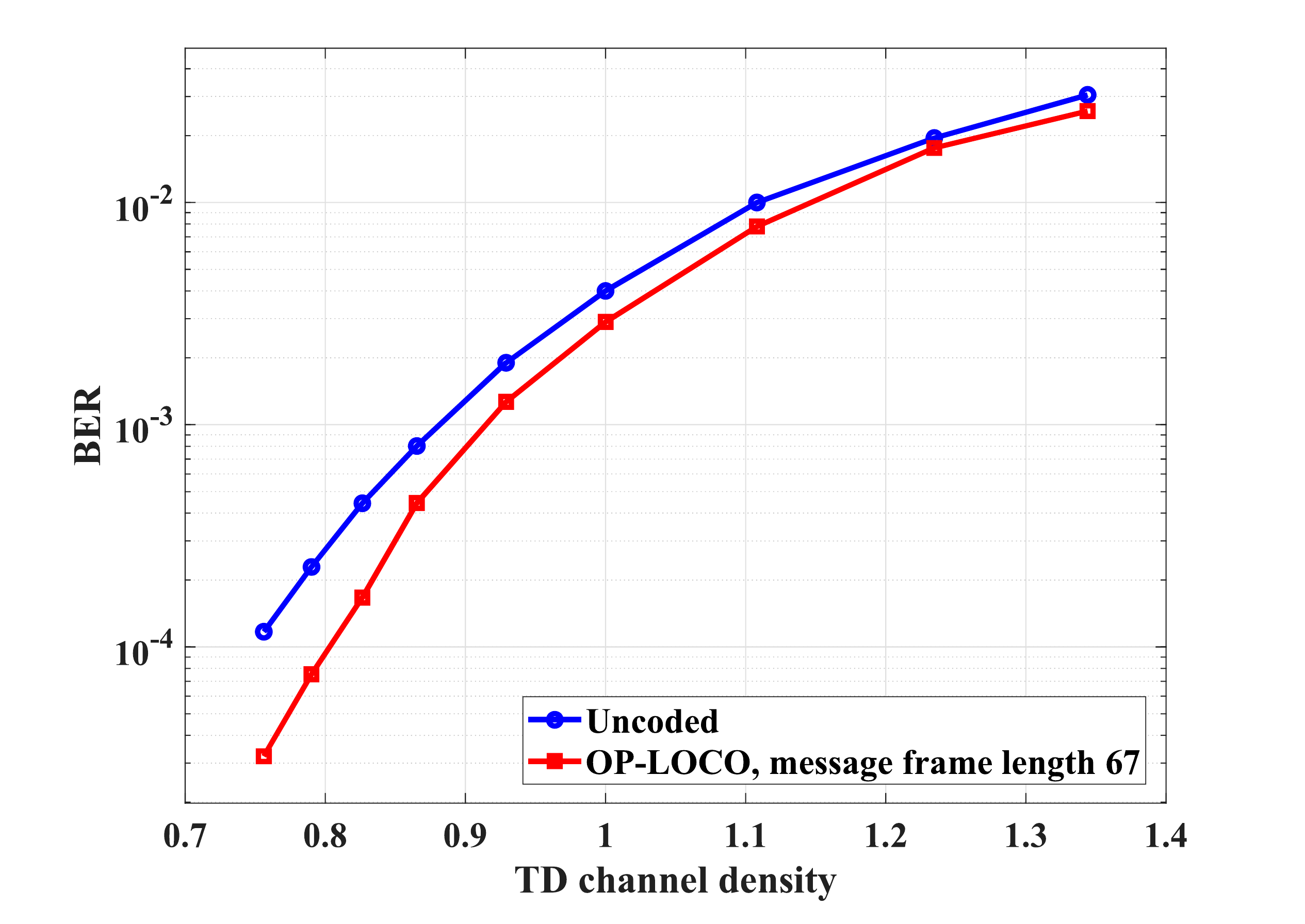}
\vspace{-0.8em}
\caption{FER (left) and BER (right) comparisons between the uncoded setting and the setting adopting the OP-LOCO code $\mathcal{OPC}^8_{23}$ ($m = 23$).}
\label{fig_5}
\vspace{-0.3em}
\end{figure}

In the simulations, we sweep the TD channel density $D_{\textup{TD}}$ given in (\ref{eqn_tddensity}). This is performed as follows. The parameters $PW_{50,\textup{CT}}$ and $PW_{50,\textup{DT}}$ are fixed at $20.00$ nm and $14.00$ nm, respectively. The parameter $TW$ is swept between $17.25$ nm and $24.50$ nm, while the parameter $BP$ is swept between $12.08$ nm and $17.15$ nm. We keep the ratio $TW/BP$ the same at all sweep points according to:
\begin{equation}\label{eqn_sameratio}
\frac{TW}{BP} = \frac{PW_{50,\textup{CT}}}{PW_{50,\textup{DT}}} = \frac{10}{7}.
\end{equation}
Thus, and using (\ref{eqn_tddensity}), the TD density $D_{\textup{TD}}$ is swept between $1.3437$ and $0.6664$ as shown in Table~\ref{table_5}, Fig.~\ref{fig_4}, Fig.~\ref{fig_5}, and the following discussions. Observe that the range of the TD density simulated will be notably higher in a TDMR system with equalization, detection, and most importantly, LDPC coding customized for magnetic recording \cite{ahh_prc, ahh_scmr, shayan_tdmr, ahh_md}.

The input to the channel is $3 \times m$ grids of coded bits with their bridging columns after signaling is applied. The output from the channel is these $3 \times m$ grids with the bridging columns after Voronoi media noise/interference is applied, taking into account the aforementioned protection of the upper and lower tracks in each group of $3$ down tracks. Mathematically, the channel effect is equivalent to applying the TD convolution between the $3 \times m$ input grids with their bridging columns and the $3 \times 3$ read-head impulse response with media noise incorporated.

\textbf{Reading setup:} For each $3 \times m$ grid out of the channel, where bridging columns are ignored, hard decision is performed based on the value at each entry; if the value is less than or equal to zero, the bit is read as $0$, while if the value is greater than zero, the bit is read as $1$. Each column of $3$ bits is then converted into a GF$(8)$ symbol according to the mapping-demapping in (\ref{eqn_gf8map}). Each $8$-ary sequence of length $m$ is then checked for constraint satisfaction. If the constraint is violated, a frame error is counted. Otherwise, the OS-LOCO or OP-LOCO codeword $\widehat{\bold{c}}$ passes through Algorithm~\ref{alg_osdec} or Algorithm~\ref{alg_opdec} to decode the corresponding binary message $\widehat{\bold{b}}$. If $g(\widehat{\bold{c}}) \geq 2^s$, a frame error is counted. If $\widehat{\bold{b}} \neq \bold{b}$ (same as $\widehat{\bold{c}} \neq \bold{c}$), a frame error is counted.

\begin{remark}
It is possible to use OS-LOCO and OP-LOCO codes to perform some error correction in a way similar to what we did with LOCO codes in \cite{ahh_loco}. The idea is that if the received $8$-ary sequence of length $m$ violates the constraint, the bit corresponding to the closest value to zero in the $3 \times m$ grid of the codeword is flipped. Then, after GF$(8)$ conversion, the constraint is checked again. If it is now satisfied, one-symbol error correction was performed. 
\end{remark}

Now, we discuss the bit error statistics of the uncoded setting. Table~\ref{table_5} shows the percentage of bit errors resulting from SIS patterns and the percentage of bit errors resulting from PIS patterns out of all bit errors we collected at different TD channel densities. The bit error rate (BER) is also shown in the table at each TD density point. The percentage of bit errors resulting from SIS patterns, named SIS errors in the table, ranges between $4.6\%$ to $21.4\%$. The percentage of bit errors resulting from PIS patterns, named PIS errors in the table, ranges between $44.9\%$ to $93.1\%$. At TD density $D_{\textup{TD}} = 0.7901$ and below, more than $90\%$ of the collected bit errors are PIS errors. Moreover, as the TD density decreases, the percentage of SIS errors (except for one table point) and PIS errors consistently increases. Consequently, Table~\ref{table_5} further motivates introducing constrained codes to eliminate SIS and PIS patterns in practical TDMR systems.

Next, we discuss the performance gains achieved by our LOCO codes. We start with OS-LOCO codes. OS-LOCO codes can improve performance at lower TD channel densities. While OP-LOCO codes have a performance advantage over OS-LOCO codes at all TD densities, OS-LOCO codes have a rate advantage over OP-LOCO codes at the same length. As mentioned above, we use the OS-LOCO code $\mathcal{OSC}^8_{23}$ to improve performance in the TDMR system. According to our simulations, at TD density $D_{\textup{TD}} = 0.6664$, the frame error rate (FER) $= 6.66 \times 10^{-4}$ for the uncoded setting, while the FER $= 5 \times 10^{-4}$ for the setting adopting $\mathcal{OSC}^8_{23}$ to eliminate SIS patterns. The FER is measured from either the messages or the codewords.

\begin{table}
\caption{Bit Error Statistics of the Uncoded Setting in a TDMR System at Different TD Channel Densities}
\vspace{-0.5em}
\centering
\scalebox{1.00}
{
\begin{tabular}{|c|c|c|c|}
\hline
\makecell{TD density} & \makecell{BER} & \makecell{SIS errors}  & \makecell{PIS errors} \\
\hline
$1.2346$ & $1.95 \times 10^{-2}$ & $4.6\%$ & $44.9\%$ \\
\hline
$1.1080$ & $1.00 \times 10^{-2}$ & $5.6\%$ & $53.7\%$ \\
\hline
$1.0000$ & $4.00 \times 10^{-3}$ & $7.8\%$ & $63.9\%$ \\
\hline
$0.9290$ & $1.90 \times 10^{-3}$ & $9.3\%$ & $76.0\%$ \\
\hline
$0.8653$ & $8.02 \times 10^{-4}$ & $10.0\%$ & $84.4\%$ \\
\hline
$0.8264$ & $4.43 \times 10^{-4}$ & $16.5\%$ & $87.2\%$ \\
\hline
$0.7901$ & $2.29 \times 10^{-4}$ & $20.2\%$ & $91.7\%$ \\
\hline
$0.7561$ & $1.17 \times 10^{-4}$ & $18.9\%$ & $92.3\%$ \\
\hline
$0.7243$ & $4.23 \times 10^{-5}$ & $21.4\%$ & $93.1\%$ \\
\hline
\end{tabular}}
\label{table_5}
\vspace{-0.3em}
\end{table}

We then discuss the gains of OP-LOCO codes. We generate four performance plots, FER/BER versus TD channel density, comparing the uncoded setting with the setting adopting OP-LOCO codes in the TDMR system. In Fig.~\ref{fig_4}, the OP-LOCO code $\mathcal{OPC}^8_{18}$ (message frame length $=52$ bits) is used, while in Fig.~\ref{fig_5}, the OP-LOCO code $\mathcal{OPC}^8_{23}$ (message frame length $=67$ bits) is used. The FER is measured from either the messages or the codewords as both give the same result. The BER is measured directly at the output of the channel, i.e., right after applying the hard decision in the reading setup.

In Fig.~\ref{fig_4}, and at TD density $0.7243$, the FER $= 1.80 \times 10^{-3}$ and the BER $= 4.23 \times 10^{-5}$ for the uncoded setting, while the FER $= 8.00 \times 10^{-4}$ and the BER $= 1.76 \times 10^{-5}$ for the setting adopting the OP-LOCO code $\mathcal{OPC}^8_{18}$. This means we have a performance gain of up to $0.35$ (resp., $0.38$) of an order of magnitude in FER (resp., BER) solely by applying $\mathcal{OPC}^8_{18}$. In Fig.~\ref{fig_5}, and at TD density $0.7561$, the FER $= 5.30 \times 10^{-3}$ and the BER $= 1.17 \times 10^{-4}$ for the uncoded setting, while the FER $= 1.20 \times 10^{-3}$ and the BER $= 3.22 \times 10^{-5}$ for the setting adopting the OP-LOCO code $\mathcal{OPC}^8_{23}$. This means we have a performance gain of up to $0.65$ (resp., $0.56$) of an order of magnitude in FER (resp., BER) solely by applying $\mathcal{OPC}^8_{23}$. These gains are quite significant given that no error-correcting code, particularly no LDPC code, is applied in this system, demonstrating the importance of applying high rate constrained codes eliminating PIS patterns in TDMR systems.

Additionally, there are two important observations from these two figures. First, the performance gain increases as the length $m$ of the OP-LOCO code increases. This is primarily because as the length increases the rate increases, which in turn increases the energy per input message bit since we scale by $\sqrt{R^{\textup{n}}_{\textup{LOCO}}}$ as mentioned in the writing setup. Second, the performance gain increases as the TD density decreases. This is primarily because of the observation collected from Table~\ref{table_5} that the percentage of PIS errors increases as the TD density decreases.

\begin{remark}
At high TD channel densities, errors may result from incomplete PIS (IPIS) patterns. These are patterns where the complementary bits surround the central bit at only three out of the four positions with Manhattan distance $1$ from the center. While we can use the general method to design LOCO codes forbidding IPIS patterns along with PIS patterns, this is not a recommended idea. The reason is that such codes will suffer from a notable rate loss while offering a quite limited performance gain compared with OP-LOCO codes. The rate loss is attributed to the very high number of IPIS patterns, which is $2 \left [ \binom{4}{3} \times 2^{4} \right ] = 2 \times 64 = 128$ out of $512$.
\end{remark}

We end this section with a brief comparison with the available TD constrained codes in the literature. This comparison can be summarized in the following points:
\begin{enumerate}
\item There are many papers in the literature discussing TD constrained coding. Bounds on the capacity of TD RLL codes were discussed in \cite{kato_TCon} and \cite{siegel_TCon}. Explicit coding techniques to stuff bits into a TD grid such that certain RLL constraints are satisfied in both directions were presented in \cite{sharov_TCon} and \cite{halevy_TD}. While such papers are dealing with an important technical question, applying TD RLL codes to TDMR systems is quite inefficient rate-wise since many patterns forbidden by TD RLL codes are not detrimental in such systems.

\item Coding techniques to stuff bits into a TD grid such that isolation patterns, particularly PIS patterns, are forbidden were introduced in \cite{mohsen_tdmr} and \cite{pituso_tdmr}. TD constrained codes resulting from these techniques offer notably higher code rates for TDMR systems than TD RLL codes. However, these techniques are not customized for a TDMR system with a wide read head such as the one we adopt \cite{chan_tdmr, shayan_tdmr}. Wide read heads are attractive because they notably increase the speed of reading. In this TDMR system, PIS patterns need to be removed only within each group of three down tracks. Thus, while the highest achievable rate in \cite{mohsen_tdmr} is $0.9238$, the highest achievable normalized rate of an OP-LOCO code is $0.9710$ from (\ref{eqn_capoploco}). To the best of our knowledge, our paper \cite{bd_tdmr} was the first to present (NS-LOCO) codes for TDMR systems with wide heads.

\item While coding techniques based on bit stuffing produce constrained codes with rates approaching or achieving capacity, they have an important shortcoming. These techniques do not offer explicit ways to convert unconstrained input messages into codewords, nor do they offer explicit ways to convert codewords back to the unconstrained messages \cite{halevy_TD, mohsen_tdmr}. On the contrary, our LOCO codes for TDMR systems (like all our LOCO codes) offer a simple systematic mapping-demapping from an unconstrained message to a constrained codeword via the integer index, and vice versa.
\end{enumerate}

\section{Ideas to Reduce Complexity}\label{sec_compl}

In this section, we introduce coding schemes adopting near-optimal (rate-wise) LOCO codes for TDMR systems to eliminate the SIS and PIS patterns. In particular, the new coding schemes incur a small rate loss to achieve some complexity and error propagation reduction compared with the optimal LOCO codes introduced in Section~\ref{sec_opt}.

Because of their fixed length, LOCO codes do not suffer from codeword to codeword error propagation.\footnote{Because of their fixed length, LOCO codes also allow parallel encoding and decoding, unlike FSM-based constrained codes \cite{ahh_loco}.} However, error propagation can happen on the input message level. In particular, one symbol error in the codeword may result in multiple bit errors in the message because of the wrong index \cite{ahh_loco}, which limits the performance gains of constrained coding on the message BER level for large lengths at high densities. That is one reason why we used moderate message lengths for our OS-LOCO and OP-LOCO codes in Section~\ref{sec_gains}.

The idea is simply to develop LOCO codes defined over GF$(4)$ instead of GF$(8)$, and we first introduced it in \cite{bd_tdmr}. The steps of our coding scheme adopting NS-LOCO (resp., NP-LOCO) codes, in brief, are:
\begin{itemize}
\item We specify a mapping-demapping between GF$(8)$ and GF$(4)$: $2$ GF$(8)$ symbols $\longleftrightarrow$ $1$ GF$(4)$ symbol.
\item We design a LOCO code defined over GF$(4)$ based on the set of forbidden patterns $\mathcal{T}$ and this mapping-demapping.
\item While encoding, we divide the stream of input bits into chunks of length $s+m+1$ (resp., $s+m$) bits each.
\item We encode each $s$ bits (the message) in the chunk via the LOCO code into $m$ $4$-ary symbols (the codeword), and bridge by $1$ more symbol.
\item We use the remaining $m+1$ (resp., $m$) bits in the chunk to decide which $8$-ary symbol to write for each $4$-ary symbol according to the mapping-demapping. This means $1/3$ (resp., $m/[3(m+1)]$, which is almost $1/3$) of the data to be written is unconstrained.
\item The decoding is the same procedure performed in the reverse direction.
\end{itemize}

Having LOCO codes defined over GF$(4)$ instead of GF$(8)$ enables lower complexity \cite{ahh_qaloco, bd_tdmr}, and also enables lower error propagation as smaller message lengths are possible. Since the LOCO codes are defined over GF$(4)$ here, $\alpha$ will again be a primitive element of GF$(4)$ while $\psi$ is defined in this section as a primitive element of GF$(8)$. That is:
\begin{equation}\label{eqn_psi}
\textup{GF}(4) \triangleq \{0, 1, \alpha, \alpha^2\}, \textup{ } \textup{GF}(8) \triangleq \{0, 1, \psi, \psi^2 , \dots, \psi^7\}.
\end{equation}
The mapping-demapping between GF$(8)$ symbols and $3 \times 1$ columns of bits to write in the TDMR grid is in (\ref{eqn_gf8map}) with $\psi$ replacing $\alpha$.

\subsection{Near-Optimal Square LOCO Codes}

We start with our near-optimal square LOCO (NS-LOCO) codes, which are codes preventing the SIS patterns shown in Fig.~\ref{fig_1} (along with other patterns) within each group of three adjacent down tracks. These $2$ SIS patterns map to the $2$ GF$(8)$ patterns $0\psi0$ and $\psi^6\psi^4\psi^6$. We adopt the following GF$(8)$ $\longleftrightarrow$ GF$(4)$ mapping-demapping:
\begin{align}\label{eqn_mapnsloco}
\{\psi, \psi^4\} &\longleftrightarrow 0, \hspace{+3.0em} \{1, \psi^5\} \longleftrightarrow 1, \nonumber \\
\{\psi^2, \psi^3\} &\longleftrightarrow \alpha, \hspace{+2.8em} \{0, \psi^6\} \longleftrightarrow \alpha^2.
\end{align}
Based on this mapping-demapping, an NS-LOCO code should forbid the pattern $\alpha^20\alpha^2$, which covers the $2$ SIS patterns (and more). The FSTD, adjacency matrix, and capacity derivations are in \cite{bd_tdmr}. We care about the capacity of the coding scheme, including the unconstrained part of data to be written. Thus, and using \cite{bd_tdmr}, the capacity $C$, in input bits per coded symbol, and the normalized capacity $C^{\textup{n}}$ are:
\begin{equation}\label{eqn_capnsloco}
C = 1.9780 + 1 = 2.9780 \textup{ and } C^{\textup{n}} = \frac{1}{3} C = 0.9927,
\end{equation}
which means that the capacity loss compared with the optimal case from (\ref{eqn_caposloco}) is $0.54\%$.

Denote an NS-LOCO code of length $m$ by $\mathcal{NSC}^4_m$. The definition of the code is exactly the definition of a generic LOCO code, which is Definition~\ref{def_genloco}, with $q=4$, $\mathcal{C}^q_m = \mathcal{NSC}^4_m$, and $\mathcal{T}$ given by:
\begin{equation}\label{eqn_tnsloco}
\mathcal{T} = \mathcal{NS}^4 \triangleq \{\alpha^20\alpha^2\}.
\end{equation}
Both $\bold{c}$ in $\mathcal{C}^q_m = \mathcal{NSC}^4_m$ and $g(\bold{c})$ are used as they were in Section~\ref{sec_gen}. The cardinality of $\mathcal{NSC}^4_m$ is $N_q(m) = N_4(m)$.

Since NS-LOCO codes were already introduced in \cite{bd_tdmr}, we just state the outcome of each step of the general method. Having said that, Steps~3, 4, and 5 give new insights about NS-LOCO codes.\vspace{+0.7em}

\textbf{Step~1)} We end up with three (final) groups covering all the NS-LOCO codewords in $\mathcal{NSC}^4_m$: \textit{Group~1}, which contains all the codewords starting with $\beta_1$, $\beta_1 \in \{0, 1, \alpha\}$, from the left, \textit{Group~2}, which contains all the codewords starting with $\alpha^2\beta_2$, $\beta_2 \in \{1, \alpha, \alpha^2\}$, from the left, and \textit{Group~3}, which contains all the codewords starting with $\alpha^20\beta_1$, $\beta_1 \in \{0, 1, \alpha\}$, from the left. The groups are defined for $m \geq 2$.\vspace{+0.7em}

\textbf{Step~2)} The cardinality of an NS-LOCO code $\mathcal{NSC}^4_m$ is given by:
\begin{equation}\label{eqn_nslococard}
N_4(m) = 4N_4(m-1) - N_4(m-2) + 3N_4(m-3), \text{ } m \geq 2,
\end{equation}
where the defined cardinalities are:
\begin{equation}\label{eqn_nslococdef}
N_4(-1) \triangleq \frac{1}{3}, \text{ } N_4(0) \triangleq 1, \text{ and } N_4(1) \triangleq 4.
\end{equation}\vspace{+0.2em}

\textbf{Step~3)} We end up with two (final) cases for $c_i$ based on $c_i$ and its preceding symbols: a \textit{special case} for $c_{i+1} c_i = \alpha^2\beta_2$ and the \textit{typical case}. The typical case is the case when the special cases is not enabled and $c_i \neq 0$.\vspace{+0.7em}

\textbf{Steps~4 and 5)} Theorem~\ref{thm_nslocorule} gives the encoding-decoding rule of an NS-LOCO code $\mathcal{NSC}^4_m$. Recall that $a_i \triangleq \mathcal{L}(c_i)$.

\begin{theorem}\label{thm_nslocorule}
Let $\bold{c}$ be an NS-LOCO codeword in $\mathcal{NSC}^4_m$. The relation between the lexicographic index $g(\bold{c})$ of this codeword and the codeword itself is given by:
\begin{equation}\label{eqn_nslocorule}
g(\bold{c}) = \sum_{i=0}^{m-1} \left [ (a_i - y_{i,1}) N_4(i) + 3y_{i,1} N_4(i-1) \right ],
\end{equation}
where $y_{i,1}$ is specified as follows:
\begin{align}\label{eqn_nslocordef}
y_{i,1} &= 1 \text{ if } c_{i+1} c_i = \alpha^2\beta_2, \text{ } \beta_2 \in \{1, \alpha, \alpha^2\}, \text{ and } y_{i,1} = 0 \text{ otherwise}.
\end{align}
\end{theorem}

\begin{IEEEproof}
The proof operates on the two cases in Step~3 to find the contribution $g_i(c_i)$. The details are left to the interested reader for brevity.
\end{IEEEproof}\vspace{+0.2em}

\textbf{Step~6)} Bridging here differs from \cite{bd_tdmr}. We bridge in NS-LOCO codes with one GF$(4)$ symbol, which is converted eventually to one column of three bits, between each two consecutively written codewords as follows:
\begin{itemize}
\item If the RMS of a codeword and the LMS of the next codeword are both $1$'s, bridge with $\alpha$, i.e., $\psi^2$ or $\psi^3$ in GF$(8)$.
\item If this is not the case, bridge with $1$, i.e., $1$ or $\psi^5$ in GF$(8)$.
\end{itemize}
The mapping-demapping in (\ref{eqn_mapnsloco}) and that in (\ref{eqn_gf8map}) illustrate what is written for bridging. There are two available options for the bridging column of three bits, and an input bit makes the selection in a way similar to what happens with codeword symbols. This bridging is efficient in terms of low added redundancy, and optimal in terms of maximum protection of edge symbols. With our bridging, the maximum number of consecutive $3 \times 1$ columns with no transition after writing via the coding scheme involving an NS-LOCO code $\mathcal{NSC}^4_m$ is $m+1$.

Given our bridging method, the rate of the coding scheme involving an NS-LOCO code $\mathcal{NSC}^4_m$, in input bits per coded symbol, and the normalized rate are:
\begin{equation}\label{eqn_nslocorate}
R^{\textup{sch}}_{\textup{NS-LOCO}} = \frac{s}{m+1} + 1 = \frac{\lfloor N_4(m) \rfloor}{m+1}+1, \textup{ } R^{\textup{sch,n}}_{\textup{NS-LOCO}} = \frac{1}{3} \left [\frac{\lfloor N_4(m) \rfloor}{m+1} + 1 \right ].
\end{equation}
It is easy to deduce that NS-LOCO codes achieve the capacity of an $\mathcal{NS}^4$-constrained code. Encoding and decoding algorithms can be built in a way similar to what is in \cite{bd_tdmr}.

\begin{remark}
While designing NS-LOCO codes, we opted to use simple bridging. It is important to note that it is possible to bridge for NS-LOCO codes with one symbol out of the set $\{1, \alpha\}$ that is picked based on one input bit. Thus, the following notable normalized rate gain can be achieved:
\begin{equation}\label{eqn_nsrgain}
\overline{R}^{\textup{sch,n}}_{\textup{NS-LOCO}} - R^{\textup{sch,n}}_{\textup{NS-LOCO}} = \frac{1}{3(m+1)}.
\end{equation}
Some changes for self-clocking and modifications to the encoding-decoding algorithms will be required.
\end{remark}

\subsection{Near-Optimal Plus LOCO Codes}

\begin{figure}
\vspace{-0.5em}
\center
\includegraphics[trim={2.3in 0.7in 2.3in 0.7in}, width=3.5in]{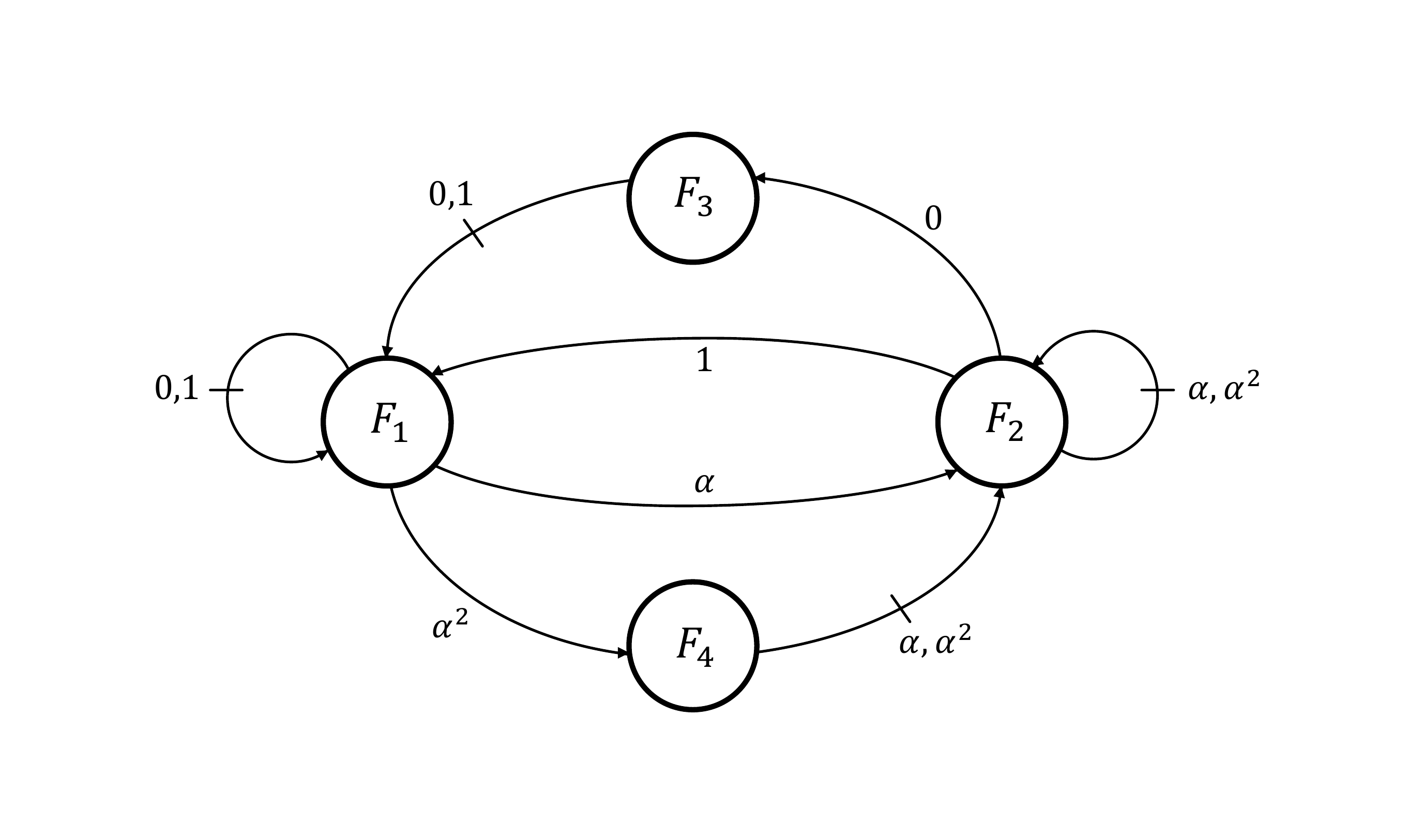}
\vspace{-0.5em}
\caption{An FSTD representing an infinite $\mathcal{NP}^4$-constrained sequence (patterns in $\mathcal{NP}^4$ are prevented).}
\label{fig_6}
\vspace{-0.5em}
\end{figure}

We move on to our near-optimal plus LOCO (NP-LOCO) codes, which are codes preventing the PIS patterns shown in Fig.~\ref{fig_2} (along with other patterns) within each group of three adjacent down tracks. These $32$ PIS patterns map to the $32$ GF$(8)$ patterns in (\ref{eqn_toploco}). We adopt the following GF$(8)$ $\longleftrightarrow$ GF$(4)$ mapping-demapping:
\begin{align}\label{eqn_mapnploco}
\{\psi^3, \psi^4\} &\longleftrightarrow 0, \hspace{+3.77em} \{0, 1\} \longleftrightarrow 1, \nonumber \\
\{\psi^5, \psi^6\} &\longleftrightarrow \alpha, \hspace{+2.8em} \{\psi, \psi^2\} \longleftrightarrow \alpha^2.
\end{align}
As we shall see shortly, this mapping-demapping makes the analysis simpler. Based on this mapping-demapping, an NP-LOCO code should forbid the $8$ patterns $\overline{\beta}_1\alpha^2\beta_1$, for all $\overline{\beta}_1, \beta_1 \in \{0, 1\}$, and $\overline{\beta}_20\beta_2$, for all $\overline{\beta}_2, \beta_2 \in \{\alpha, \alpha^2\}$, which covers the $32$ PIS patterns (and more). The FSTD of an infinite $4$-ary constrained sequence in which these $8$ patterns are prevented is in Fig.~\ref{fig_6}. The corresponding adjacency matrix is:
\begin{gather*}\label{eqn_nplocoadj}
\bold{F}=
\begin{bmatrix}
2 & 1 & 0 & 1\\
1 & 2 & 1 & 0\\
2 & 0 & 0 & 0\\
0 & 2 & 0 & 0
\end{bmatrix}.
\end{gather*}
We care about the capacity of the coding scheme, including the unconstrained part of data to be written. Thus, the capacity $C$, in input bits per coded symbol, and the normalized capacity $C^{\textup{n}}$ are:
\begin{equation}\label{eqn_capnploco}
C = \log_2(\lambda_{\textup{max}}(\bold{F})) + 1 = \log_2 3.5616 + 1 = 2.8325 \textup{ and } C^{\textup{n}} = \frac{1}{3} C = 0.9442,
\end{equation}
which means that the capacity loss compared with the optimal case from (\ref{eqn_capoploco}) is $2.76\%$.

Denote an NP-LOCO code of length $m$ by $\mathcal{NPC}^4_m$. The definition of the code is exactly the definition of a generic LOCO code, which is Definition~\ref{def_genloco}, with $q=4$, $\mathcal{C}^q_m = \mathcal{NPC}^4_m$, and $\mathcal{T}$ given by:
\begin{equation}\label{eqn_tnploco}
\mathcal{T} = \mathcal{NP}^4 \triangleq \{\overline{\beta}_1\alpha^2\beta_1, \overline{\beta}_20\beta_2, \textup{ } \forall \overline{\beta}_1, \beta_1 \in \{0, 1\} \textup{ and } \forall \overline{\beta}_2, \beta_2 \in \{\alpha, \alpha^2\}\}.
\end{equation}
Both $\bold{c}$ in $\mathcal{C}^q_m = \mathcal{NPC}^4_m$ and $g(\bold{c})$ are used as they were in Section~\ref{sec_gen}. The cardinality of $\mathcal{NPC}^4_m$ is $N_q(m) = N_4(m)$.

Now, we will apply the steps of the general method to find out how to encode and decode NP-LOCO codes using a simple encoding-decoding rule.\vspace{+0.7em}

\textbf{Step~1)} Using the patterns in $\mathcal{NP}^4$, we determine initial groups of $\mathcal{NPC}^4_m$ as shown below.
\begin{itemize}
\item For the patterns $0\alpha^2\beta_1$, $\beta_1 \in \{0, 1\}$, there is an initial group having all the codewords starting with $0\alpha^2\beta_2$, $\beta_2 \in \{\alpha, \alpha^2\}$, from the left. There are three more initial groups having all the codewords starting with $0\beta_3$, a group for each $\beta_3 \in \{0, 1, \alpha\}$, from the left. There are three more initial groups having all the codewords starting with non-zero symbols, a group for each $\beta_4 \in \{1, \alpha, \alpha^2\}$, from the left. We do the same for the patterns $1\alpha^2\beta_1$, $\beta_1 \in \{0, 1\}$.
\item For the patterns $\alpha^20\beta_2$, $\beta_2 \in \{\alpha, \alpha^2\}$, there is an initial group having all the codewords starting with $\alpha^20\beta_1$, $\beta_1 \in \{0, 1\}$, from the left. There are three more initial groups having all the codewords starting with $\alpha^2\beta_4$, a group for each $\beta_4 \in \{1, \alpha, \alpha^2\}$, from the left. There are three more initial groups having all the codewords starting with non-$\alpha^2$ symbols, a group for each $\beta_3 \in \{0, 1, \alpha\}$, from the left. We do the same for the patterns $\alpha0\beta_2$, $\beta_2 \in \{\alpha, \alpha^2\}$.
\end{itemize}

After operating on these initial groups, we end up with four (final) groups covering all the NP-LOCO codewords in $\mathcal{NPC}^4_m$: \textit{Group~1}, which contains all the codewords starting with $0$ from the left, \textit{Group~2}, which contains all the codewords starting with $1$ from the left, \textit{Group~3}, which contains all the codewords starting with $\alpha$ from the left, and \textit{Group~4}, which contains all the codewords starting with $\alpha^2$ from the left. The groups are defined for $m \geq 2$.

Group~1 is further partitioned into two subgroups: \textit{Subgroup~1.1}, which contains all the codewords starting with $0\beta_3$ from the left, and \textit{Subgroup~1.2}, which contains all the codewords starting with $0\alpha^2\beta_2$ from the left. The same partitioning to subgroups applies to Group~2. Additionally, Group~4 is further partitioned into two subgroups: \textit{Subgroup~4.1}, which contains all the codewords starting with $\alpha^20\beta_1$ from the left, and \textit{Subgroup~4.2}, which contains all the codewords starting with $\alpha^2\beta_4$ from the left. The same partitioning to subgroups applies to Group~3.\vspace{+0.7em}

\textbf{Step~2)} Theorem~\ref{thm_nplococard} gives the cardinality of an NP-LOCO code.

\begin{theorem}\label{thm_nplococard}
The cardinality of an NP-LOCO code $\mathcal{NPC}^4_m$ is given by:
\begin{equation}\label{eqn_nplococard}
N_4(m) = 3N_4(m-1) + 2N_4(m-2), \text{ } m \geq 2,
\end{equation}
where the defined cardinalities are:
\begin{equation}\label{eqn_nplococdef}
N_4(0) \triangleq 2 \text{ and } N_4(1) \triangleq 4.
\end{equation}
\end{theorem}

\begin{IEEEproof}
We note that an NP-LOCO code $\mathcal{NPC}^4_m$ is symmetric because of the nature of forbidden patterns in $\mathcal{NP}^4$. Thus, we only derive a recursive cardinality formula for Group~1 of $\mathcal{NPC}^4_m$, then multiply by $4$. We work on Group~1 of $\mathcal{NPC}^4_m$.

As for Subgroup~1.1, each codeword starting with $0\beta_3$ from the left in this subgroup corresponds to a codeword in $\mathcal{NPC}^4_{m-1}$ that starts with the same $\beta_3$ from the left such that they share the remaining $m-2$ RMSs. This correspondence is bijective. Since $\beta_3$ is in $\{0, 1, \alpha\}$ and the code $\mathcal{NPC}^4_{m-1}$ is symmetric, the cardinality of Subgroup~1.1 is:
\begin{equation}\label{eqn_nplocogr1}
N_{4,1.1}(m) = \frac{3}{4} N_4(m-1).
\end{equation}

As for Subgroup~1.2, each codeword starting with $0\alpha^2\beta_2$ from the left in this subgroup corresponds to a codeword in $\mathcal{NPC}^4_{m-2}$ that starts with the same $\beta_2$ from the left such that they share the remaining $m-3$ RMSs. This correspondence is bijective. Since $\beta_2$ is in $\{\alpha, \alpha^2\}$ and the code $\mathcal{NPC}^4_{m-2}$ is symmetric, the cardinality of Subgroup~1.2 is:
\vspace{-0.1em}\begin{equation}\label{eqn_nplocogr2}
N_{4,1.2}(m) = \frac{2}{4} N_4(m-2) = \frac{1}{2} N_4(m-2).
\end{equation}

Using (\ref{eqn_nplocogr1}) and (\ref{eqn_nplocogr2}), the cardinality of Group~1 in $\mathcal{NPC}^4_m$ then is:
\begin{equation}\label{eqn_nplocogr3}
N_{4,1}(m) = \sum_{i=1}^2 N_{4,1.i}(m) = \frac{3}{4} N_4(m-1) + \frac{1}{2} N_4(m-2).
\end{equation}
From (\ref{eqn_nplocogr3}) and using the symmetry of the code, the cardinality of $\mathcal{NPC}^4_m$ is:
\begin{equation}\label{eqn_nplocogr4}
N_4(m) = \sum_{i=1}^4 N_{4,i}(m) = 4 N_{4,1}(m) = 3N_4(m-1) + 2N_4(m-2), \textup{ } m \geq 2. \nonumber
\end{equation}

As for the defined cardinalities, it is clear that $N_4(1) \triangleq 4$. We also know that $N_4(2) = 4^2 = 16$. Consequently, and using the proved (\ref{eqn_nplococard}),
\begin{equation}\label{eqn_nplocogr5}
16 = 3 \times 4 + 2N_4(0) \implies N_4(0) \triangleq 2.
\end{equation}
Computing the defined cardinalities completes the proof.
\end{IEEEproof}\vspace{+0.7em}

\textbf{Step~3)} We now specify the special cases. Using the patterns in $\mathcal{NP}^4$, we determine initial special cases for the NP-LOCO code $\mathcal{NPC}^4_m$ as shown below.
\begin{itemize}
\item For the patterns $0\alpha^2\beta_1$, $\beta_1 \in \{0, 1\}$, the only initial special case is $c_{i+2} c_{i+1} c_i = 0\alpha^2\beta_2$, $\beta_2 \in \{\alpha, \alpha^2\}$. We do the same for the patterns $1\alpha^2\beta_1$, $\beta_1 \in \{0, 1\}$.
\item For the patterns $\alpha^20\beta_2$, $\beta_2 \in \{\alpha, \alpha^2\}$, the only initial special case is $c_{i+1} c_i = \alpha^2\beta_4$, $\beta_4 \in \{1, \alpha, \alpha^2\}$. We do the same for the patterns $\alpha0\beta_2$, $\beta_2 \in \{\alpha, \alpha^2\}$.
\end{itemize}

We end up with three (final) cases for $c_i$ based on $c_i$ and its preceding symbols: a \textit{special case} for $c_{i+2} c_{i+1} c_i = \overline{\beta}_1\alpha^2\beta_2$, a \textit{special case} for $c_{i+1} c_i = \overline{\beta}_2\beta_4$, and the \textit{typical case}. Recall that $\overline{\beta}_1 \in \{0, 1\}$ and $\overline{\beta}_2 \in \{\alpha, \alpha^2\}$, while the rest of variables are specified above. The typical case is simply the case when neither of the two special cases is enabled and $c_i \neq 0$. As usual, the priority of a case increases as its sequence length increases.\vspace{+0.7em}

\textbf{Steps~4 and 5)} Theorem~\ref{thm_nplocorule} gives the encoding-decoding rule of an NP-LOCO code $\mathcal{NPC}^4_m$. Recall that $a_i \triangleq \mathcal{L}(c_i)$.

\begin{theorem}\label{thm_nplocorule}
Let $\bold{c}$ be an NP-LOCO codeword in $\mathcal{NPC}^4_m$. The relation between the lexicographic index $g(\bold{c})$ of this codeword and the codeword itself is given by:
\begin{equation}\label{eqn_nplocorule}
g(\bold{c}) = \sum_{i=0}^{m-1} \left [ \frac{1}{4} (a_i -2y_{i,1} - y_{i,2}) N_4(i+1) + \frac{1}{2}y_{i,2} N_4(i) \right ],
\end{equation}
where $y_{i,1}$ and $y_{i,2}$ are specified as follows:
\begin{align}\label{eqn_nplocordef}
y_{i,1} &= 1 \text{ if } c_{i+2} c_{i+1} c_i = \overline{\beta}_1\alpha^2\beta_2, \text{ } \overline{\beta}_1 \in \{0, 1\},\beta_2 \in \{\alpha, \alpha^2\}, \text{ and } y_{i,1} = 0 \text{ otherwise}, \nonumber \\
y_{i,2} &= 1 \text{ if } c_{i+1} c_i = \overline{\beta}_2\beta_4, \text{ } \overline{\beta}_2 \in \{\alpha, \alpha^2\},\beta_4 \in \{1, \alpha, \alpha^2\} \textup{ s.t. } y_{i,1} = 0, \text{ and } y_{i,2} = 0 \text{ otherwise}.
\end{align}
\end{theorem}

\begin{IEEEproof}
First, we perform Step~4 of the method. We aim at computing the contribution of each NP-LOCO codeword symbol $c_i$ to the codeword index $g(\bold{c})$ for the three final cases, i.e., $g_{i,i_{\textup{c}}}(c_i)$ for all $i_{\textup{c}}$.

We start off with the typical case, which we index by $i_{\textup{c}} = 1$. The contribution of $c_i$ to $g(\bold{c})$ in this case is the number of codewords in $\mathcal{NPC}^4_m$ starting with $c_{m-1} c_{m-2} \dots c_{i+1} c'_i$ from the left such that $c'_i < c_i$. This number is the number of codewords in $\mathcal{NPC}^4_{i+1}$ starting with $c'_i$, for all $c'_i < c_i$, from the left. Thus, and using symmetry, we can write $g_{i,1}(c_i)$ as:
\begin{equation}\label{eqn_nplocorule1}
g_{i,1}(c_i) = \sum_{j=1}^{a_i} N_{4,j}(i+1) = a_i N_{4,1}(i+1) = \frac{1}{4} a_i N_4(i+1).
\end{equation}

Next, we study the special case characterized by $c_{i+2} c_{i+1} c_i = \overline{\beta}_1\alpha^2\beta_2$, $\overline{\beta}_1 \in \{0, 1\}$ and $\beta_2 \in \{\alpha, \alpha^2\}$, which we index by $i_{\textup{c}} = 2$. The contribution of $c_i$ to $g(\bold{c})$ in this case is the number of codewords in $\mathcal{NPC}^4_m$ starting with $c_{m-1} c_{m-2} \dots c_{i+3} \overline{\beta}_1\alpha^2 c'_i$ from the left such that $c'_i < c_i = \beta_2$. This number is the number of codewords in $\mathcal{NPC}^4_{i+1}$ starting with $c'_i$, for all $c'_i < c_i$ such that $c'_i \notin \{0, 1\}$, from the left. Thus, and using symmetry, we can derive $g_{i,2}(c_i)$ as follows:
\begin{equation}\label{eqn_nplocorule2}
g_{i,2}(c_i) = \sum_{j=1}^{a_i-2} N_{4,j}(i+1) =  \frac{1}{4} (a_i-2) N_4(i+1).
\end{equation}

Next, we study the special case characterized by $c_{i+1} c_i = \overline{\beta}_2\beta_4$, $\overline{\beta}_2 \in \{\alpha, \alpha^2\}$ and $\beta_4 \in \{1, \alpha, \alpha^2\}$, which we index by $i_{\textup{c}} = 3$. The contribution of $c_i$ to $g(\bold{c})$ in this case is the number of codewords in $\mathcal{NPC}^4_m$ starting with $c_{m-1} c_{m-2} \dots c_{i+2} \overline{\beta}_2 c'_i$ from the left such that $c'_i < c_i = \beta_4$. Looking from the right, these codewords correspond to codewords in $\mathcal{NPC}^4_{i+1}$. We divide such codewords in $\mathcal{NPC}^4_{i+1}$ into two portions. The first portion has the codewords in $\mathcal{NPC}^4_{i+1}$ starting with $c'_i$, for all $c'_i < c_i$ such that $c'_i \neq 0$, from the left. Let the number of codewords in this portion be $g'_{i,3}(c_i)$. Thus, and using symmetry, we can derive $g'_{i,3}(c_i)$ as follows:
\begin{equation}\label{eqn_nplocorule3}
g'_{i,3}(c_i) = \sum_{j=1}^{a_i-1} N_{4,j}(i+1) =  \frac{1}{4} (a_i-1) N_4(i+1).
\end{equation}
The second portion has the codewords in $\mathcal{NPC}^4_{i+1}$ starting with $c'_i = 0$ from the left. Let the number of codewords in this portion be $g''_{i,3}(c_i)$. From the set of forbidden patterns $\mathcal{NP}^4$, we know that $\overline{\beta}_20$ in an NP-LOCO codeword has to be followed by $\beta_1 \in \{0, 1\}$. Thus, and aided by (\ref{eqn_nplocogr2}), we can derive $g''_{i,3}(c_i)$ as follows:
\begin{equation}\label{eqn_nplocorule4}
g''_{i,3}(c_i) = \frac{2}{4} N_4(i) = \frac{1}{2} N_4(i).
\end{equation}
Using (\ref{eqn_nplocorule3}) and (\ref{eqn_nplocorule4}), we get:
\begin{equation}\label{eqn_nplocorule5}
g_{i,3}(c_i) = g'_{i,3}(c_i) + g''_{i,3}(c_i) = \frac{1}{4} (a_i-1) N_4(i+1) + \frac{1}{2} N_4(i).
\end{equation}\vspace{+0.2em}

Now, we are ready to perform Step~5 of the method. We want to combine the different contributions for all cases into one expression, which is the NP-LOCO encoding-decoding rule.

We need only two merging variables: $y_{i,1}$, for the case indexed by $i_{\textup{c}} = 2$, and $y_{i,2}$, for the case indexed by $i_{\textup{c}} = 3$ (lower priority). If the two variables are zeros, the typical case contribution is switched on.

Now, we pick the merging function $f^{\textup{mer}}_0 (\cdot) = \frac{1}{4} (a_i -2y_{i,1} - y_{i,2})$ for $N_4(i+1)$. We also pick the merging function $f^{\textup{mer}}_1 (\cdot) = \frac{1}{2}y_{i,2}$ for $N_4(i)$. Observe that the values of these merging functions at different cases are quite consistent with (\ref{eqn_nplocorule1}), (\ref{eqn_nplocorule2}), and (\ref{eqn_nplocorule5}). Observe also that if $c_i = 0$, this means $a_i = y_{i,1} = y_{i,2} = 0$, which in turn means $f^{\textup{mer}}_0 (\cdot) = f^{\textup{mer}}_1 (\cdot) = 0$.

Using these two merging functions, the unified expression representing the contribution of a symbol $c_i$ to the codeword index $g(\bold{c})$ can be written as:
\begin{align}\label{eqn_nplocorule6}
g_i(c_i) &= f^{\textup{mer}}_0 (\cdot) N_4(i+1) + f^{\textup{mer}}_1 (\cdot) N_4(i) \nonumber \\
&= \frac{1}{4} (a_i -2y_{i,1} - y_{i,2}) N_4(i+1) + \frac{1}{2}y_{i,2} N_4(i).
\end{align}
The encoding-decoding rule (\ref{eqn_nplocorule}) of an NP-LOCO code follows from (\ref{eqn_nplocorule6}).
\end{IEEEproof}\vspace{+0.7em}

\textbf{Step~6)} We bridge in NP-LOCO codes with one GF$(4)$ symbol, which is converted eventually to one column of three bits, between each two consecutively written codewords as follows:
\begin{itemize}
\item If the $\textup{RMSs} - \textup{LMSs}$ are $\beta_1\alpha^2 - 0\beta_2$, bridge with one no-writing symbol $z$, i.e., one $3 \times 1$ column with no writing.
\item \textit{Else if the $\textup{RMS(s)} - \textup{LMS(s)}$ are $\beta_1\alpha^2 - 0\beta_1$, $\alpha^2 - \beta_4$, $\beta_2\alpha^2 - 0\beta_1$, or $1 - 1$, bridge with $\alpha$, i.e., $\psi^5$ or $\psi^6$ in GF$(8)$.}
\item Else if the $\textup{RMS(s)} - \textup{LMS(s)}$ are $\beta_2\alpha^2 - 0\beta_2$, $0 - \beta_3$, or $\alpha - \alpha$, bridge with $1$, i.e., $0$ or $1$ in GF$(8)$.
\item For any other scenario, bridge with $\alpha$, i.e., $\psi^5$ or $\psi^6$ in GF$(8)$.
\end{itemize}
The second item above in italic could be removed as it is included it in the last item. The mapping-demapping in (\ref{eqn_mapnploco}) and that in (\ref{eqn_gf8map}) illustrate what is written for bridging. There are two available options for the bridging column of three bits except for the first case above. In (\ref{eqn_mapnploco}), the left (resp., right) symbol is picked if the input bit is $0$ (resp., $1$). This bridging is optimal in terms of minimum added redundancy, and it offers near-maximum protection of edge symbols. With our bridging, the maximum number of consecutive $3 \times 1$ columns with no transition after writing via the coding scheme involving an NP-LOCO code $\mathcal{NPC}^4_m$ is $m$. For this maximum to be achieved, the $m$ additional bits in a relevant chunk of size $s+m$ bits should be all $0$'s or all $1$'s.

Given our bridging method, the rate of the coding scheme involving an NP-LOCO code $\mathcal{NPC}^4_m$, in input bits per coded symbol, and the normalized rate are:
\begin{equation}\label{eqn_nplocorate}
R^{\textup{sch}}_{\textup{NP-LOCO}} = \frac{s+m}{m+1} = \frac{\lfloor N_4(m) \rfloor + m}{m+1}, \textup{ } R^{\textup{sch,n}}_{\textup{NP-LOCO}} = \frac{\lfloor N_4(m) \rfloor + m}{3(m+1)}.
\end{equation}
It is easy to deduce that NP-LOCO codes achieve the capacity of an $\mathcal{NP}^4$-constrained code. Encoding and decoding algorithms can be built as shown in previous sections, and they are omitted for brevity.

\begin{example}\label{example_5}
This example illustrates the decoding process of coding scheme adopting an NP-LOCO code. Consider a scheme adopting the NP-LOCO code $\mathcal{NPC}^4_6$ ($m=6$). Using (\ref{eqn_nplococard}) and (\ref{eqn_nplococdef}), we get $N_4(0) \triangleq 2$, $N_4(1) \triangleq 4$, $N_4(2) = 16$, $N_4(3) = 56$, $N_4(4) = 200$, $N_4(5) = 712$, and $N_4(6) = 2536$. Consider the following read sequence after hard decision and $8$-ary conversion $0\psi\psi^6\psi^21\psi^5$ (level-equivalent $027316$). The bridging column is ignored in NP-LOCO codes, and we assume the sequence received is error-free. The first step in decoding is to decide the input bits used for selection along with the written $4$-ary codeword. Using (\ref{eqn_mapnploco}), we can deduce that the selection bits for $0\psi\psi^6\psi^21\psi^5$ are $001110$, and that the written codeword is $\bold{c} = c_5 c_4 c_3 c_2 c_1 c_0 = 1\alpha^2\alpha\alpha^21\alpha$ in $\mathcal{NPC}^4_6$. The case indexed by $i_{\textup{c}} = 1$ applies for $c_5$, $c_4$, and $c_0$. The case indexed by $i_{\textup{c}} = 2$ applies for $c_3$. The case indexed by $i_{\textup{c}} = 3$ applies for $c_2$ and $c_1$. Consequently, and using (\ref{eqn_oplocorule}), we get:
\begin{align}
g(\bold{c}=1\alpha^2\alpha\alpha^21\alpha) &= \left [ \frac{1}{4} \times N_4(6) \right ] + \left [ \frac{1}{4} \times 3N_4(5) \right ] + \left [ \frac{1}{4} \times 0 \times N_4(4) \right ] + \left [ \frac{1}{4}\times 2N_4(3) + \frac{1}{2} N_4(2) \right ]  \nonumber \\
&\hspace{+1.0em}+ \left [ \frac{1}{4}\times 0 \times N_4(2) + \frac{1}{2} N_4(1) \right ] + \left [ \frac{1}{4} \times 2N_4(1) \right ] \nonumber \\
&= 634 + 534 + 0 + 36 + 2 + 2 = 1208, \nonumber
\end{align}
which is consistent with the codeword index produced by the program we wrote to exhaustively generate and lexicographically order all NP-LOCO codewords in $\mathcal{NPC}^4_m$, with $m=6$ here. It corresponds to the binary message $10010111000$ ($s = 11$). Thus, the final decoded binary input stream is $10010111000\textup{ }001110$.
\end{example}

To demonstrate how near-optimal codes can be used to further reduce complexity and error propagation, we give an example. Suppose that the required normalized rate is around $0.88$, and we aim at eliminating PIS patterns in a TDMR system. The OP-LOCO code with $m = 10$ and $s = 29$ achieves a normalized rate of $0.8788$. On the other hand, a coding scheme adopting the NP-LOCO code with $m = 13$ and $s = 24$ achieves a normalized rate of $0.8810$ for the scheme. Thus, the scheme adopting the NP-LOCO code achieves approximately the same rate at a reduced adder size, i.e., reduced complexity and reduced error propagation, compared with the OP-LOCO code. Schemes adopting near-optimal codes lose this advantage as the rates get higher, and they have a gap to capacity, which justifies why we present both optimal and near-optimal codes.

\begin{remark}
While our general method allows the code designer to build a LOCO code for any finite set of forbidden patterns, the mapping-demapping used between binary and $q$-ary forbidden patterns may be used to simplify the analysis when applicable. One example has already been given in the analysis of our NP-LOCO codes. Another example is for our NS-LOCO codes: we can change the mapping-demapping in (\ref{eqn_mapnsloco}) such that the forbidden pattern is $0\alpha^20$ while keeping the mapping-demapping in (\ref{eqn_gf8map}) as it is. In this case, the analysis becomes simpler, and the encoding-decoding rule of the NS-LOCO code becomes:
\begin{equation}\label{eqn_nslocorule2}
g(\bold{c}) = \sum_{i=0}^{m-1} (a_i - y_{i,1}) N_4(i),
\end{equation}
where $N_4(i)$ is obtained recursively using (\ref{eqn_nslococard}), and $y_{i,1}$ is specified as follows:
\begin{align}\label{eqn_nslocordef2}
y_{i,1} &= 1 \text{ if } c_{i+2} c_{i+1} c_i = 0\alpha^2\beta_2, \text{ } \beta_2 \in \{1, \alpha, \alpha^2\}, \text{ and } y_{i,1} = 0 \text{ otherwise}.
\end{align}
However, we do not alter the mapping-demapping in certain cases either to demonstrate the strength of the general method or for consistency with prior work like \cite{bd_tdmr}. This has no effect on the complexity of the encoding-decoding algorithms.
\end{remark}\vspace{+0.2em}

Given the promising results presented in this paper regarding applying novel LOCO codes in TDMR systems, one interesting future direction is pairing efficient multi-dimensional constrained codes with high performance multi-dimensional graph-based (LDPC) codes \cite{ahh_md} in modern storage devices to further increase density and lifetime gains. Observe that all the new LOCO codes presented in this paper are reconfigurable.

\section{Conclusion}\label{sec_conc}

We introduced a general method to systematically design constrained codes based on lexicographic indexing, collectively named LOCO codes. The method reveals the secret arithmetic of forbidden/allowed patterns in these constrained codes. In particular, it starts from the finite set of forbidden patterns to find the cardinality of the code recursively and derive an encoding-decoding rule that links the index to the codeword. We gave two examples from the literature to show how the general method works. We used the general method to design optimal constrained codes preventing isolation patterns in TDMR systems, named OS-LOCO and OP-LOCO codes. OS-LOCO and OP-LOCO codes are capacity achieving, are simple, and they notably improve performance with very limited redundancy. We applied OP-LOCO codes to a practical TDMR system, and demonstrated significant FER and BER gains even though no error-correcting code was applied. We introduced coding schemes adopting near-optimal codes that can be used in TDMR systems to prevent isolation patterns and further reduce complexity. We suggest that our general method will be a tool to support the evolution of modern, multi-dimensional magnetic and electronic storage systems. Moreover, our method can also be valuable to various data transmission systems.

\section*{Acknowledgment}\label{sec_ack}

We would like to thank Mohsen Bahrami and Prof. Bane Vasic for providing the TDMR model that we modified and used to generate the results in Section~\ref{sec_gains}.


\end{document}